\documentclass[12pt]{article}

\usepackage{amsmath,amssymb,graphicx,color} %use drftcite in draftmode otherwise
\usepackage{epsf}

\newcommand{\Tr}{\mathrm{Tr~}}

\newcommand{\et}{{\it et al.}}
\newcommand{\eg}{{\it e.g.}}

\renewcommand{\theequation}{\thesection.\arabic{equation}}

\def\to{\rightarrow}

\newcommand\as{\alpha_{\mathrm{S}}}

\def\beq{\begin{equation}}
\def\eeq{\end{equation}}
\def\beeq{\begin{eqnarray}}
\def\eeeq{\end{eqnarray}}
\def\be{\begin{eqnarray}}
\def\ee{\end{eqnarray}}
\def\beal{\begin{align}}
\def\eeal{\end{align}}

\newcommand{\la}{\langle}
\newcommand{\ra}{\rangle} 
\def\nn{\nonumber}
\def\b0{b_0}

 \def\b{\beta}  \def\d{\delta}
\def\e{\epsilon} \def\f{\phi} \def\g{\gamma} \def\h{\eta}
   \def\l{\lambda} \def\m{\mu}
  \def\p{\pi}  \def\r{\rho}
\def\s{\sigma}   \def\x{\xi} 
\def\D{\Delta} \def\F{\Phi}

\def\w{\omega}

     \def\cm{{\cal M}} 
\def\co{{\cal O}} \def\cp{{\cal P}}

\def\cm{{\cal M}}

\def\NLO{next-to-leading order}

\def\ID{1 \kern -.45 em 1}

\def\pt{{p_{T}}}
\def\ptmin{{p_{T}^{min}}}
\def\ptr{p_{T} R}
\def\ttbar{t\bar{t}}
\def\gev{\, \mbox{GeV}}
\def\tev{\, \mbox{TeV}}

\def\invfb{\mbox{fb}^{-1}}
\def\massmin{140 \, \mbox{GeV}}
\def\massmax{210 \, \mbox{GeV}}

\def\slap#1#2{\setbox0=\hbox{$#1{#2}$}
    #2\kern-\wd0{\hbox to\wd0{\hfil$#1{/}$\hfil}}}

\begin{document}
\begin{titlepage}

\vskip -1.5cm 
\vspace*{-1.7cm} 
\hfill$\vcenter{
\hbox{\footnotesize \hspace*{2.5cm}   YITP-SB-08-37; WIS/17/08-SEPT-DPP} }$
%\begin{flushright}
%\today
%\end{flushright}

\vskip 1.9365cm

\begin{center}
{\huge \bf  Top Jets at the LHC }
\vskip.42cm
\end{center}

\begin{center}
{\bf {Leandro G. Almeida}$^a$, {Seung J. Lee}$^{a,b}$, {Gilad Perez}$^{a,b}$, \\ {Ilmo Sung}$^a$ and {Joseph Virzi}$^c$} \\
\end{center}
\vskip 8pt

\begin{center}
{\footnotesize $^{a}$ {\it C. N.Yang Institute for Theoretical Physics, Stony Brook University,\\
Stony Brook, NY 11794-3840, USA}}\\

\vspace*{0.3cm}

{\footnotesize $^{b}$ {\it Department of Particle Physics,  Weizmann Institute of Science, Rehovot 76100, Israel}}\\

\vspace*{0.3cm}

{\footnotesize $^c$ {\it University of California; LBNL,
Physics Division, 1 Cyclotron Rd., Berkeley, CA 94720, USA}}
%
%{\tt  lalmeida@insti.physics.sunysb.edu, slee@max2.sunysb.edu, sung@insti.physics.sunysb.edu, gilad@insti.physics.sunysb.edu, sterman@insti.physics.sunysb.edu, jvirz@lbl.gov}

\end{center}

\vglue 0.3truecm

\begin{abstract}
\vskip 3pt \noindent

We investigate the reconstruction of high $p_T$ hadronically-decaying top quarks at the Large Hadron Collider.
One of the main challenges in identifying energetic top quarks is that the decay products become increasingly collimated.
This reduces the efficacy of conventional reconstruction methods that exploit the topology of the top quark decay chain.
We focus on the cases where the decay products of the top quark are reconstructed as a single jet, a ``top-jet".
The most basic ``top-tag" method based on jet mass measurement is considered in detail.
To analyze the feasibility of the top-tagging method,
both theoretical and experimental aspects of the large QCD jet background contribution are examined.
Based on a factorization approach, 
we derive a simple analytic approximation for the shape of the QCD jet mass spectrum.
We observe very good agreement with the Monte Carlo simulation.
We consider high-$p_T$ $t \bar{t}$ production in the Standard Model as an example,
and show that our theoretical QCD jet mass distributions can efficiently characterize the background via sideband analyses.
We show that with 25~fb$^{-1}$ of data, our approach allows us to resolve top-jets with $p_T$ $\ge$ 1 TeV, from the QCD background,
and about 1.5~TeV top-jets with 100~fb$^{-1}$, without relying on $b$-tagging.
To further improve the significance we consider jet shapes (recently analyzed in 0807.0234 [hep-ph]), 
which resolve the substructure of energy flow inside cone jets.
A method of measuring the top quark polarization by using the transverse momentum of the bottom quark is also presented.
The main advantages of our approach are: 
(i) the mass distributions are driven by first principle calculations, instead of relying solely on Monte Carlo simulation;
(ii) for high $p_T$ jets ($p_T\ge$1~TeV), IR-safe jet shape variables are robust against detector resolution effects.
Our analysis can be applied to other boosted massive particles such as the electroweak gauge bosons and the Higgs.
\end{abstract}

\end{titlepage}

\newpage

%%%%%%%%%%%%%%%%%%%%%%%%%%%%%%%%%%%%%%%%%%%%%%%%%%%%%%%%%%%%%%%%%%%%%%%%%%%%%
%%%%%%%%%%%%%%%%%%%%%%%%%%%%%%%%%%%%%%%%%%%%%%%%%%%%%%%%%%%%%%%%%%%%%%%%%%%%%
%%%%%%%%%%%%%%%%%%%%%%%%%%%%%%%%%%%%%%%%%%%%%%%%%%%%%%%%%%%%%%%%%%%%%%%%%%%%%

\section{Introduction}
\label{intro} \setcounter{equation}{0} \setcounter{footnote}{3}

The Large Hadron Collider (LHC) is expected to uncover some of the most interesting mysteries of nature. 
We expect to probe the underlying principles of electroweak symmetry breaking (EWSB)
and what stabilizes the weak scale against radiative corrections from unknown microscopic dynamics.
Due to its large mass, the top quark induces the most severe contributions to the Higgs quadratic divergence.
Furthermore, in almost every known natural model of EWSB, the top sector plays a crucial role in breaking the EW symmetry.
Thus, the top sector might hold a key to a new physics (NP) discovery. 
Many interesting models of EWSB predict new particles with mass $\sim$ TeV scale. 
In several known examples, the new particles decay into highly boosted top quark pairs ($pp \to X \to t\bar{t}$), 
or other decay chains containing a single top quark ($pp \to X \to t\,Y$).
%Other examples include cascade decay chains involving heavy supersymmetric top quarks (stops) or TeV scale resonances
%decaying preferentially to third generation quarks.
In addition, the Standard Model (SM) predicts that the LHC will produce more than $10^5$ top quarks with $\pt \ge 1 \tev$,
significantly enhancing our ability to study high $\pt$ tops and resolving beyond the SM dynamics.

Top quarks decay dominantly into hadronic final states ($t \rightarrow bW \rightarrow bq\bar q$) with a branching ratio $\sim 2/3$, 
providing potentially enhanced statistics.
In the present work, we focus on highly boosted top quarks (decaying through the hadronic channel), and on the dominant QCD jet background.
We refer to a top quark that decays hadronically as a {\it hadronic top}.
For moderately boosted top quarks ($\pt \sim 500$ GeV), 
conventional top quark reconstruction methods, which exploit the decay chain topology, remain adequately efficient (see {\it e.g.}~\cite{TDRs}).
As the top quark $\pt$ approaches $1 \tev$, the situation significantly changes~\cite{Agashe:2006hk,Fitzpatrick:2007qr,Baur:2007ck,LHCnotes}.\footnote{For earlier works in the case of boosted EW bosons see also~\cite{Butterworth:2002tt}.}
The average separation of the top quark decay products approaches the limits of reliable jet reconstruction (cone size $R \sim 0.4$),
and starts to encroach upon the detector resolution ($R \sim 0.1$).
As a result, the efficiency of conventional reconstruction methods drops quickly.
The performance of $b$-tagging and light jet rejection is expected to drop substantially in this kinematic regime.
At present, there is very little published data on $b$-tagging at $\pt \sim 1 \tev$~\cite{BTagging}.
We perform our analysis without accounting for the possible benefits of $b$-jet identification.\footnote{The possibility of $b$-tagging jets,
when the top quark reconstructs to 2 (or more) jets, one of which has a mass $\sim M_W$ and the sum of the two jets has a mass $\sim m_t$, is outside the main focus of this paper.}

We turn our focus away from this family of ``conventional" reconstruction methods. 
We examine the situation where the decay products of at least one top quark are reconstructed as a single jet, or {\it top-jet}.
In semileptonic $\ttbar$ events, for example, the leptonic top may still be reconstructed via semi-conventional reconstruction methods, 
giving up on lepton isolation cuts~\cite{Agashe:2006hk}, see also~\cite{Baur:2007ck,Baur:2008uv}. These methods call for further extensive study due to expected reducible backgrounds
and uncertainties related to the ability to measure the collimated semi-leptonic top mass (dileptonic $\ttbar$ events are also analyzed in~\cite{Bai:2008sk}).
Hadronic top, on the other hand, will give rise to a top-jet.
There will still be some small, but non-negligible, number of $\ttbar$ events where one of the top quarks reconstructs as a top-jet,
but the other top quark can be reconstructed via conventional methods (or semi-conventional methods where one of the tops is manifested as a two-jet object).
In this paper, we focus on the top-jet itself as a means of identifying $\ttbar$ events.
The main reasoning behind that is as follows:
\begin{itemize}
\item[(i)] We find that for $\pt > 1\,$TeV the majority of hadronic-tops are manifested as top-jets, even for cone size as small as $R=0.4$. Thus, it is clear
that our tools will be applicable  for a wide range of top momenta.
\item[(ii)] The distributions and shapes of both background and signal can be understood via first principle calculations as shown in this study and in Ref.~\cite{us}.
It may allow for a cleaner analysis, in the sense that a more direct contact between actual data (expected to arrive soon) and the microscopical theory can be made.
\end{itemize}

Apart from substructure, to leading order, top jets provide four pieces of information, namely its energy, two angles and mass (just as any QCD jet, ignoring the possibility of $b$-tagging).
Without a mass cut, the QCD jet background swamps the hadronic top signal by orders of magnitude.
The most basic tagging method after giving up conventional methods is to use the jet mass 
as a discriminator between the QCD background and the hadronic top signal;
the high-$\pt$ top-jet mass distribution should peak around the top mass while the QCD jet mass distribution peaks near zero.
However, using a jet mass as a discriminator is more complicated for several reasons. 
Due to radiation, QCD jets acquire a large tail in the mass distribution.
The cross section for acquiring large jet mass, for example near the top mass, increases substantially with $\pt$ and cone size.
Top-jets also broaden due to radiation, hardening their jet mass distribution.\footnote{
For a detailed recent study see~\cite{tscet} and references therein.} 
Furthermore, a finite jet reconstruction cone size will not always capture all the daughters of the top quark decay chain,
thus softening its mass distribution.
The net effect is a smearing of the expected naive, broadened, Breit-Wigner distribution for the top jet mass distribution.
Detector effects further smear the distribution, making the above idealized description unrealistic.

Nevertheless, jet mass cuts should retain some rejection power against the QCD background~\cite{
Skiba:2007fw, Holdom:2007ap, Agashe:2007ki}.
Our study addresses this issue in both quantitative and qualitative manner, 
by considering the experimental and theoretical aspects of the analysis.
On the theoretical front, based on a factorization approach, 
we derive a simple approximation for the shape of the QCD jet mass spectrum.
We demonstrate that there is good agreement between our simple analytic predictions and Monte Carlo (MC) results. 
We are able to compute from first principle various features related to a jet mass cut.
We evaluate its significance in the form of a semi-analytical expression for the rejection power and show that it is independent of pseudorapidity.
We provide a quantitative study of the distribution of the signal and background, 
via MadGraph/MadEvent~\cite{Maltoni:2002qb,Stelzer:1994ta, Alwall:2007st}(MG/ME) and Sherpa~\cite{Sherpa}.
We consider the detector resolution by using transfer functions~\cite{JoeVirziAtlas}, 
smearing jets according to a profile obtained from full Geant4 Atlas simulation.
Transfer functions provide a versatile mechanism to explore such effects as shifts in jet energy scale (JES), etc.

We apply the results of our studies to analyze boosted SM top quark pair production, an important discovery channel for 
NP~\cite{Agashe:2006hk, Fitzpatrick:2007qr, Baur:2007ck, Baur:2008uv, Frederix:2007gi, Barger:2006hm, Han:2008gy,gravitons,Han:2008xb}. 
To put results into perspective, we use both $25 \, \invfb $ and $100 \, \invfb $ of integrated luminosity as reference luminosities.
At this time, these correspond to many years of data taking.
We show that using single- and double-tagging methods with our jet functions (defined below) to analyze jet mass distributions,
we can significantly separate the Standard Model $\ttbar$ signal from the QCD background.
Our theoretical QCD jet mass distributions can efficiently characterize the background via sideband analyses.
With 25~fb$^{-1}$ of data, our approach allows us to resolve 1~TeV top-jets from the QCD background,
and about 1.5~TeV top-jets with 100~fb$^{-1}$, if we exploit the kinematics of the so-called ``away'' side of the event, without relying on $b$-tagging. 
The essence of the away side mass cut is that it preferentially keeps the $\ttbar$ signal over the background.
We analyze the mass distribution in more detailed manner, as simple counting methods are inadequate.
As described above, the $\ttbar$ signal is expected to exhibit pronounced structure near the top quark mass.
In order to resolve this ``peak" against the QCD background,
we need to understand the shape of both the $\ttbar$ signal and the QCD background.
%In principle, we have a theoretical handle on the shape of the $\ttbar$ signal, 
%but for the purpose at hand, it suffices to use MC to provide this shape.
To characterize the background we perform a sideband analysis to reduce contamination by the signal.
Our theory-driven ansatz for the QCD background is an admixture of quark- and gluon-jet functions,
the coefficients of which we analyze by fitting in the sidebands (outside the top mass window).
We interpolate the results of the fit into the top mass window $\left( \massmin \le m_J \le \massmax \right)$.
Armed with {\it{shapes}} for the signal and background, we fit them into the data to obtain the normalization constants.
These normalization constants are the magnitude of the signal and background.
The errors associated with the normalization provides a measure of the significance of the measurement.

To further improve the significance we consider jet shapes~\cite{us}, which resolve substructure
of energy flow inside cone jets.
In a companion paper~\cite{us}, we explore the possibility that, requiring a large jet mass, perturbative predictions
for jet shapes differ between jets that originate from the decay of heavy particles, and those
which result from the showering of light quarks and gluons. With such additional handles, we might have a chance to distinguish
boosted $\ttbar$ signal from the QCD background even at a smaller integrated luminosities.\footnote{
There are other approaches dealing with a similar situation in a different perspective in recent literature~\cite{LHCnotes,lookinside,Kaplan:2008ie,Thaler:2008ju}.}
We discuss jet substructure later in the text.

We turn our attention to the use of $b$-jets as spin analyzers for the top quarks.
For highly boosted top quarks, chirality is approximately equal to helicity and is conserved to a good approximation.
Information about the top chirality is encoded in the angular distribution of the decay products~\cite{Agashe:2006hk,ttbarspin,ttbarspinmore}.
Naively, one would argue that for hadronic tops this information is inaccessible due to collimation 
and the absence of leptons which are known to be good spin analysers~\cite{ttbarspin,ttbarspinmore}.
We explore the possibility of using $\pt$ of the $b$-quark for measuring the top quark polarization, which is important for exploring NP.
For this, we explore the case when at least one of the boosted top quark can be resolved into more than two jets.
We also consider the possibility of using $\pt$ of the lepton for measuring the top quark polarization for semi-leptonically decaying tops.

This work has two main focal points, namely QCD jet mass distributions and hadronic $\ttbar$ signal,
and is structured as follows.
In the next section, we discuss the MC generation and detector simulation.
In section~\ref{section_QCD} we focus on highly boosted QCD jets.
The jet mass distribution is examined numerically, via MC methods, and analytically, via jet functions.
The salient points of the jet functions are introduced, leaving detailed derivations for the Appendix.
Section~\ref{section_ttbar} discusses the top-jet signal.
In section~\ref{section_analysis}, we compare high $\pt$ hadronic $\ttbar$ events with QCD jets.
In section~\ref{section_substructure} we discuss jet shapes~\cite{us},
which can be used as additional discriminants against the background.
Section~\ref{top_polarization} discusses the hadronic top quark polarization by using the transverse momentum of the bottom quark. 
We conclude in section~\ref{section_conclusions}.

\section{Event Simulation} 
\label{section_setup}

\subsection{Monte Carlo Generation} 
The Sherpa~\cite{Sherpa} and MG/ME (version 4)~\cite{Maltoni:2002qb, Alwall:2007st} MC generators were used to produce $\ttbar$ and QCD jet events,
with parameters appropriate to the LHC.
To effect partonic level cuts during the generation of QCD jets $\left( \pt \left( \ge 1\,\, \mbox{parton} \right) \ge 800 \, \gev \right)$, 
we used customized code provided by the Sherpa authors applicable to Sherpa V1.1.0.
For technical reasons, $\ttbar$ events were generated using Sherpa version 1.0.9, whereas QCD jet events were generated with Sherpa version 1.1.0.
MG/ME interfaces to Pythia V6.4 (for parton shower and fragmentation)~\cite{Sjostrand:2006za}. 
For jet reconstruction, we used SISCone V1.3~\cite{Salam:2007xv} for both Sherpa and MG/ME.
Cross sections are calculated to leading order.
Jets are defined via the cone algorithm~\cite{cone_jet_algorithm} with $R = 0.4$ and $R = 0.7$,
referred to as C4 and C7, respectively.
Jets have $\pt > 50 \gev$ and $|\eta| \le 2$. 
At the hard scatter level, final state partons are required to have $\pt \ge 20 \gev$.
For MG/ME events, the final state partons have $ | \eta | \le 4.5 $.

We do not account for pile-up effects nor characterize the underlying event.
Efficiencies for triggering and reconstruction of jets at these energies are very close to unity;
the corrections are negligible and are not considered.
The strong coupling constant was allowed to run.
Throughout the analysis, we used Sherpa V1.0.9 with CTEQ6M parton distribution functions (PDF)~\cite{CTEQ6ML}. 
Comparisons to MG/ME were made whenever appropriate, and also occasionally to Pythia (version 8.1)~\cite{Sjostrand:2007gs} for $2\rightarrow 2$ process without matching.
In such cases, the distinct curves are marked accordingly.
The events used in the analysis were inclusive, i.e.~$pp \to \ttbar(j)$ and $pp \to jj(j)$, with matching (see~\cite{Alwall:2007fs} for a detailed discussion): modified MLM~\cite{Mangano:2002ea} for MG/ME and CKKW~\cite{Catani:2001cc} for Sherpa. 

\subsection{Cross Sections}
\label{section:cross_section}
In table~\ref{t:cross_section} we present cross sections for producing final state (hadronic level) jets with $\pt \ge 1 \tev$ for the different MC simulations.
There are large uncertainties in the cross sections, due to differences between the MLM and CKKW matching, between MC generators, and between PDFs.
It is outside the scope of this paper to explore the reasons behind these differences. 
We estimate a 100\% systematic uncertainty associated with the $\ttbar$ cross section,
and a 20\% systematic uncertainty in the QCD jet cross section.

\begin{table}[h]
\begin{center}
\begin{tabular}{|c|c|c|c|c|}
\hline
Process & Generator & PDF & Matching & Cross Section \\ 
\hline
$pp \rightarrow \ttbar (j)$ & SHERPA 1.0.9 & CTEQ6M & CKKW & 135 fb\\
$pp \rightarrow \ttbar (j)$ & SHERPA 1.1.2 & CTEQ6M & CKKW & 149 fb\\
$pp \rightarrow \ttbar (j)$ & MG/ME 4 & CTEQ6M & MLM & 68 fb\\
$pp \rightarrow \ttbar (j)$ & MG/ME 4 & CTEQ6L & MLM & 56 fb\\
$pp\rightarrow \ttbar$ & Pythia 6.4 & CTEQ6L & - & 157 fb\\
$pp\rightarrow \ttbar$ & Pythia 8.1 & CTEQ6M & - & 174 fb\\
\hline
$pp \rightarrow jj(j) $ & SHERPA 1.1.0 & CTEQ6M & CKKW & 10.2 pb \\
$pp \rightarrow jj(j) $ & MG/ME 4& CTEQ6L & MLM & 8.54 pb \\
$pp \rightarrow jj(j) $ & MG/ME 4& CTEQ6M & MLM & 9.93 pb \\
$pp \rightarrow jj$ & Pythia 6.4 & CTEQ6L & - & 13.7 pb\\
$pp \rightarrow jj$ & Pythia 8.1 & CTEQ6M & - & 13.3 pb\\
\hline
\end{tabular}
\caption{\label{t:cross_section}Cross sections for producing final state $R=0.4$ leading cone jets with $\pt \ge 1 \tev$ and $| \eta | \le 2$.
Generation level cuts were imposed as follows. 
Final state partons from the hard scatter were required to have $\pt \ge 20 \gev$.
For MG/ME, final state partons have $ | \eta | \le 4.5$. 
Processes with a trailing $(j)$ suffix indicate that $2 \to 2$ and $2 \to 3$ processes are represented.
}
\end{center}
\end{table}

\subsection{Modelling Detector Effects}
A transfer function, trained with full ATLAS detector simulation on high $\pt$ jet and high $\pt \,\, \ttbar$ samples, 
was used to map particle level jets (Atlas truth jet reconstruction) onto a full simulation model~\cite{JoeVirziAtlas}.
Transfer functions work by feeding back the differences between the target collection (Full Simulation) and the source collection (Truth Jets).
The differences and efficiencies are stored as distributions, in the form of histograms, and binned in $\pt$ and $\eta$.
We refer to the collection of the smearing distributions as a {\it{transfer function}}.
It is important to note that transfer functions are applicable on events with similar jet multiplicity and topology.
We applied the transfer function (trained on Atlas truth jets) to SISCone truth jets, 
which preserve the salient characteristics of the Atlas truth jets.
We used the transfer function to effect $\pt$ and mass smearing, but not reconstruction efficiency.
At the energies considered in this paper, reconstruction efficiency is very close to unity.
In summary, the results of the transfer function should be viewed simply as realistic detector smearing.

In this paper, a jet is transferred as follows.
The transverse momentum and mass of truth-level jets are smeared according to the appropriate distribution.
For the purposes of modeling the effects of the JES, the means of the $\pt$ distributions are shifted accordingly,
without cross correlation to the mass smearing.
This is a subtle point.
Depending on the reconstruction mechanism, reported jet masses may depend proportionally on the JES; a JES shift results in a jet mass shift.
In our study of the effects of the JES, we do not make a correlation between the $\pt$ and mass distributions.
This effect is much smaller, and such precision is not warranted in these studies.

\begin{figure}[fhptb]
\begin{tabular}{cc}
\includegraphics[width=.48\hsize]{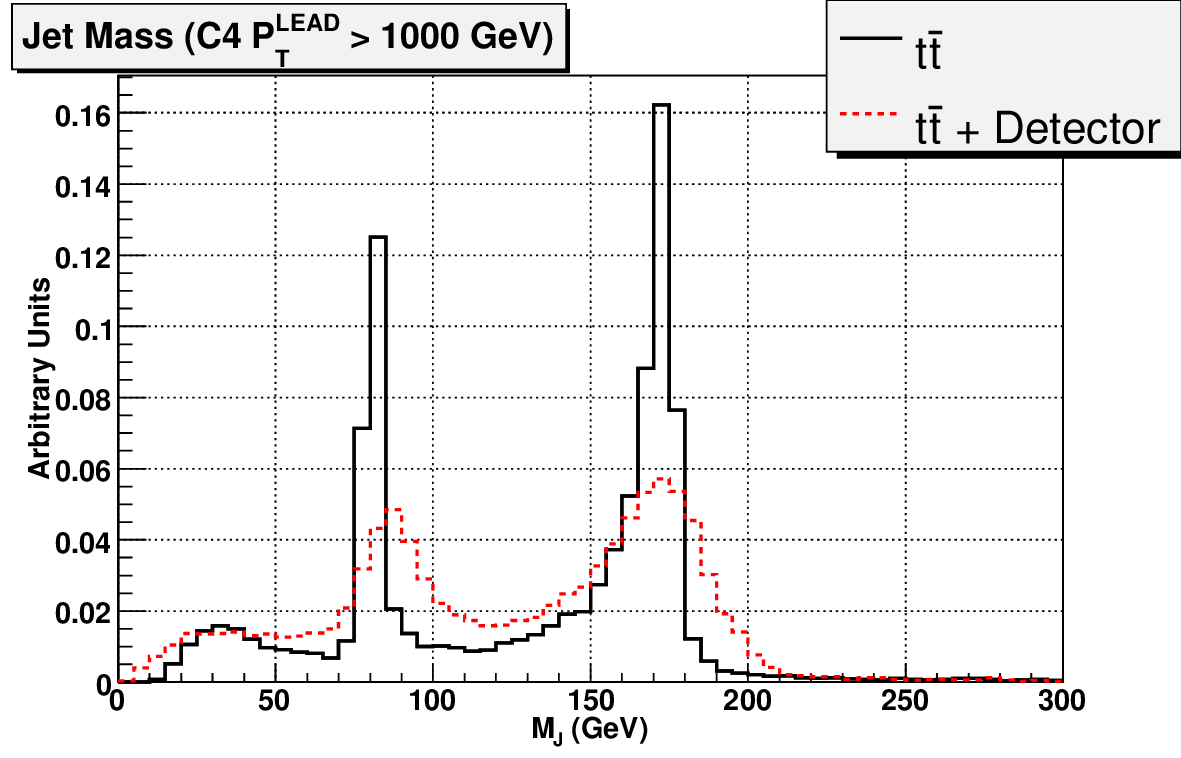}  &
\includegraphics[width=.48\hsize]{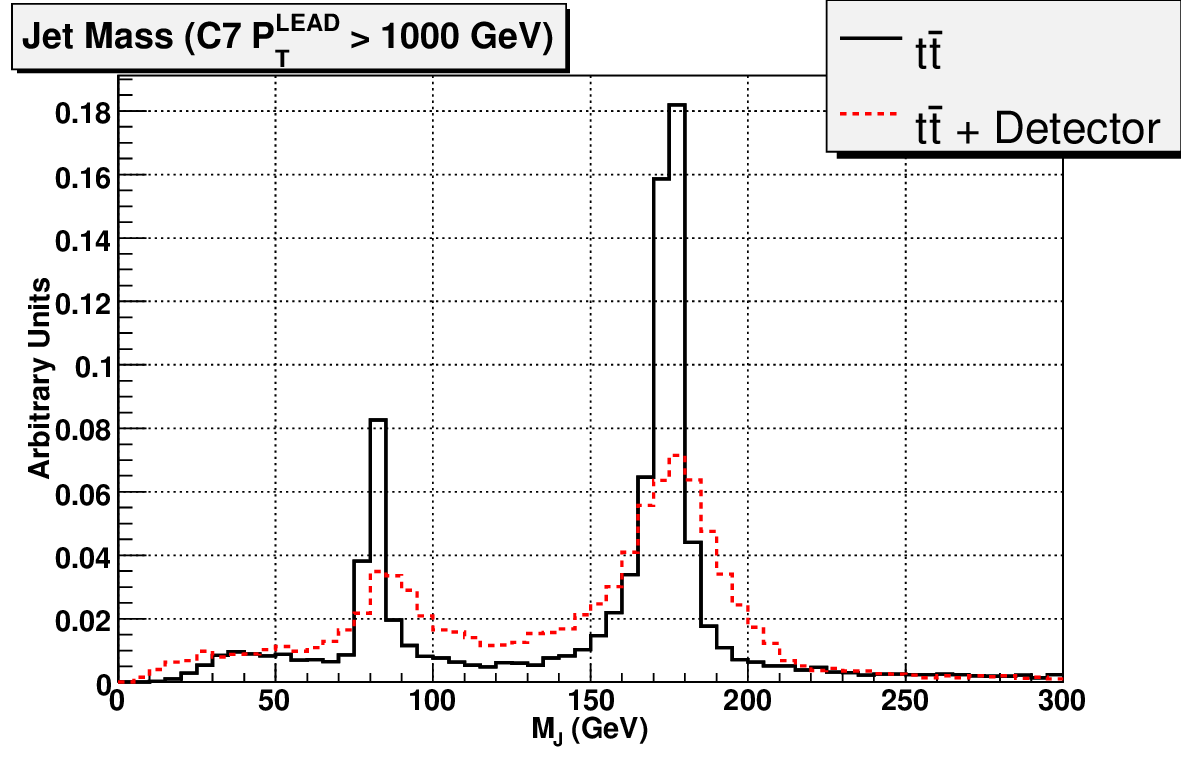} \\ 
\end{tabular}
\caption{
We compare the mass distribution of the leading jet $\left( \pt^{lead} \ge 1000 \gev \right)$ for the $\ttbar$ 
signal with (the red dotted curve) and without (the black solid curve) leading detector effects.
% The transfer function is used to analyze the leading effects of detector resolution.
The plot on the left corresponds to C4 jets; the plot on the right corresponds to C7 jets.
} \label{p:ttbar_detector_effects}
\end{figure} 

In Fig.~\ref{p:ttbar_detector_effects}, 
% we see the effects of the detector smearing, using the transfer function, on the $\ttbar$ signal,
we compare the $\ttbar$ jet mass distributions for C4 and C7 jets, with and without detector smearing, for $ \pt^{lead} \ge 1000 \gev$.
% GP: I've added this part to explain the figure.
We see, as expected, that due to the finite cone size even the top jet mass distribution is far from the naive Breit-Wigner shape.
In cases where the outgoing $b$ quark is outside the cone, we expect that the top jet mass to be peaked around the $W$ mass. 
In cases where one of the quarks from the $W$ decay is outside the cone we expect
a smooth distribution with a typical invariant mass of roughly $m_t/\sqrt 2$, etc.
These effects are present even at the truth level, without detector effects.
The black curve shows a smooth distribution with a spurious peak around the $W$ mass.
The red curve demonstrates how the detector effects further smear the top jet mass distribution.
% We plot the leading jet mass distributions,  for the $\ttbar$ signal.
% The plot on the left corresponds to C4 jets; the plot on the right corresponds to C7 jets.

\section{QCD Jet Background}
\label{section_QCD} \setcounter{equation}{0} \setcounter{footnote}{1}

If jet mass methods are to be viable, we must be able to characterize the dominant QCD jet background~\cite{Ellis:2007ib}.
One of the primary points in this work is that we are able to understand the QCD jet background {\emph{analytically}}  as well as through MC simulations.
In this section, we present the summary of our analytic calculations of the QCD jet mass distribution based on the factorization formalism~\cite{Collins:1989gx,Berger:2003iw}, 
which is presented in the Appendix. 
We compare our theoretical prediction with simulated MC data.
Note that the final states, which induce the jet masses, simulated by MC event generators are much more complicated (due to radiation, showering etc.)
than our simple two body final states. Yet, as we shall see, we can consistently describe the simulated MC data.

\subsection{Analytic Prediction} 

We are interested in looking at the following processes: 
$$H_a(p_a) +H_b(p_b) \to J_1(m_{J_1}^2,p_{1,T},R) + X  $$
$$H_a(p_a) +H_b(p_b) \to J_1(m_{J_1}^2,p_{1,T},R) + J_2(m_{J_2}^2,p_{2,T},R) + X  $$
where, $H_i$ are the initial hadrons, $p_i$ being the corresponding momenta, 
and the final states include jets in the direction of the outgoing partons of the underlying process, with a fixed jet mass, 
$m_{J_i}$,  ``cone size"   $ R^2 = \D \h^2 + \D \f^2 $ and tranverse momenta, $p_{i,T}$.

We begin with the factorized hadronic cross section for single inclusive jet processes, 
\beeq
\frac{d \s_{ H_A H_B \to J_1 X} (R)}{dp_T d m_{J} d \h } &=&  \sum_{abc} \int  d x_a \, d x_b \, \f_{a} (x_a) \, \f_{b} (x_b)  \frac{d \hat{\s}_{ab\to cX} }{d p_T dm_J d\h}(x_a, x_b, p_T, \h, m_{J},R)\, , \nn \\
\eeeq
which in the limit of small $R$, we can further factorize into (see Appendix B), 
\beeq
\frac{d \s_{ H_A H_B \to J_1 X} (R) }{dp_T d m_{J} d \h } &=&  \sum_{abc} \int  d x_a \, d x_b \, \f_a(x_a) \, \f_b(x_b)  H_{ab\to cX} (x_a, x_b, p_T, \h,R)  \nn \\
 && \times J_1^{c} (m_{J},p_T,R) . \label{eq:fac2}
\eeeq  
The factorization and renormalization scales are chosen to be $p_T$, $\f_i$ is the PDF for the initial hadrons, 
$H_{ab\to cX}$ denotes the perturbative cross section, 
and $J^{c}$ denotes jet functions, whose matrix elements are defined in Appendix  
\ref{sec:Jetfunction} (see \eg~\cite{Reviews} for recent reviews and references therein).
Furthermore the $J^{c}$s are, by definition, normalized as
\beq
\int d m_J \, J^{c} =1\,.
\eeq 
We have used the fact that the jet functions do not depend on $\eta$ in the leading expansion (see Appendix \ref{sec:Jetfunction}). 
Therefore, we can write Eq.~(\ref{eq:fac2}) for the hadronic cross section as
\beq
\frac{d \s (R)}{d p_T d m_{J} }  =\sum_c J^c (m_{J},p_T,R) \, \frac{d\hat{ \s}^{c}(R)}{d p_T} \, ,
\label{eqn:facJptmin}
\eeq
where $c$ represents the flavour of the jet, and where
\beq  
\frac{d \hat{\s}^c (R)}{d p_T}  = \sum_{ab} \int  d x_a \, d x_b \, \f_a \, \f_b  \int d \h \int d m_J \,\, \frac{d \hat{\s}_{ab \to cX} (R)} {dp_T d m_{J}  d \h } \label{sigmapT} \, .
\eeq
We employ the jet functions given in the Appendix by Eqs.~(\ref{eqn:jfq}) and (\ref{eqn:jfg}),  for fixed jet mass and $R$ at the next-to-leading order (NLO)  with running coupling effects. As we will see below, these results are consistent with the MC data for sufficiently large ($m_J \ge  {\cal{O}}(100 \gev)$) jet masses.

At the lower end of the jet mass spectrum, where $m_J \ll p_T R$, the jet mass distribution is dominated by higher order corrections and non-perturbative physics~\cite{Reviews},
which are beyond the scope of our work, as our interest lies in the region of high jet mass.  
We note this causes complications when trying to predict the moments of the mass distributions, such as the mean and RMS, unless we introduce a lower cutoff on the mass.

In the Appendix, we provide the full NLO result for the jet function in term of $\theta_S$, the angle of the softer particle with respect to the jet axis. These exact results can be approximated by the eikonal approximation introduced in Appendix B as
\beeq
J^{(eik),c} (m_J,p_T,R) &=&\as (p_T)  \frac{  4\,  C_c}{\pi m_J} \log\left( \frac{1}{z} \tan\left(\frac{R}{2}\right)   \sqrt{4-z^2} \right) \\
 &\simeq& \as (p_T)  \frac{  4\,  C_c}{\pi m_J} \log\left(\frac{R\,p_T}{m_J} \right)\, \nonumber,
\label{eq:eiko3}
\eeeq
where $\as (p_T) $ is the strong coupling constant at the appropriate scale,  $z= \frac{m_J}{p_T}$, $c$ represents the flavour of the parton which initiated the jet and $C_c$ equals $C_F=4/3$ for quarks, and $C_A=3$ for gluons. 
These expressions agree with the full NLO jet functions to the level of about $1\%$ and $ 10\%$ for quark and gluon initiated jets in the region of the top mass window, 
respectively (checked for $R = 0.4$ and $0.7$ and $\pt\gtrsim 1\,$TeV).

\begin{figure}[top]
\begin{center}
\includegraphics[width=.7\hsize]{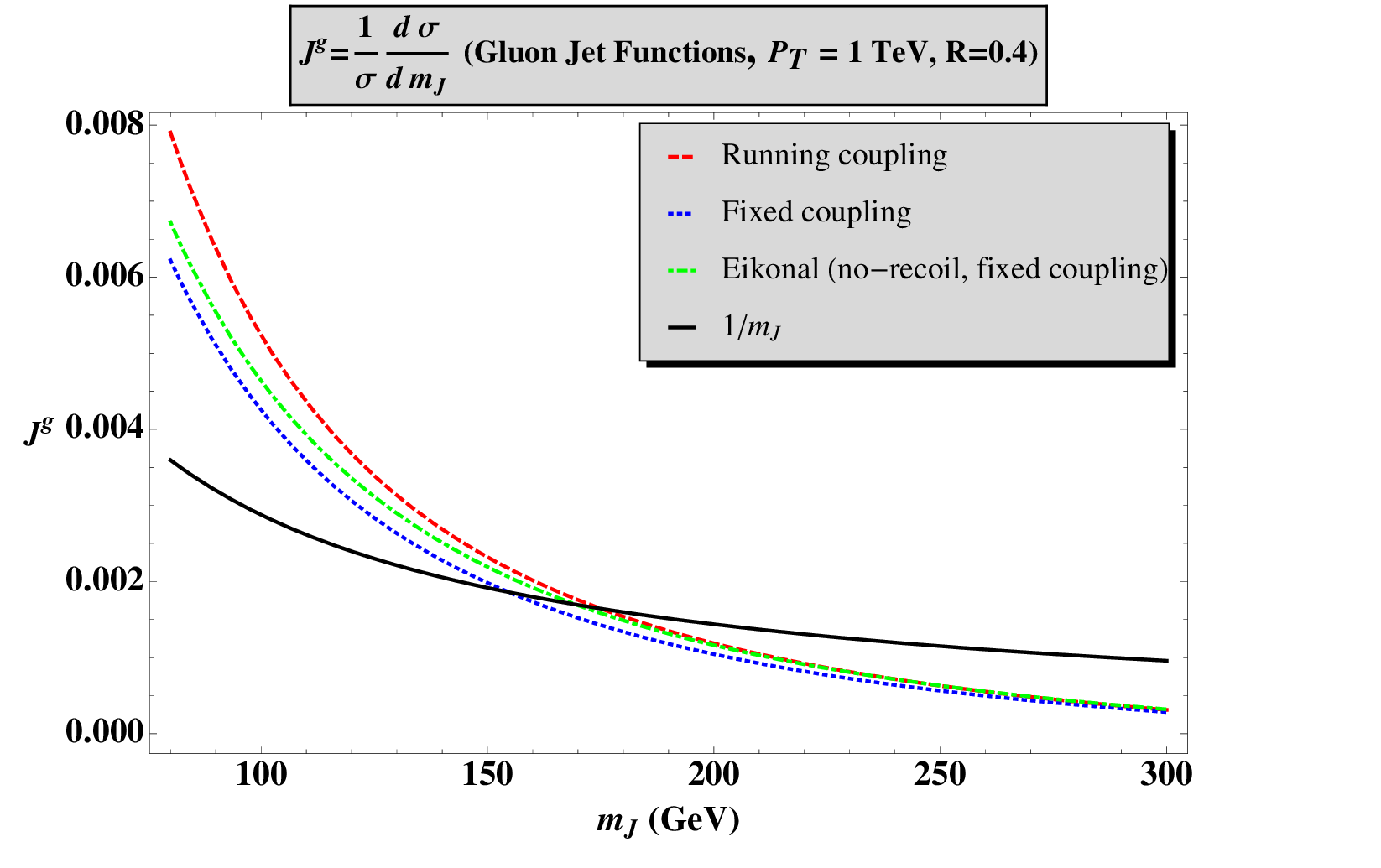}
\caption{
%Leo
Various theoretical gluon-jet mass distributions, along with a $1/m_J$ curve, are plotted for $\pt =1 \tev$ and $R=0.4$. 
Plotted are the jet mass distribution from (\ref{eqn:jfg}) with running (red, dashed), and fixed (blue, dotted) coupling, along with the eikonal jet function (green, dashed-dotted) with fixed coupling. 
For the jet functions with no running the scales were chosen be $p_T$.}
\label{p:jet_mass_theory_curves} 
\end{center}
\end{figure}

We can interpret the jet function as a probability density functions for a jet with a given $\pt$ to acquire a mass between $m_J$ and 
$m_J+\delta m_J$. 
Our rather simple treatment is valid for the higher end of the jet mass spectrum (above $m_J\sim {\cal{O}}(100 \gev)$), 
where NLO perturbative calculation captures the dominant physics. 
%%%%Leo
In Fig.~\ref{p:jet_mass_theory_curves} we show the gluon jet mass distribution from (\ref{eqn:jfg}) with running (red, dashed), and fixed (blue, dotted) coupling, 
along with the eikonal jet function (green, dashed-dotted) with fixed coupling. The fixed scales are chosen to be $p_T$.
For reference we  also superimpose  in the Fig. a $1/m_J$ curve which has the same dimension as that of our jet functions
and is roughly of the form of the soft function ({\it cf} Appendix B). 
It is remarkable that our theory curves are significantly different from simple $1/m_J$ curve whose normalization is chosen such that
this curve overlaps with our theory curves around the top mass. 
This indicates that logarithmic factor is very important in our theory prediction.
Note that at lower masses the running is much harder than the fixed cases since the configurations associated with this mass region have lower $k_T$ (the radiated gluon momenta), 
leading to a larger $\as$. 
Also, the eikonal approximation is equivalent to a no recoil approximation, thus resulting overall in a harder process than the result in Eq.~(\ref{eqn:jfg}) at fixed scales.
%%%
%%%

For the purpose of comparing the mass distributions obtained from jet functions and the MC simulations, 
Eq.~(\ref{sigmapT}) can be matched to $(d \sigma^c (R)/d p_T)_{\rm MC}$ obtained from MC, leading to the following relation,
\beeq
{\frac{d\sigma^{c}_{pred}(R)}{ dp_T d m_{J} }} & = & J^{c} \left( m_J, \pt, R \right)  \left(\frac{d \sigma^{c} \left( R \right)}{ d \pt} \right)_{\rm MC}\,,
\label{eqn:predicted_curves_pt}
\eeeq
for the prediction of quark and gluon jet mass distribution based on perturbative calculated jet functions, Eqs.~(\ref{eqn:jfq}) and (\ref{eqn:jfg}). 
Note, however, that this would require us to split the MC output in terms of the parton flavours $c$,
%GP I've expanded it why we don't do this, it's not because we're lazy ...
which for realistic simulation leads to ambiguities especially when matching is used. 
Therefore, for our analysis, instead, we use the analytic result to suggest bounds for the ``data'' distribution from the MC. 
There is, however, no a posteriori way to determine the flavour which initiated the jet (as with real data).
Thus, we write
\beeq
{\frac{d\sigma_{pred}(R)}{ dp_T d m_{J} }}_{upper\, \,bound} & = & J^{g} \left( m_J, \pt, R \right) \sum_c  \left(\frac{d \sigma^{c} \left( R \right)}{ d \pt} \right)_{\rm MC}\,,
\label{eqn:predicted_curves_upper} \\
{\frac{d\sigma_{pred}(R)}{ dp_T d m_{J} }}_{lower \,\,bound} & = & J^{q} \left( m_J, \pt, R \right) \sum_c  \left(\frac{d \sigma^{c} \left( R \right)}{ d \pt} \right)_{\rm MC}\,,
\label{eqn:predicted_curves_lower} 
\eeeq
exploiting the fact that $J^{g} > J^{q}$ in the region of high jet mass, as can be seen in Eq.~(\ref{eq:eiko3}).
%%%%%%%%%%%%%%

\subsection{Jet Function, Theory vs. MC Data}
\label{section_theory_vs_mc}

\begin{figure}
\begin{tabular}{cc}
\includegraphics[width=.44\hsize]{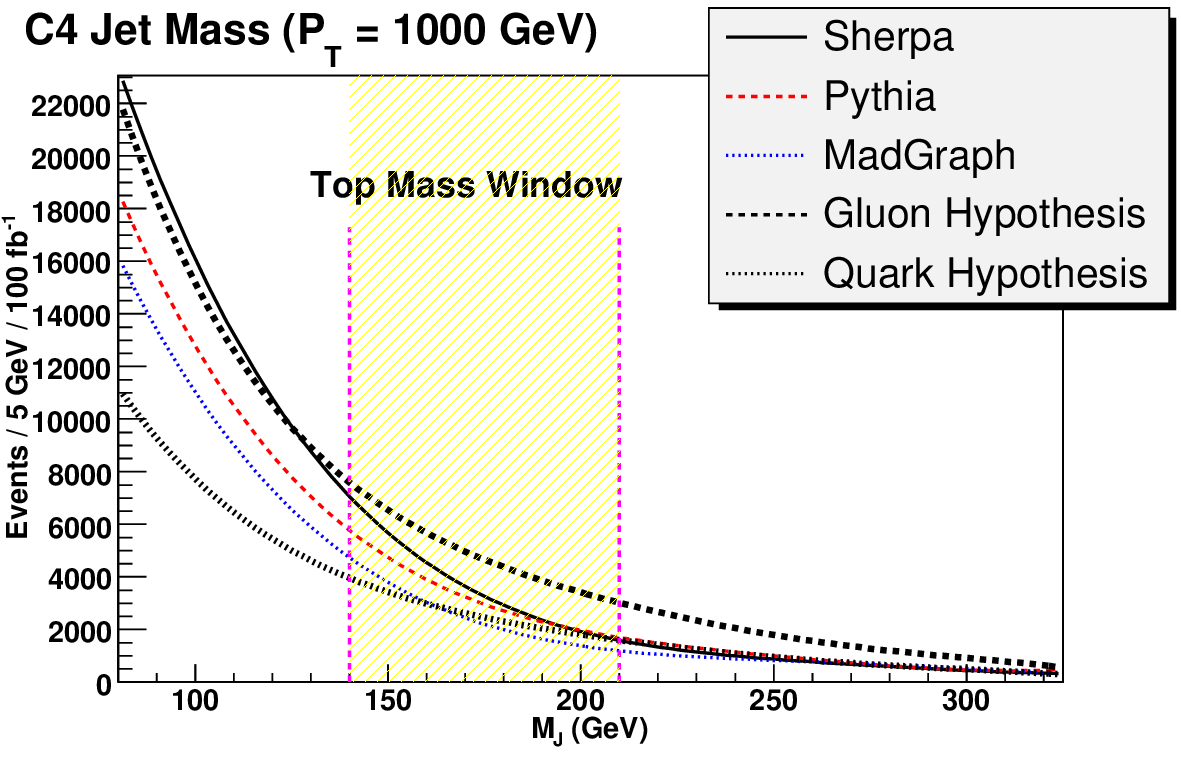}  &
\includegraphics[width=.44\hsize]{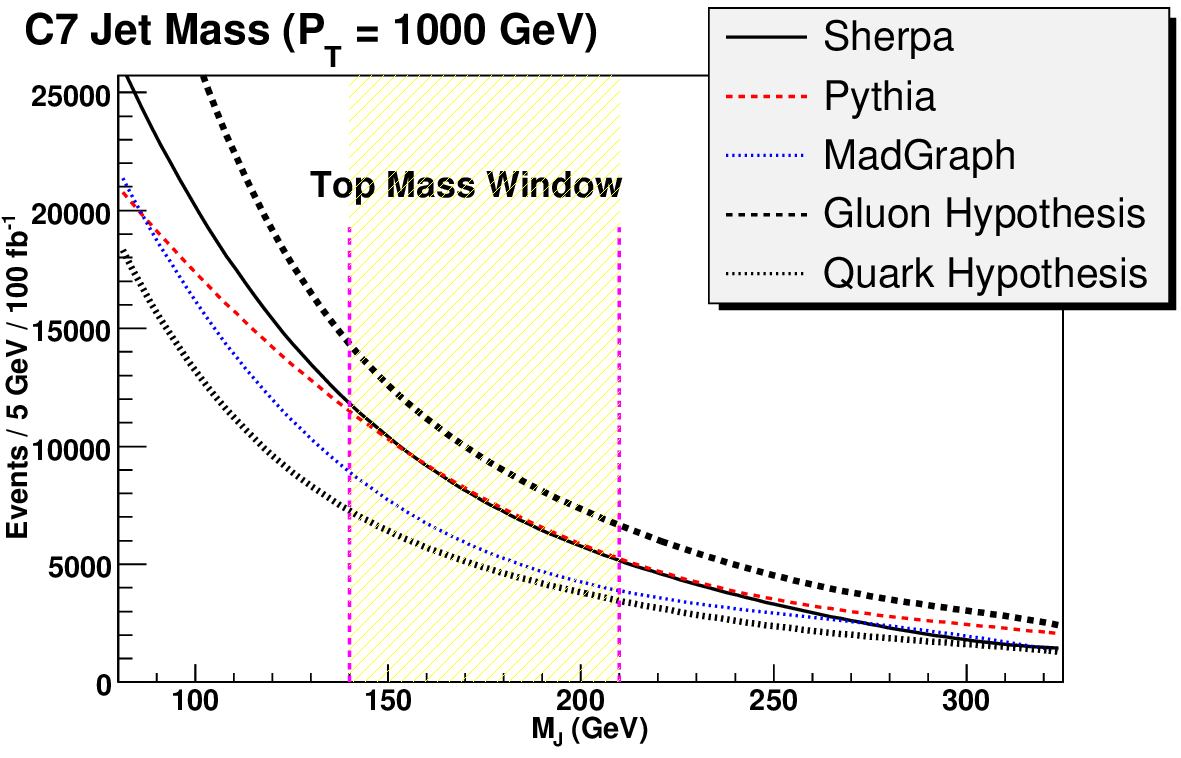} \\ 
\includegraphics[width=.44\hsize]{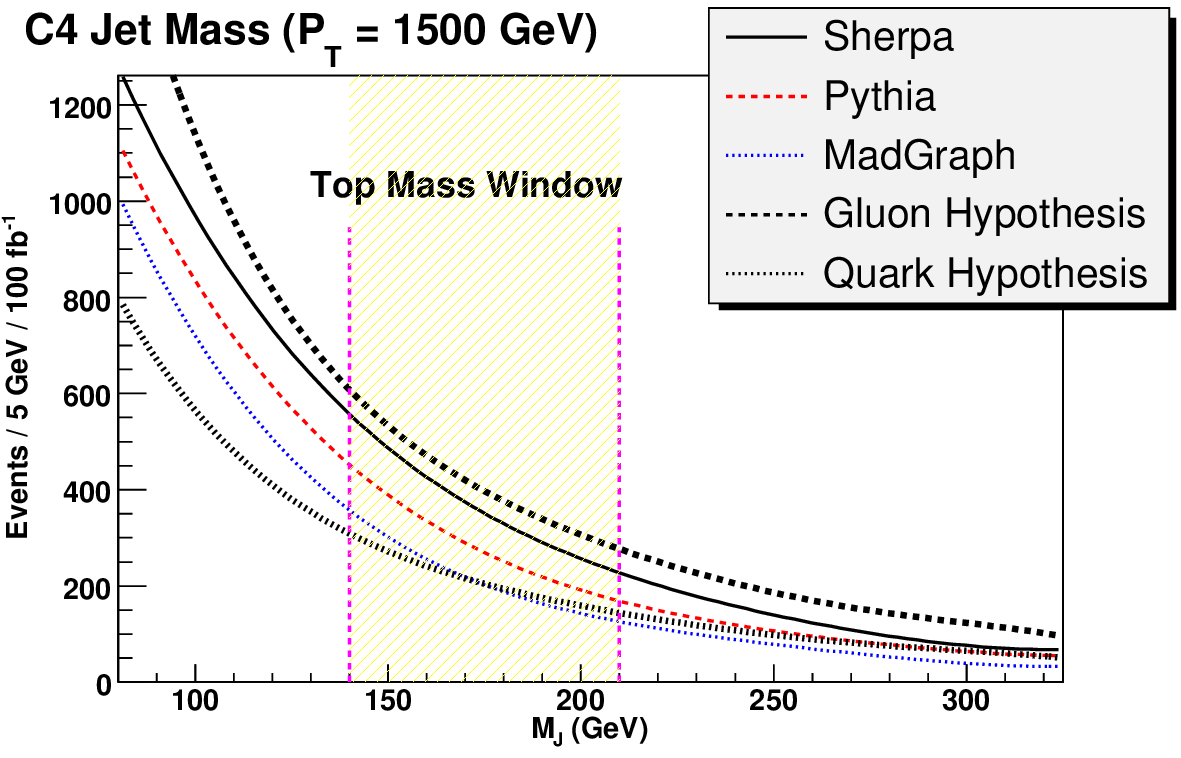}  &
\includegraphics[width=.44\hsize]{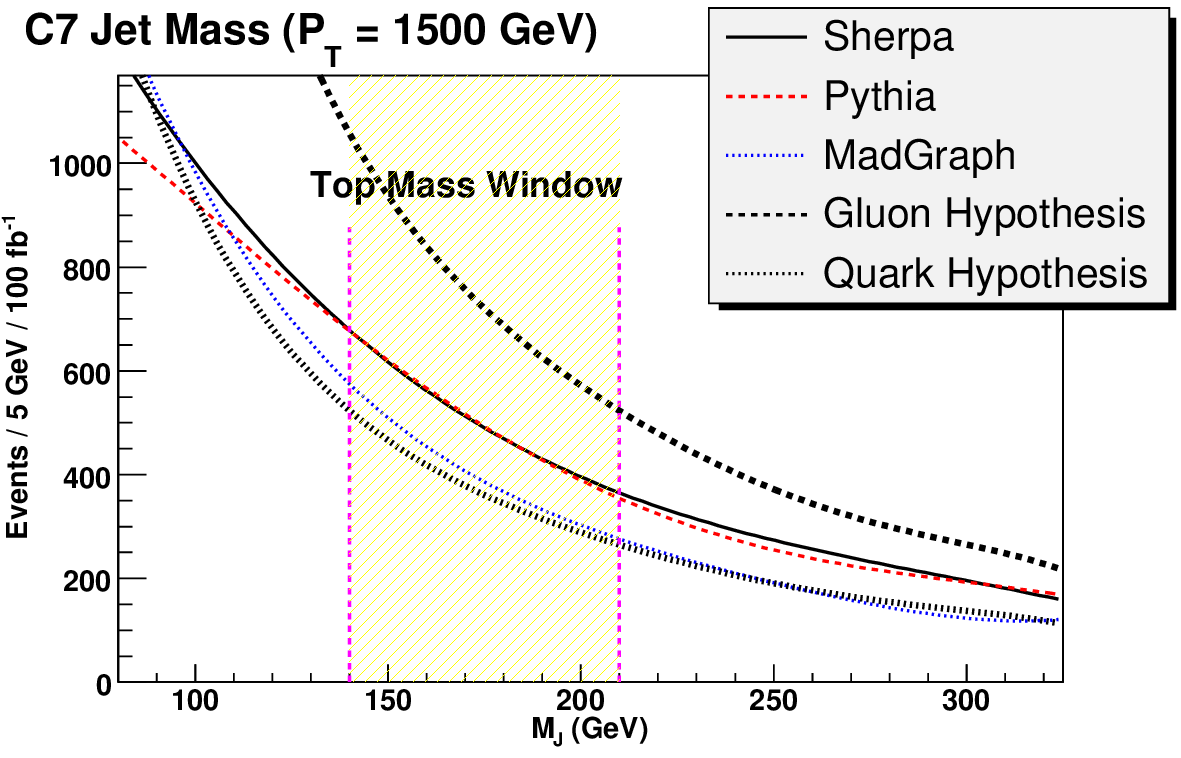} \\ 
\end{tabular}
\caption{
The jet mass distributions for Sherpa, Pythia and MG/ME are plotted for different $\pt$ and jet cone sizes.
The quark and gluon mass distributions from the jet functions are overlaid, using  Eqs.~(\ref{eqn:predicted_curves_upper}) and (\ref{eqn:predicted_curves_lower}).
The upper left plot corresponds to $950 \gev \le \pt \le 1050 \gev$ and $R=0.4$.
The upper right plot corresponds to $950 \gev \le \pt \le 1050 \gev$ and $R=0.7$.
The lower left plot corresponds to $1450 \gev \le \pt \le 1550 \gev$ and $R=0.4$.
The lower right plot corresponds to $1450 \gev \le \pt \le 1550 \gev$ and $R=0.7$.
}
\label{p:jet_mass_theory_vs_data} 
\end{figure}
In this section, we compare a set of theory-based bounds for the jet mass distribution to the mass distribution obtained via MC event generators.
This part contains one of our main results, where we demonstrate that our theoretical predictions are in agreement with the MC data.
In Fig.~\ref{p:jet_mass_theory_vs_data},  we compare the quark and gluon jet mass distributions from Eqs.~(\ref{eqn:predicted_curves_upper}) 
and (\ref{eqn:predicted_curves_lower}) to the distributions from different MC generators (MG/ME, Sherpa and Pythia).
We perform this comparison at fixed $p_T$, since we are interested in the relative shapes of these distributions around the top mass window.
Note that above $m_J\sim {\cal{O}}(100 \gev)$, the {\it{shapes}} of three MC distributions are very similar. 
Sherpa and MG/ME distributions interpolate between the quark jet function (lower bound) 
and the gluon jet function (upper bound) as expected.
%GP: i've moved this from the caption to the text.
For the purposes of comparing shapes, Pythia and MG/ME are rescaled so that their total cross sections agree with Sherpa.
This cross section scaling does not affect the predictive quality of the theory curves,
since it affects both sides of Eqs.~(\ref{eqn:predicted_curves_upper}) and (\ref{eqn:predicted_curves_lower}).
The scaling allows us to present the results from the different event generators on a single plot. 
%SBU addition, very important:
Note, as mentioned before, that for $m_J \ll p_T R$, higher order corrections will contribute, pushing the distribution down, with a Sudakov-like suppression, which can be seen in the lower mass region for $p_T=1.5$ TeV and $R=0.7$. 

In a typical experimental setup, a lower cut over $\pt$ will be assumed and the distributions
will be integrated above that $\ptmin$ cut. Thus we can integrate over the appropriate region on 
Eq.~(\ref{eqn:predicted_curves_pt}), which leads to the analog of Eqs. (\ref{eqn:predicted_curves_upper}) and (\ref{eqn:predicted_curves_lower}) for the $p_T$-integrated jet mass cross section,
\beeq
{d\sigma^c_{pred}(R) \over d m_{J}} & = & \int_{p_T^{min}}^{\infty}\, d\pt \, J^{c} \left(m_J, \pt, R \right)   \sum_{c'} \left( \frac{d \sigma^{c'} ( R)}{ d \pt} \right)_{\rm MC}\,,
\label{eqn:predicted_curves}
\eeeq
where $J^{c}$ is defined as before. The MC differential cross section is obtained by summing over the contributions from both quark and gluon jets. 
Therefore, the cross section's shape is characterized by an admixture of quark and gluon jets and should interpolate between the two curves, $c=q$ and $g$.
\begin{figure}[fhptb]
\begin{tabular}{cc}
%\hspace*{-1cm}
\includegraphics[width=.44\hsize]{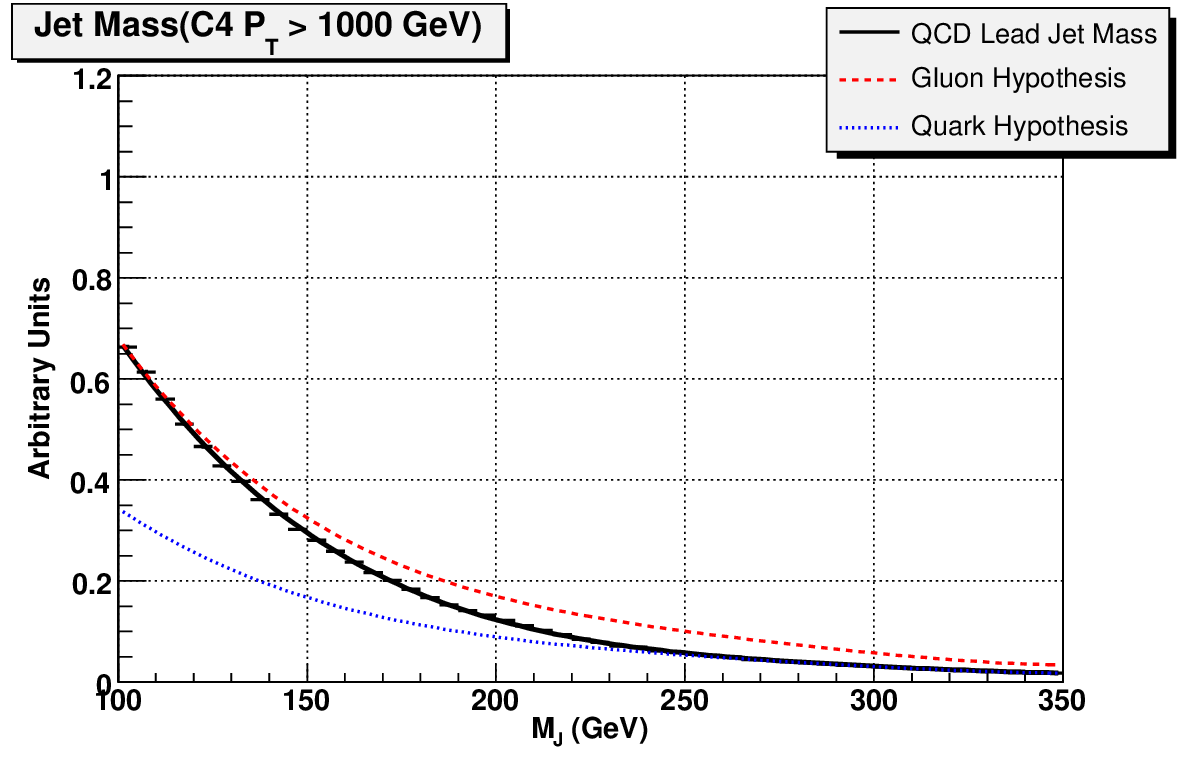}  &
\includegraphics[width=.44\hsize]{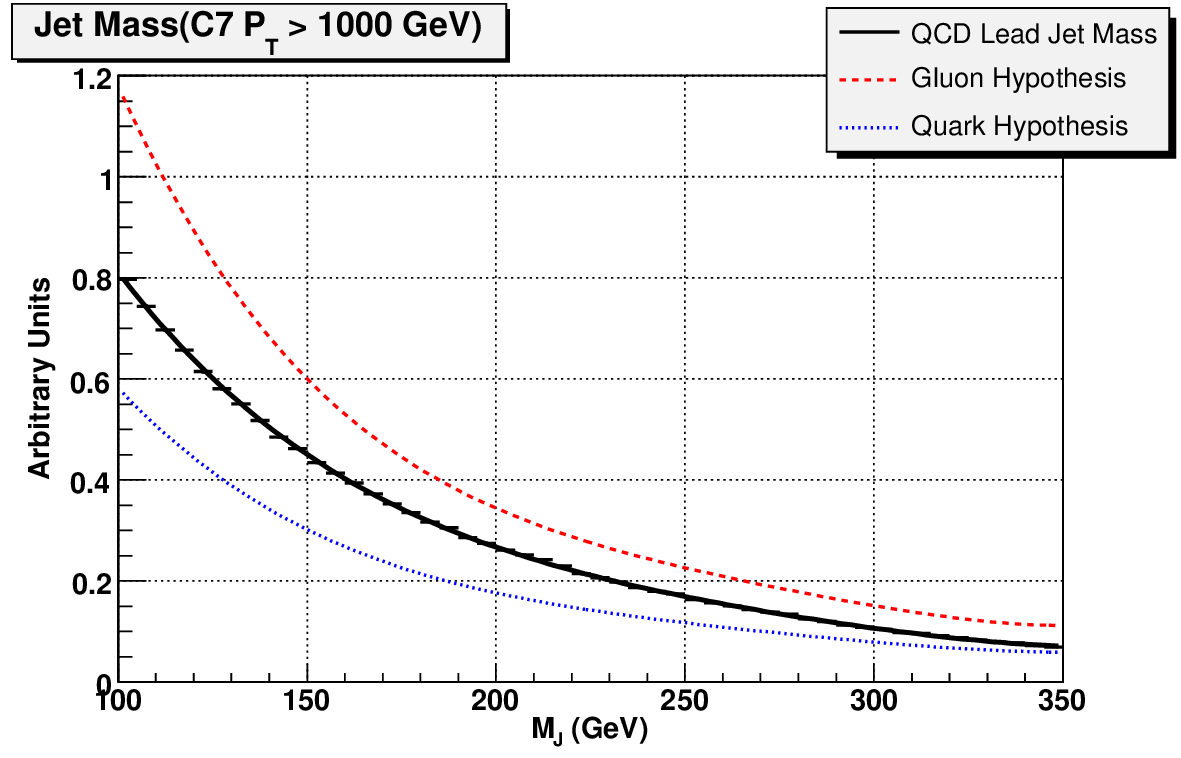}  \\  
%\hspace*{-1cm}
\includegraphics[width=.44\hsize]{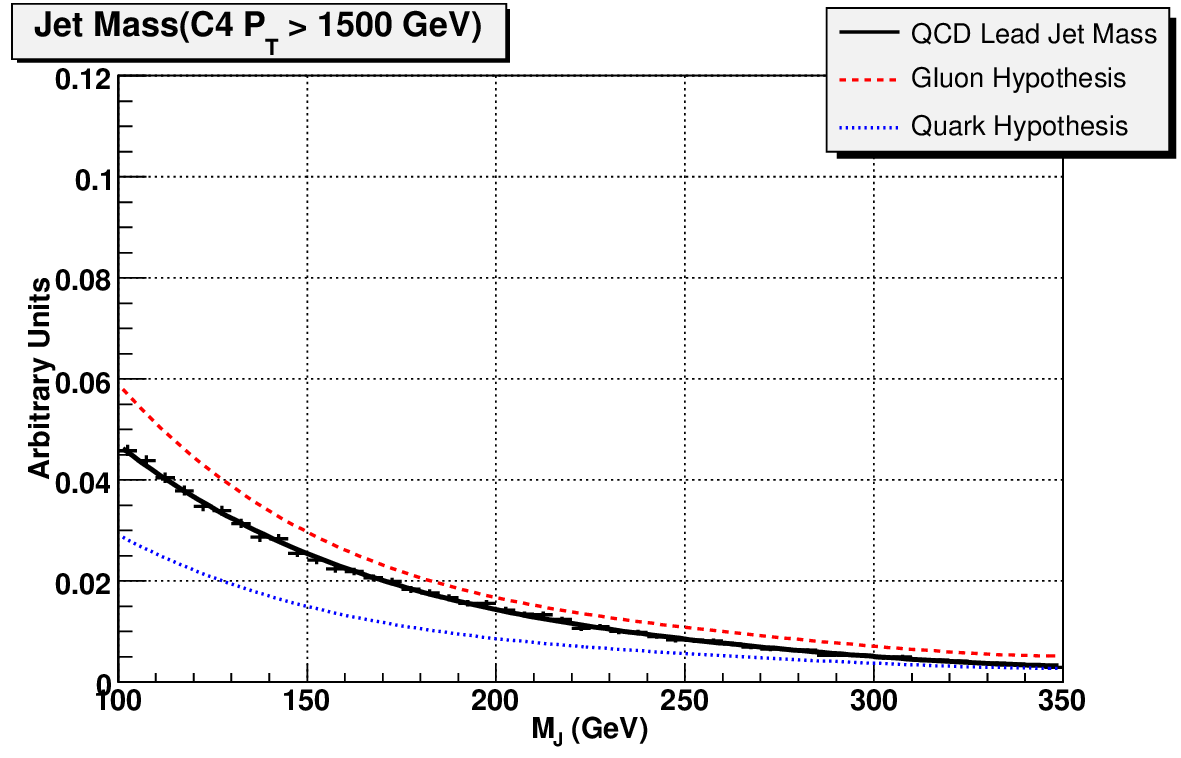}  &
\includegraphics[width=.44\hsize]{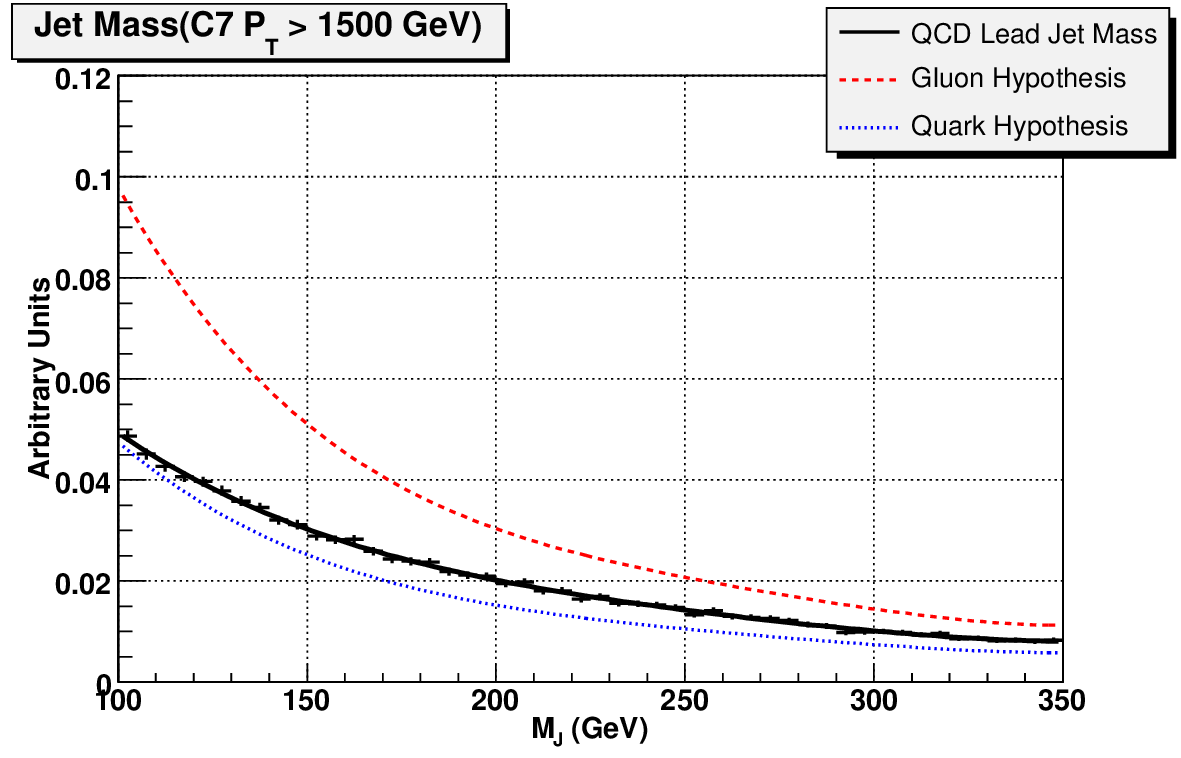}  \\  
\end{tabular}
\caption{Comparison between the theoretical jet mass distributions and MC leading jet mass distribution from Sherpa.
The minimum $\pt$ and cone size are indicated on the plots.
A gluon (quark) hypothesis is the prediction made if the entire contribution were from gluon (quark) jets (cf Eq.~(\ref{eqn:predicted_curves})).
} \label{p:dsdm_dijet}
\end{figure} 
In Fig.~\ref{p:dsdm_dijet}, we compare  leading jet mass distribution for events where the leading jet has $\pt \ge 1 \tev$ obtained from Sherpa.  The quark and gluon curves, obtained from Eq.~(\ref{eqn:predicted_curves}), with use of the jet functions in Eqs.~(\ref{eqn:jfq}) and (\ref{eqn:jfg}),  correspond to the cases where the lead jets are all quark or gluons jets, respectively.

As before, we find the bounds for the total cross section
\beeq
\s(R)_{upper \,\,bound} & = & \int_{p_T^{min}}^{\infty}d\pt \sum_c \left( \frac{d \sigma^c \left(R \right)}{ d \, \pt} \right)_{\rm MC}\, \int_{140 \, GeV}^{210 \, GeV} J^{g} \left( m_J, \pt, R \right) dm_J   \label{eqn:xsecup}\, ,\\
\s(R)_{lower \,\,bound}& = & \int_{p_T^{min}}^{\infty} d\pt \sum_c \left( \frac{d \sigma^c \left(R \right)}{ d \, \pt} \right)_{\rm MC} \, \int_{140 \, GeV}^{210 \, GeV} J^{q} \left( m_J, \pt, R \right) dm_J \, . \label{eqn:xsecdown}
\eeeq
In table~\ref{t:qcd_test}, we refer to the gluon and quark jets from the results in Eqs.~(\ref{eqn:xsecup}) and (\ref{eqn:xsecdown}), respectively.
The numbers in the table were calculated as follows. 
From a MC sample corresponding to 100 $\invfb$ of data, we extracted the number of events with C4 lead jet $\pt \ge 1000(1500) \gev$ and $\massmin < m_J < \massmax$, the top mass window.
We repeated this exercise for C7 jets. The {\it{data}} column contains these results. 

\begin{table}[fhptb]
\begin{center}
\begin{tabular}{|l|c|c|c|c|c|}
\hline
$\pt^{lead}$ cut & Cone Size & Data & Quark hypothesis & Gluon hypothesis   \\ 
\hline
1000 GeV & C4 &  113749 & 70701 & 135682  \\
1000 GeV & C7 &  197981 & 131955 & 260045  \\
1500 GeV & C4  & 10985 & 6513 & 12785  \\
1500 GeV & C7 & 13993 & 11164 & 22469  \\
\hline
\end{tabular}%
\caption{\label{t:qcd_test} Comparison of Sherpa MC data to predictions of pure-quark and pure-gluon hypothesis, for the number of events with leading jet with mass between 140 GeV and 210 GeV.
The data is compared to the bounds given in Eqs.~(\ref{eqn:xsecup}) and~(\ref{eqn:xsecdown}). The statistics reflect 100 $\invfb$ of integrated luminosity.
}
\end{center}
\end{table}

\subsubsection{Fractional Fake Rate}

With the theoretical machinery discussed in the previous section,
we are able to make a prediction of the rate at which QCD jets will fake the mass signature of top-jets.
We define the fractional fake rate as the fraction of jets with $\massmin \le m_J \le \massmax$, for given $\pt$ and $R$. 
We estimate the upper and lower bounds of the fractional fake rate as 
\beeq
\int^{210 \, \, GeV}_{140 \, \, GeV}\,dm_J\, J ^{q}(m_{J},p_T,R) \le \, {\rm{\, Fractional\,\,fake\,\,rate}} \,\le \int^{210 \, \, GeV}_{140 \, \, GeV} \,dm_J\, J ^{g}(m_{J},p_T,R)\, .\nn \\
\label{eqn:fake}
\eeeq

\begin{figure}[fhptb]
\begin{tabular}{cc}
\includegraphics[width=.48\hsize]{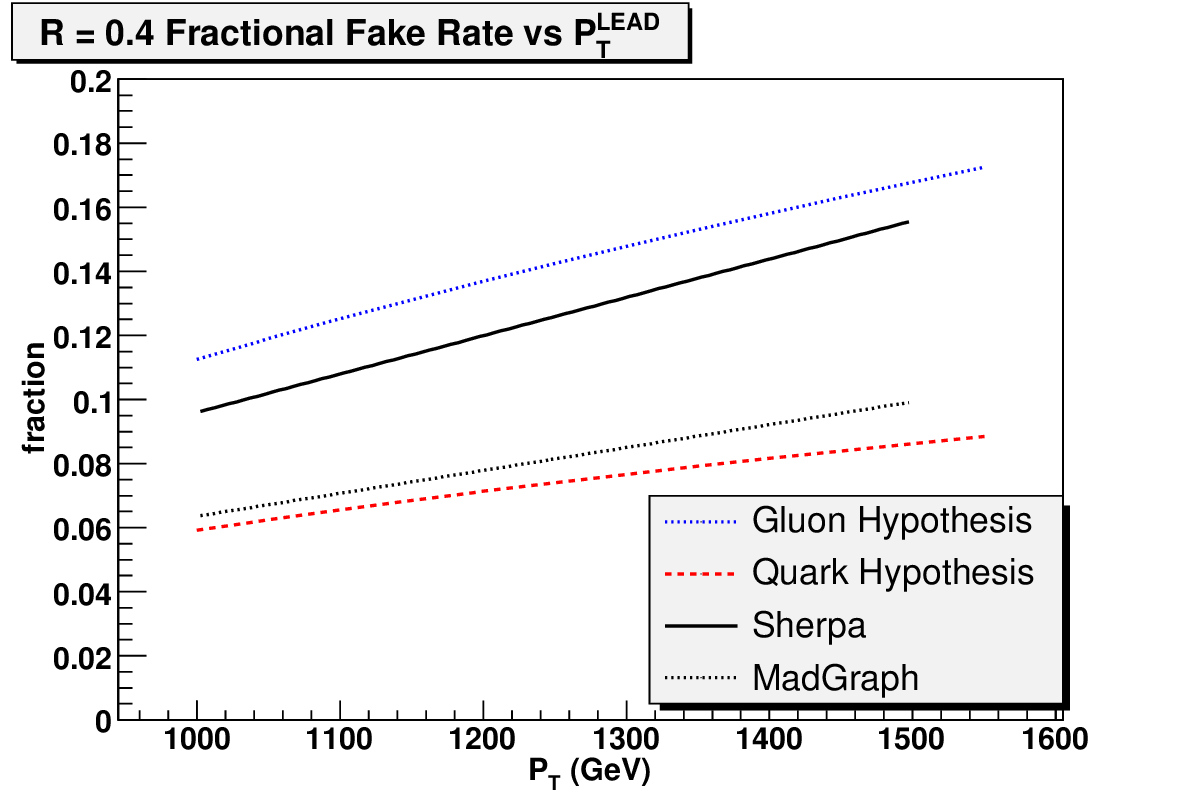}  &
\includegraphics[width=.48\hsize]{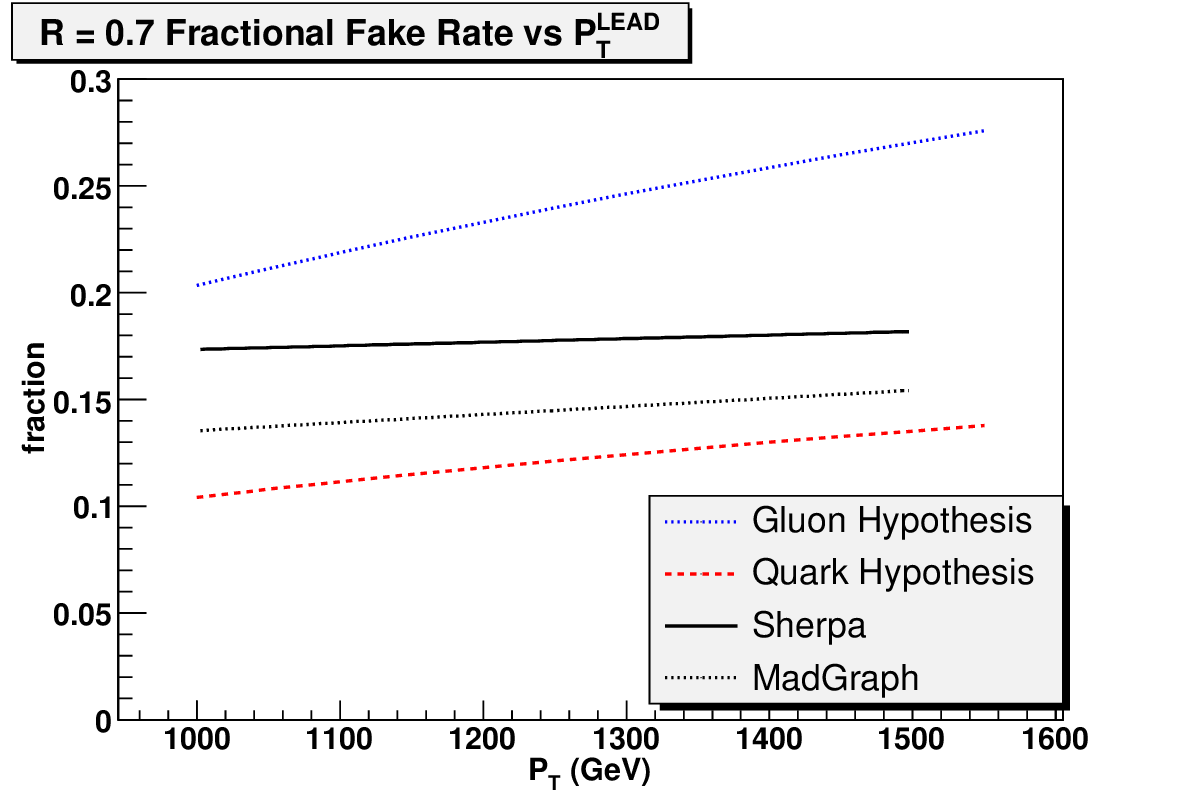} \\ 
\end{tabular}
\caption{
The fraction of jets which acquire $\massmin \le m_J \le \massmax $ as a function of the transverse momentum of the leading jet.
Quark- and gluon-hypothesis curves yield the prediction for the fractional fake rate,
if all jets were either quark- or gluon-jets, respectively.
The plot on the left corresponds to C4 jets; 
the plot on the right corresponds to C7 jets (cf Eq.~(\ref{eqn:fake})).
}\label{p:fraction}
\end{figure}

\begin{figure}[fhptb]
\begin{center}
\includegraphics[width=.74\hsize]{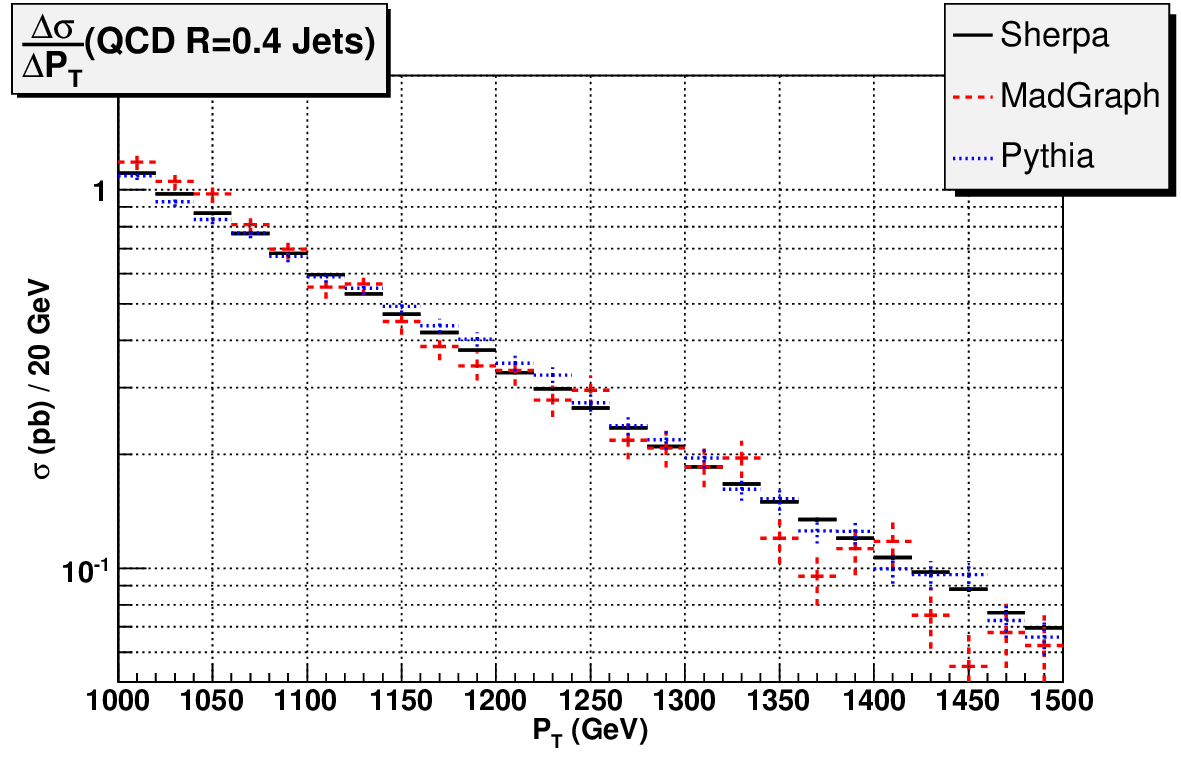}
\caption{The differential cross section for QCD ($R=0.4$) jet production with respect to the $\pt$ of the leading jet.
Sherpa, MG/ME and Pythia are represented.
} \label{p:dijet_pt}
\end{center}
\end{figure}

In Fig.~\ref{p:fraction}, we plot the fractional fake rate as a function of jet transverse momentum.
To predict the number of fakes in our sample, 
we fold the differential cross section for QCD jet production (Fig.~\ref{p:dijet_pt}) with the fractional fake rate (Fig.~\ref{p:fraction}).
% important addition from SBU:
 Again we expect a Sudakov-like suppression when $m_J \ll p_T R$, thus flatting the theoretical fractional fake rate as $p_T$ increases.
This can be seen more predominately for $R=0.7$ in Fig.~\ref{p:fraction}.

\subsubsection{Pseudorapidity Independence of the Jet Mass Distribution}
\begin{figure}[fhptb]
\begin{center}
\includegraphics[width=.74\hsize]{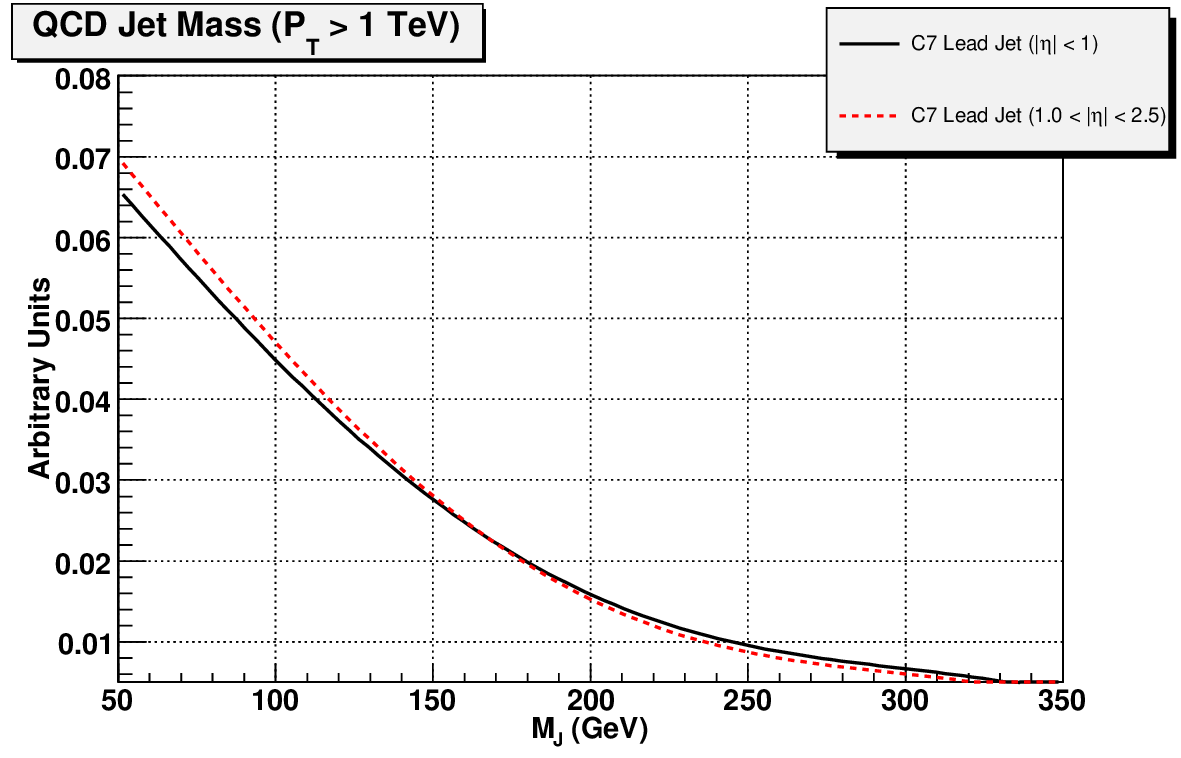}
\caption{The $R = 0.7$ jet mass distribution for central jets ($|\eta| < 1$) and for jets with $1 \le |\eta| \le 2.5$. Jets have $\pt \ge 1 \tev$. 
This plot is produced with the Sherpa MC.
} \label{p:qcd_mass_eta}
\end{center}
\end{figure}

In general, we expect that NP signals will have a pseudorapidity dependence.
Therefore, the study of pseudorapidity dependence may provide a tool for NP searches (for an interesting discussion see~\cite{Meade:2007sz}). 
In Fig.~\ref{p:qcd_mass_eta}, we plot the jet mass distributions for central and outer jets.
We observe consistency with the approximation that the distributions are to leading order, independent of pseudorapidity.

%%%%%%%%%%%%%%%%%%
%%%%%%%%%%%%%%%%%%

\section{High $\pt$ Hadronic Top Quarks}
\label{section_ttbar}

\begin{figure}[fhptb]
\begin{center}
\includegraphics[width=.74\hsize]{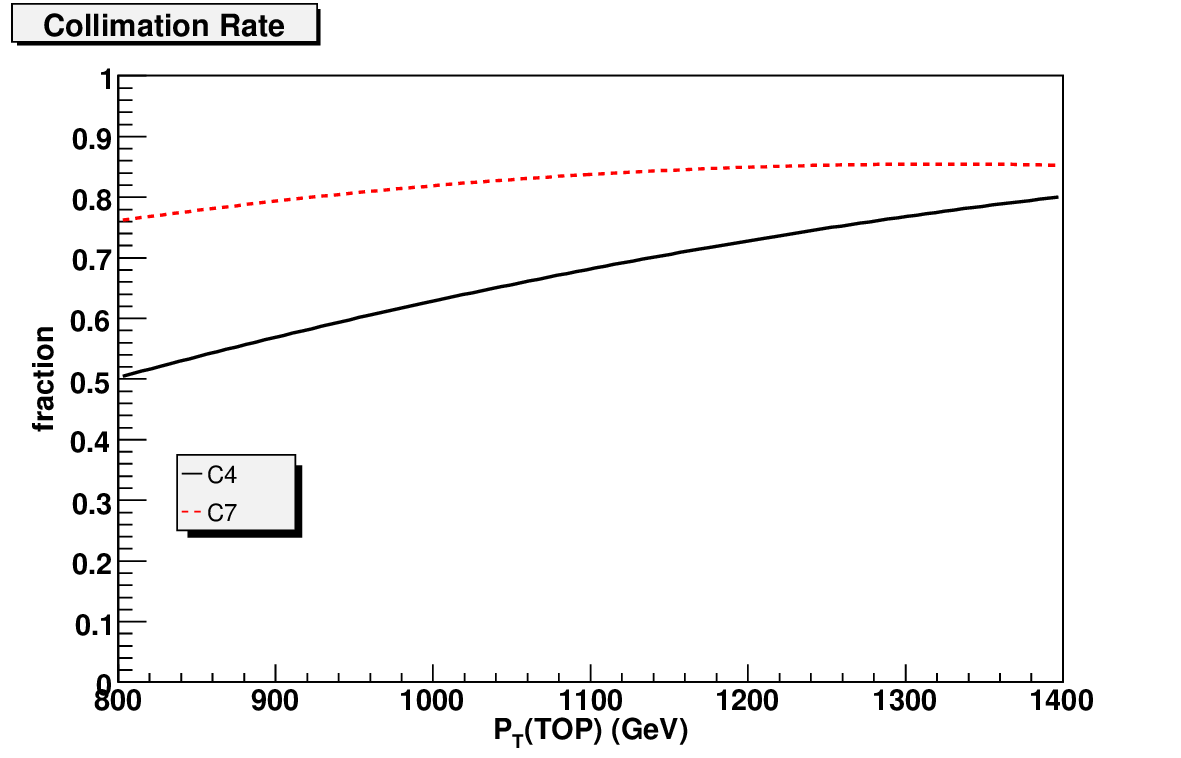}
\caption{The collimation rate for top quarks as a function of their transverse momentum, for C4 (black solid curve) and C7 (red dashed curve) jets.
Collimation rate is defined as the fraction of top quarks with $140 \gev \le m_J \le 210 \gev$.
} \label{p:collimation_rate}
\end{center}
\end{figure}

In this section, we discuss the collimation of the top quark decay products. 
In Fig.~\ref{p:collimation_rate}, 
we plot the rate of collimation as a function of the top $\pt$ 
(for a related discussion and analysis see~\cite{Baur:2007ck,LHCnotes,Baur:2008uv}).
We define collimation rate as the fraction of top quarks which reconstruct to a jet having $140 \gev \le m_J \le 210 \gev$.

To examine the efficiency of the jet mass methods, it is instructive to look at mass distributions for the signal and background.
We examine the distributions for events where the leading jet $\pt$ exceeds $1000 \gev$ and $1500 \gev$ with C4 and C7 jets.
In Fig.~\ref{p:ttbar_detector_effects}, we plot the jet mass distribution for the $\ttbar$ signal for $\pt^{lead} \ge 1000 \gev$.
The efficiency of C7 jets for capturing the hadronic top is greater than that for C4 jets.
For C4 jets, we still observe pronounced structure around the $W$-mass ($M_W$), which diminishes for C7 jets.
We also note that the peak for the C7 jets moves closer to the top mass,
indicating a higher efficiency for capturing the hadronic top.
We expect that detector effects will further smear the signal.
Fig.~\ref{p:ttbar_detector_effects} also  shows the mass distributions including leading detector effects (using transfer functions).

We note that the analysis has an inherent tension with regard to choosing the cone size for the jet.
The reconstruction cone should be sufficiently wide to capture all the daughter products of the hadronic top.
On the other hand, we need to keep the cone appropriately small to keep out the QCD jet background and other soft contamination~\cite{LHCnotes}.

We describe the gross features of the top mass distribution, without providing a detailed analytic expression for the top jet.
\footnote{For a more precise analysis in the case of $e^+ e^-$ collider and without providing a finite cone size see~\cite{tscet}.}
At next to leading order, we expect the top-jet to be broken into two contributions, $J^t_{QCD}$ and $J^t_{EW}$ (at leading order it is just given by the top bare mass).
The first contribution, $J^t_{QCD}$, is similar to that of the QCD jets. 
It is characterized by a very short time scale of ${\cal O}(10)\,$GeV and makes the top-jet mass harder. 
Using factorization, this process can be calculated by methods similar to the one discussed in the Appendix, fixing the mass of the final parton to $m_t$ and assuming it is stable. 
For our purposes, the resulting broadening is subdominant for a top mass window of $\pm 35\, \gev$.
\footnote{It is crucial to understand this behavior if one aims to improve the top mass measurement at the LHC. 
At the moment this has been studied only for lepton colliders~\cite{tscet}.}
At leading order, the second contribution, $J^t_{EW}$, is expected to be kinematical in nature, due to the weak decay of the top quark. 
The time scale is longer (of order $ \Gamma_t / \gamma_t={\cal O}(0.2)\, \gev$, where $\Gamma_t$ is the top quark width and $\gamma_t$ is the Lorentz boost).
The main effect here is top mass softening, because the jet cone may not capture all the particles from the top quark decay chain.
This kinematic effect depends solely on $m_t/(\ptr)$.
It should reduce the mass of the top-jet and is expected to exhibit structure near the $W$ mass.
Since the top jet mass softening is a kinematic process, it should be well described by simple phase space generators.
We schematically express the top jet mass as a sum of three contributions
\be
m^t_J\sim m_t+\delta m_{QCD}+\delta m_{EW}\,,
\ee
where the jet mass function can be schematically written as a convolution of three different sources
\be
\hspace*{-.5cm} & \hspace*{-.5cm} J^t(m_J, m_t, R,\pt)\sim& \int dm_{QCD}\, dm_{EW}\, dm_0 \, \delta(m_0-m_t)\, \delta(m_J -m_0 - m_{QCD}-m_{EW})\,  \times \nn \\
&& J^t_{QCD} (m_{QCD}, R,\pt)\times J^t_{EW}(m_{EW},m_t/(\ptr))\,.
\ee
The top mass is large, so we are not concerned about uncertainties in the lower jet mass spectrum.
We conclude that existing MC tools should well describe this part of our studies.

%%%%%%%%%%%%%%%%%%%%%%%%%%%%%%%%%%%%%%%%%%%%%%%%%%%%%%%%%%%%%%
%%%%%%%%%%%%%%%%%%%%%%%%%%%%%%%%%%%%%%%%%%%%%%%%%%%%%%%%%%%%%%%%%%
%section 5

%%%%%%%%%%%%%%%%%%%%%%%%%%%%%%%%%%%%%%%%%%%%%%%%%%%%%%%%%%%%%%
%%%%%%%%%%%%%%%%%%%%%%%%%%%%%%%%%%%%%%%%%%%%%%%%%%%%%%%%%%%%%%%%%%
\section{$t\bar t $ Jets vs. QCD Jets at the LHC}
\label{section_analysis}

In this section, we combine the results of the previous discussions, 
and apply them to analyze energetic SM $\ttbar$ events vis-a-vis QCD jet production at the LHC. 
The main purpose of this section is to understand how well we can discriminate our signal from the overwhelming
QCD background.
We illustrate an example analysis using the jet functions, and evaluate their performance on MC data.
Unfortunately, it is very difficult to outline a one-size-fits-all analysis.
Therefore, we perform a broad-strokes analysis that contains sufficient detail to provide general guidance.
We do not attempt to invoke advanced, but analysis-specific, procedures that could provide further refinement.
It is also important to bear in mind that the final evaluation of the jet functions, 
as precision analysis tools, can really only be done on real data.
The primary reason is that we expect the jet functions to describe physics data.
The MC distributions are, at this point, an approximation to what we believe will be LHC data.
A precision analysis will show the strains between the jet function-based shape predictions 
and the effective distribution that MC uses to generate its mass distribution.

We examine two cases in detail, both at truth-level  (no detector effects) and accounting for detector effects.
The first case, {\it{single tagging}}, consists of ``top-tagging" (requiring $\massmin \le m_J \le\massmax$) the leading jets satisfying a $\pt$ cut.
The second case, {\it{double tagging}}, consists of top-tagging the leading and subleading jets, with a $\pt$ cut only on the leading jet.

\subsection{Peak Resolution}
\label{section:peak_resolution}
In this analysis, one objective is to resolve the excess of events 
where the mass of the leading jet lies in the top mass window $\left( \massmin \le m_J \le \massmax \right)$.
It is important to note that we are not hunting for a peak; we already know its location.
The issue is that of resolving its magnitude and estimate the probability that the background 
would fluctuate to yield the observed data.
To estimate the significance of such a measurement, we perform a rudimentary analysis for resolving peaks.
We emphasize that it is misleading to estimate the significance as $S / \sqrt{B}$, 
where the signal and background are separate MC samples.
These numbers represent an unrealizable scenario, and tend to be optimistic.
In real data, there is no way to separate the signal from background with certainty.
Furthermore, at the present time, 
we cannot trust MC to provide the precise shape of QCD jet mass distributions.
Therefore, we derive our approximations to the background {\it{shape}} directly from the ``data", 
via sideband analysis (outside the top mass window). 
We use our previous knowledge of the shape of the background in the sideband region, 
to minimize the number of degree of freedoms involved in the sideband fit.
We will discuss this further in the next section.

After approximating the shape of the background in the sideband region, 
we interpolate the shape of the background into the top mass window.
The primary challenges are that our background is large and also has large uncertainties, 
which induce large uncertainties in the signal.
We discuss this in more detail at the conclusion of this section.
For the shape of the $\ttbar$ signal inside the top mass window, we use MC.
%change here
In principle,
the shape of the top mass distribution can be also derived semi-analytically,
as discussed in section~\ref{section_ttbar} (see also~\cite{tscet}).
However, to leading order we expect the MC data to provide us with a reliable shape 
(it should capture the radiation at the leading log approximation, also the, phase space, population of the top decay 
products is purely kinematical). 
For simplicity we use the simulation data for this step in our analysis.
These shapes, after normalization to unit area, are referred to as {\it{probability density functions}}.
Unfortunately, the standard acronym for probability density functions conflicts with existing usage for parton distribution functions in this paper.
To avoid confusion, we simply refer to them as {\it{shapes}}.
We use the approximate shapes for the signal and background to perform an extended maximum likelihood fit to the sample,
with jet mass distribution $F \left( m_J \right)$, thereby obtaining the background and signal normalizations. We define a jet mass distribution $F \left( m_J \right)$ as
\be
F \left( m_J \right) = N_B \times b \left( m_J \right) + N_S \times s  \left( m_J \right),
\label{eqn:totaldistribution}
\ee
where $N_B$ is the predicted background, and $N_S$ is the predicted signal in the top mass window.
$b \left( m_J \right)$ and $s \left( m_J \right)$ are used to denote the background and signal shapes, respectively.
Both $N_B$ and $N_S$ are allowed to float independently.
% The sum, $N_B + N_S$, enters the extended likelihood function as a Poisson prefactor, with the total sample size $N_0$ as the mean.

\subsubsection{Sideband Background Analysis}

\begin{figure}[fhptb]
\begin{center}
\includegraphics[width=.44\hsize]{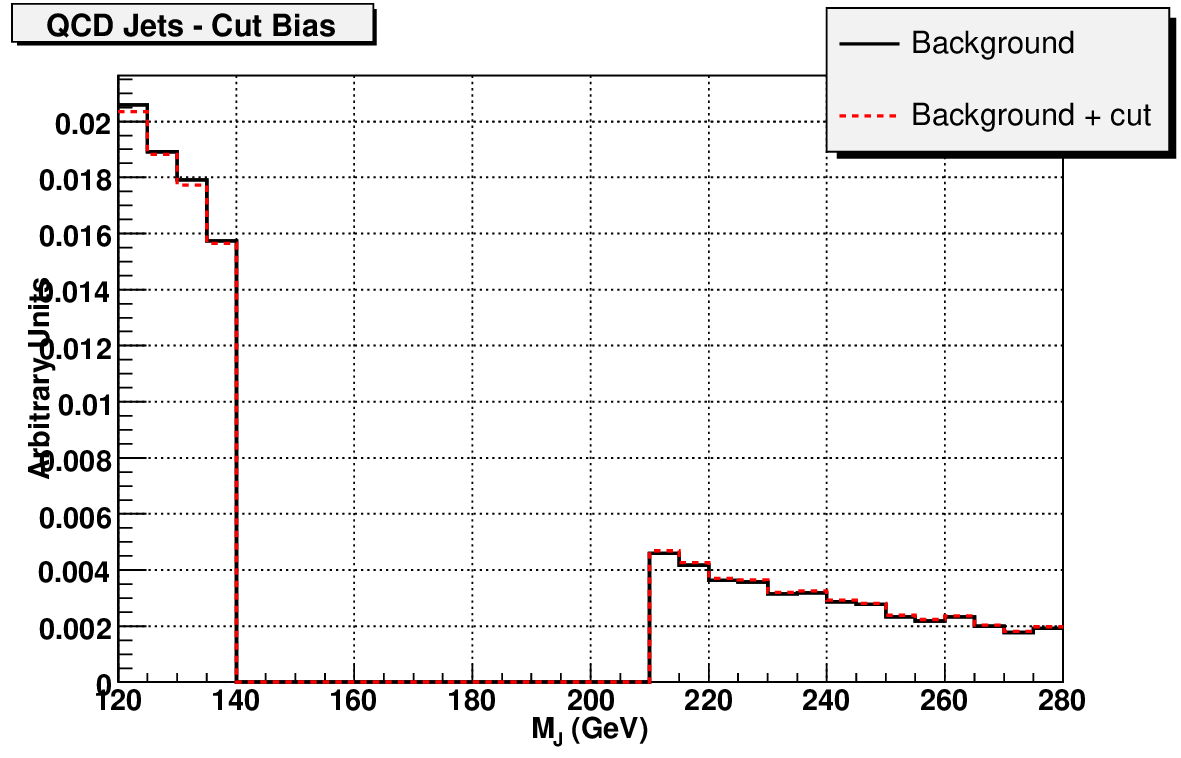}
\includegraphics[width=.44\hsize]{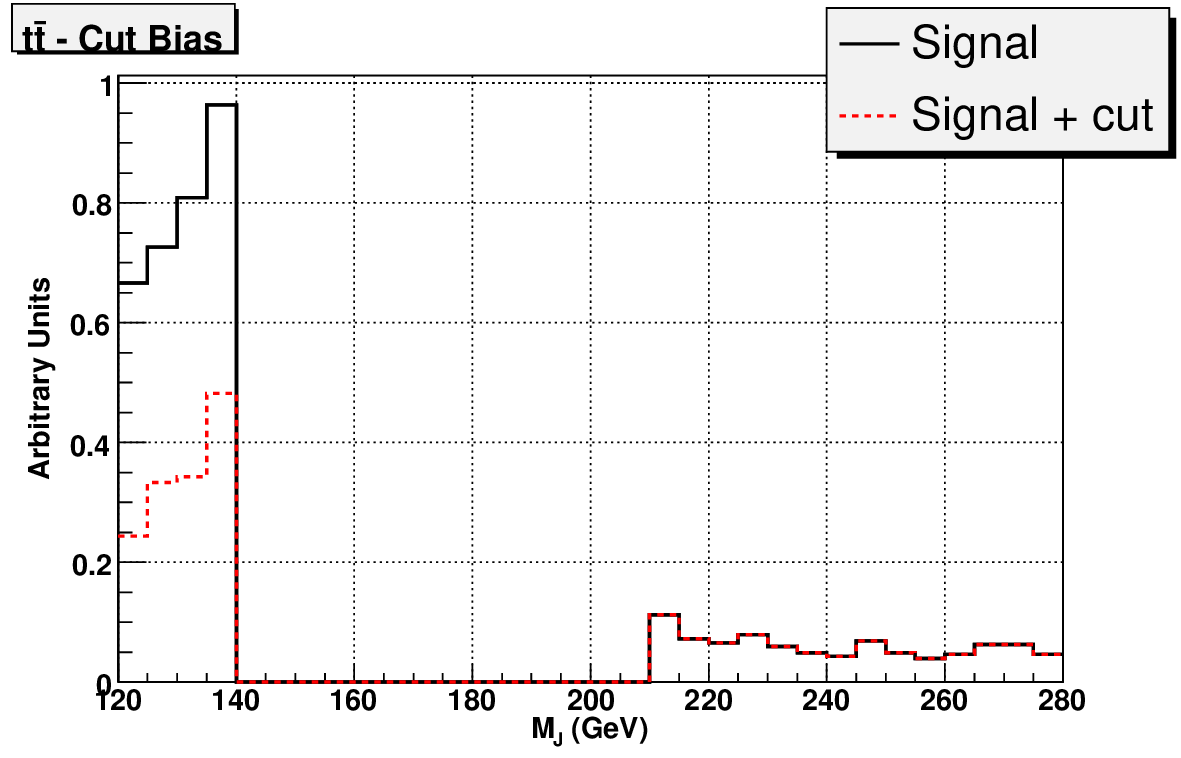} \\
\caption{\label{p:ccmass_bias}
The jet mass distributions for the $\ttbar$ and QCD jet samples in the sidebands. 
The plot on the left depicts the shape of the QCD jet sample before and after making a combined jet mass cut on $m_G$,
as described in Sec.~\ref{section:peak_resolution}.
Both curves are normalized to unit area, to show the similarity of the shapes before and after the cut.
The plot on the right depicts the effect of the combined jet mass cut on the $\ttbar$ signal.
The red (dashed) curve shows the effect of the cut relative to the original jet mass distribution (black solid curve).
Note: Unlike the left plot, these curves are not renormalized.
}
\end{center}
\end{figure}

We perform a sideband analysis in order to avoid the $\ttbar$ signal-rich region.
The basic goal is to understand the shape of the background by examining a region where there is no signal.
In the sidebands, in particular the low side, the signal contaminates the background.
In Fig.~\ref{p:ttbar_detector_effects}, we see that the $\ttbar$ signal does not vanish outside the top mass window.
Although it is small compared to the QCD background as can be seen in Fig.~\ref{p:results1}, 
this contamination substantially impacts resolution of the peak.
We attempt to purify the background in this region, 
by rejecting energetic jets consistent with originating from a top quark decay, i.e.~- signal, as follows.
For a candidate event where the leading jet passes preselection criteria, 
all jets within a cone $R = 1$ are (vectorially) added into a single combined jet.
We call this a group jet, although this definition differs slightly from that in J.~Conway, \et,  in~\cite{LHCnotes}.
If the group mass, $m_G$, of the combined jet falls within the top mass window, the candidate event is rejected.
This discriminant tends to reject events where the decay products of the top quarks are not fully collimated, 
i.e.~reconstructed as a single jet.
We must understand any biases introduced by this discriminant. 
Fig.~\ref{p:ccmass_bias} shows the effect of the $m_G$ cuts on the background and signal. The background shape is left relatively intact, but the signal is substantially diminished.

Advanced use of this $m_G$ discriminant is outside the scope of this analysis, possibly leading to more sophisticated analyses (see \eg  \ J.~Conway, \et, in~\cite{LHCnotes}). 
We simply use it to efficiently reject signal events in the sidebands, while keeping the majority of background events.

We analyze the shape of the background in the sidebands using the jet functions derived in section \ref{section_QCD}. 
We expect real QCD jets to be an admixture of quark and gluon jets.
Therefore, our Ansatz posits the admixture of quarks and gluons as a fraction.
We expect small corrections (deviations from a constant admixture) to arise from different sources.
For example, we do not consider events with a leading jet of fixed $\pt$, but rather impose a lower $\pt$ cut.
Our Ansatz for the jet mass distribution assumes the following form
\be
b ( m_J ) \propto \beta \left( m_J \right) \times J^{Q} \left( m_J; \ptmin, R \right) + \left( 1 - \beta \left( m_J \right) \right) \times J^{G} \left( m_J; \ptmin, R \right),
\ee
where $\beta \left( m_J \right)$ is a linear polynomial $\left( \beta_0 + \beta_1 \,{m_J\over \ptmin R} \right)$.
Note that with $b (m_J)$ defined above, along with Eq. (\ref{eqn:totaldistribution}), the total number of degree of freedom involved in the sideband fit is four:
$\beta_0$, $\beta_1$, $N_B$ and $N_S$.

%%%%%%%%%%%%%%%%%%%%%%%%%%%%%%%%%%%%%%%%%%%%%%%%%%%%%%%%%%%%%%%%%%%%
\subsubsection{Significance}\label{subsec_significance}
%%%%%%%%%%%%%%%%%%%%%%%%%%%%%%%%%%%%%%%%%%%%%%%%%%%%%%%%%%%%%%%%%%%%

After resolving the magnitude of the signal $\left( \ttbar \right)$ peak against that of the QCD jet background, via the methods outlined in the previous sections, 
we now discuss how to interpret those results.
Our analysis is based on  {\it{log-likelihood ratio}} method.\footnote{An excellent discussion may be found in the The Review of Particle Physics~\cite{pdg}.}
A background+signal hypothesis to describe a data sample is only meaningful if a background-only hypothesis is unlikely to describe that sample.
%In our analysis, we demand that a background-only hypothesis have less than 5\% probability, or p-value~\cite{pdg}, of obtaining a lower $\chi^2$.
% if this experiment were repeated many times.
We estimate the {\it{statistical}} significance, $n_{\sigma}$, of the peak as
\be
n_\sigma = \sqrt{2 \, \left( \log{\mathcal{L}} - \log{\mathcal{L}_0} \right) },
\label{eqn:significance}
\ee
% or, equivalently,
% \be
% \mathcal{L}_0 = \mathcal{L} \, \exp \left( - { n_\sigma^{2} \over 2} \right). \nonumber
% \ee
where $\mathcal{L}_0$ is the value of the maximized likelihood function obtained from fitting the data to the background shape alone (equivalent to setting $N_S$ to zero
in Eq.~(\ref{eqn:totaldistribution})),
and $\mathcal{L}$ is the value of the maximized likelihood function obtained from fitting the data to the background shape and signal shape.\footnote{Except in pathological cases, the significance is well approximated by $\frac{S}{\Delta S}$, 
where $S$ is the fitted signal, and $\Delta S$ is the error on $S$.}
The functional form of the likelihood function is given by
\be
\mathcal{L} = \prod_{k=1}^{N_{\rm BINS}} { \frac{ \exp \left( - F\left( m_k \right) \right) \times \left[ F\left( m_k \right) \right] ^{N_k} }{ N_k ! } },
\label{eqn:likelihood}
\ee
where we are fitting for the functional form of $F\left( m_J \right)$ as given by Eq.~(\ref{eqn:totaldistribution}).
Here, $m_k$ and $N_k$ refer to the value of the mass at the center, and the occupancy, of the $k$-th bin, respectively.

%%%%%%%%%%%%%%%%%%%%%%%%%%%%%%%%%%%%%%%%%%%%%%%%%%%%%%%%%%%%
\subsection{Single Top-Tagging}
\label{section_single_tag}
%%%%%%%%%%%%%%%%%%%%%%%%%%%%%%%%%%%%%%%%%%%%%%%%%%%%%%%%%%%%

\begin{figure}[hptb]
\begin{center}
\includegraphics[width=.44\hsize]{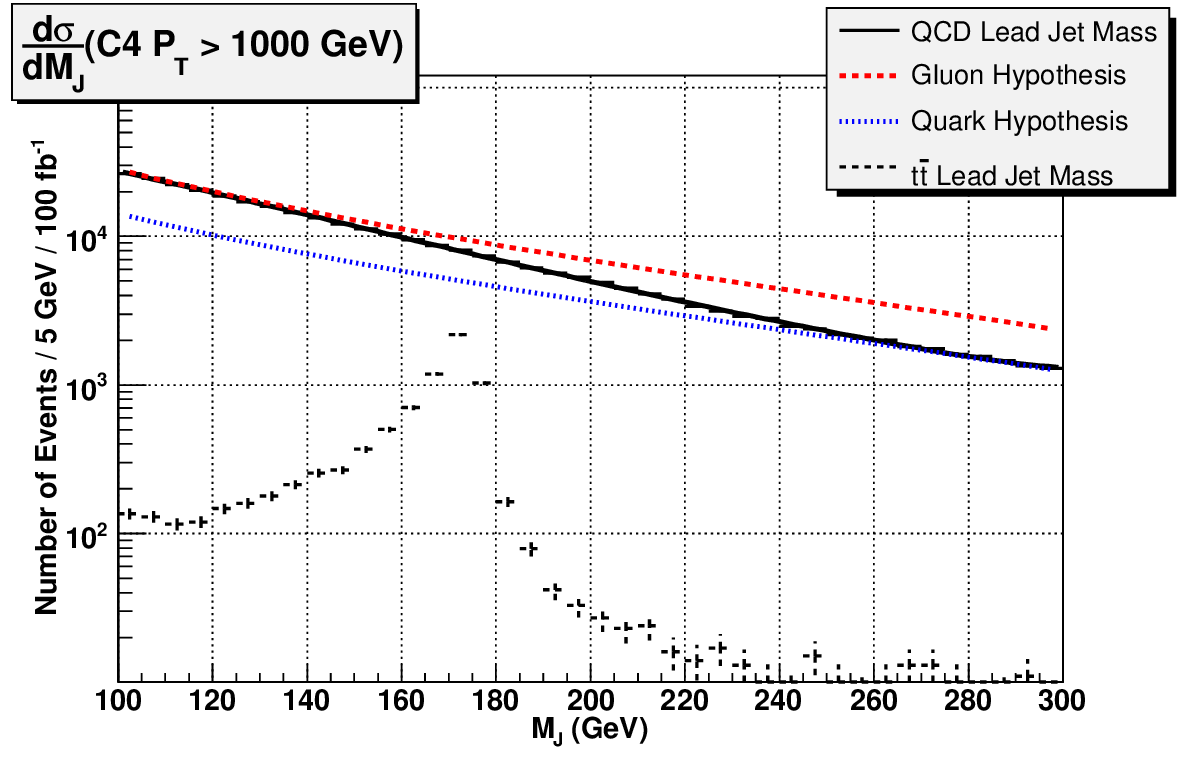}
\includegraphics[width=.44\hsize]{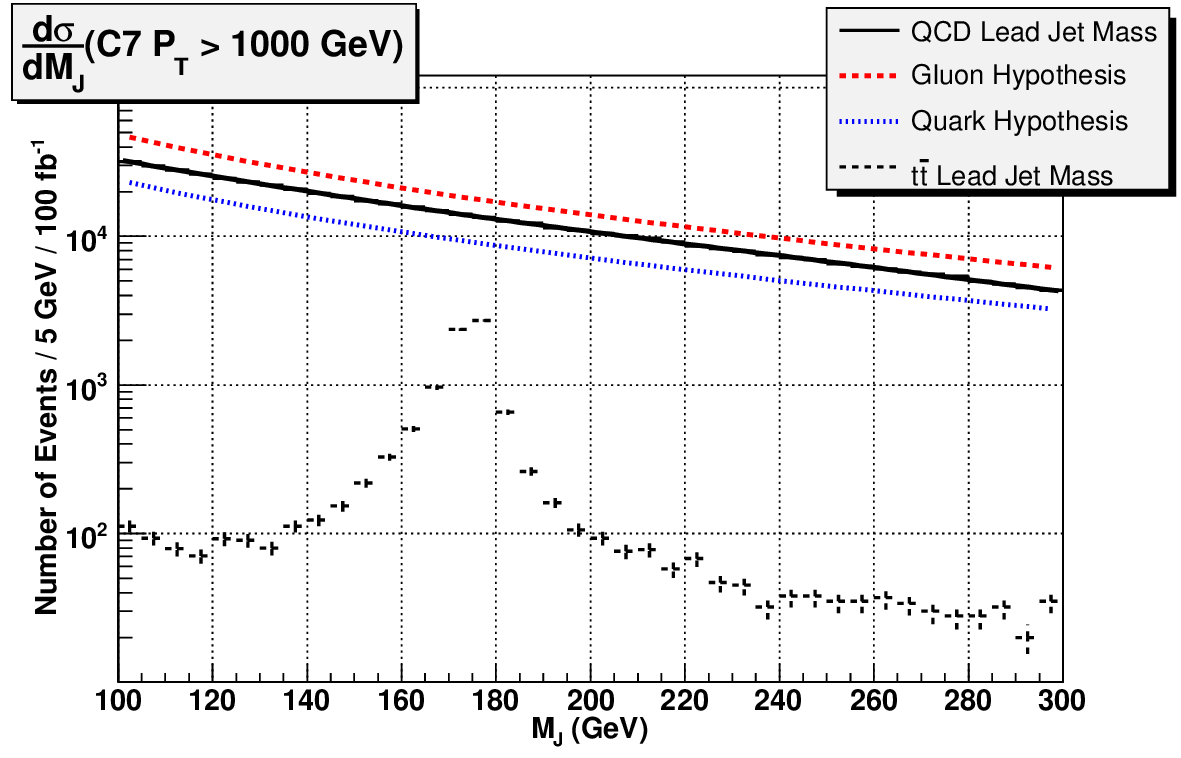} \\
\includegraphics[width=.44\hsize]{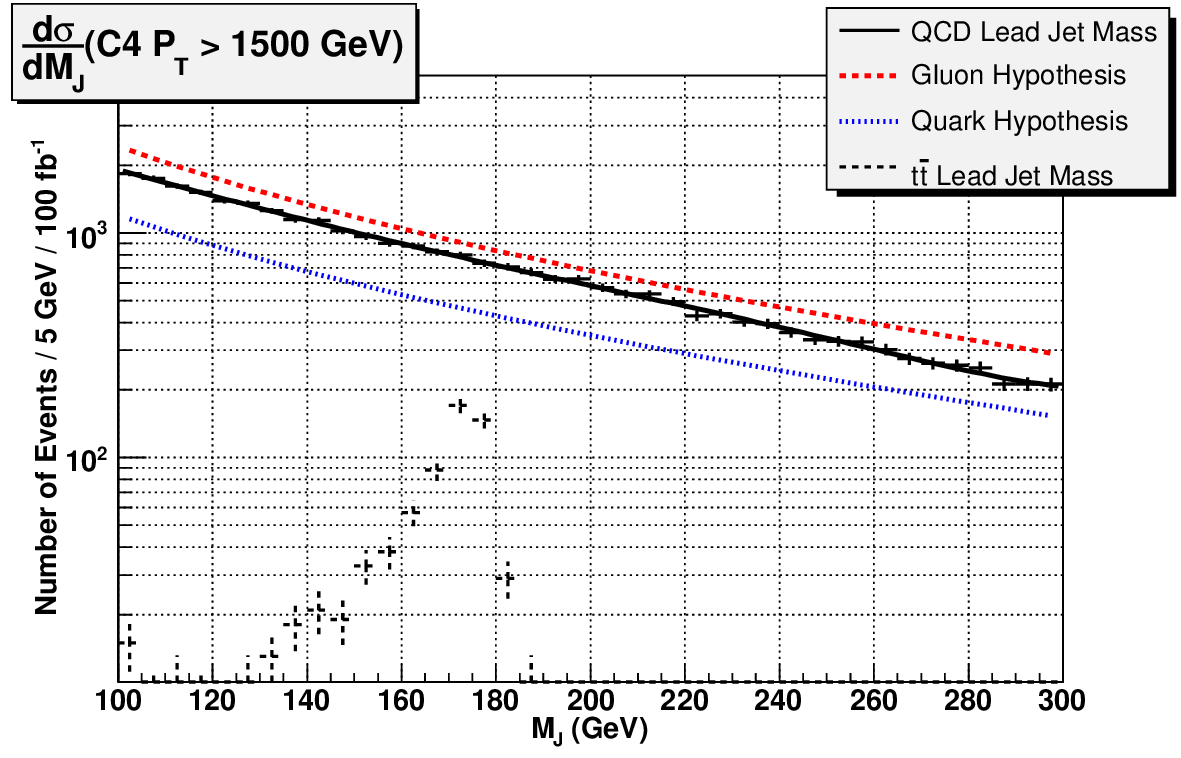}
\includegraphics[width=.44\hsize]{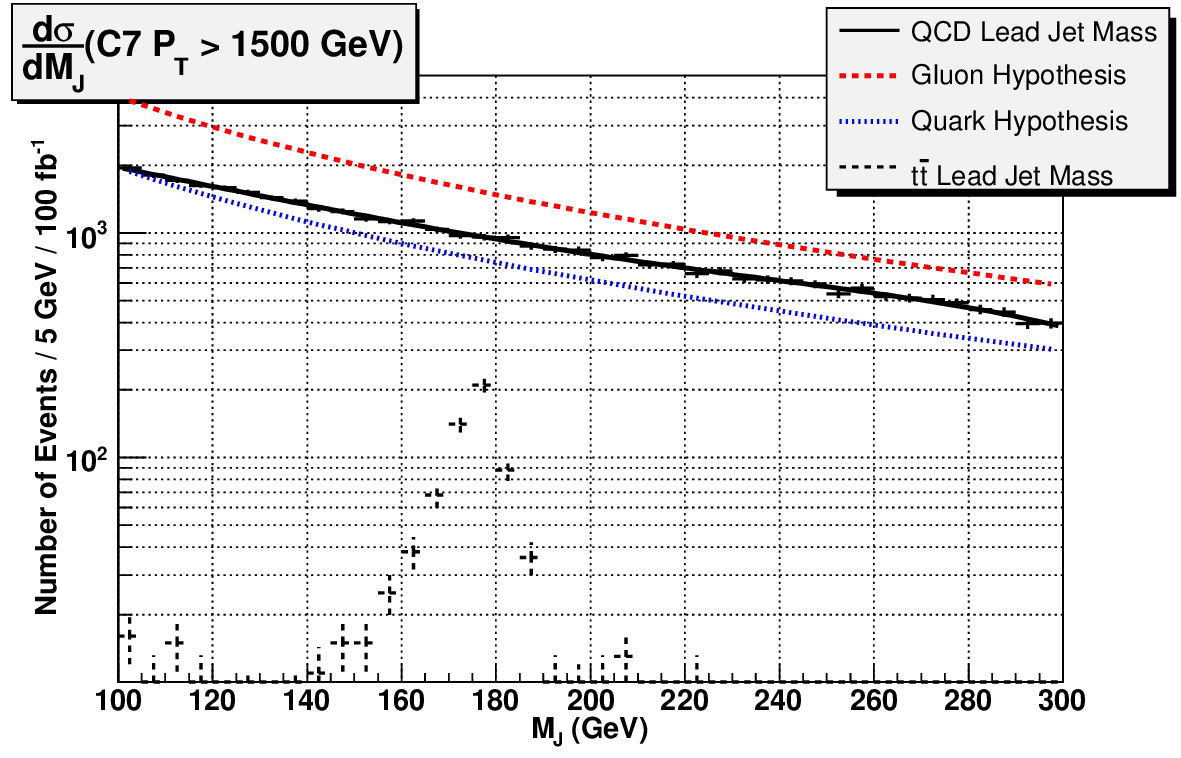} \\
\caption{\label{p:results1}The jet mass distributions for the $\ttbar$ and QCD jet samples. 
The plots on the top row correspond to a $\pt^{lead} \ge 1000 \gev$.
The plots on the bottom row correspond to a $\pt^{lead} \ge 1500 \gev$.
The plots on the left correspond to $R = 0.4$;
the plots on the right correspond to $R = 0.7$.
The theoretical bounds, Eq.~(\ref{eqn:predicted_curves}), are also plotted. 
These numbers are tabulated in table~\ref{t:single_tag}.
}
\end{center}
\end{figure}

For each of the signal ($\ttbar$) and background (QCD jets) samples, 
we preselect events with a $\pt$ cut on the leading jet.
In Fig.~\ref{p:results1} we plot the jet mass distribution including detector effects for the signal and background, including
the theoretical upper and lower bound for the background.
We show the number of events with jet mass in the range $\massmin \le m_J \le \massmax$.
For reference, the number of events for the signal and background, at the truth-level, are presented in table~\ref{t:single_tag}.
% we need to talk about this later, where we talk about omitted systematics.
%  taken from our Sherpa simulation, we emphasize that the signal to background ratio can be quite different if one to use a different PDF or event generators 
% (see for example table~\ref{t:cross_section}  in the above). 
It is clear that the background is roughly two orders of magnitude larger than the signal. 
Once we add detector effects the significance of the signal is further deteriorated.
We conclude that a simple counting method would not be effective here.
%
%%%%%%%%%%%
%%%%%%%%%%%
%
\subsubsection{Detector Effects}
Here, we repeat the truth-level procedure from above, 
accounting for the leading effects of detector resolution and $\pm 5\%$ jet energy scale.
We also tabulate the relative change in acceptance of the signal and background, due to detector resolution and energy scale, which we define as
\be
\Delta_{\rm JES} = \frac{N_{\rm JES}-N_{\rm TRUTH}}{N_{\rm TRUTH}},
\label{eqn:delta_acceptance}
\ee
where $N_{\rm JES}$ is the number of events passing the selection criteria after detector smearing and JES effects have been applied.
These results are tabulated in table~\ref{t:single_tag_xfx}, which shows how the signal and background are affected differently by smearing effects.
%{\bf SL: Add more discussion: Explain the table, and what it implies!}
We see that the net effects of the detector smearing plus the uncertainties in the JES lead to substantial uncertainties 
$\mathcal{O} \left( 10 \% - 30 \% \right)$ in the signal and background.
As anticipated, this leads to a clear failure of simple counting type analyses and calls for a different approach,
which will be introduced in the following in the form of sideband analyses and jet shapes.
\begin{table}[hptb]
\begin{center}
\begin{tabular}{|l|c|c|c|c|}
\hline
$\pt^{lead}$ cut & Cone Size & $\ttbar$ ($S$) & Background ($B$) & $S / B$ \\ 
\hline
1000 GeV & C4 & 6860 & 113749 & 0.060 \\
1000 GeV & C7 & 8725 & 197981 & 0.044 \\
1500 GeV & C4 & 630 & 10985 & 0.057 \\
1500 GeV & C7 & 689 & 13993 & 0.049 \\
\hline
\end{tabular}%
\caption{\label{t:single_tag}
Truth-level (no detector effects) results for single-tag jet mass method using, reflecting $100 \, \invfb$ of integrated luminosity.
}
\end{center}
\end{table}
\begin{table}[fhptb]
\begin{center}
\begin{tabular}{|c|c||c|c||c|c||c|c|}
\hline
$\pt^{lead}$ cut & Cone & $S$ (0\% JES) & $\Delta_{0}$ & +5\% JES & $\Delta_{5}$ & -5\% JES & $\Delta_{-5}$ \\ 
\hline
1000 GeV & C4 & 5778 & -15.8\% & 6562 & -4.3\% & 4798 & -30.1\%\\
1000 GeV & C7 & 7367 & -15.6\% & 8543 & -2.1\% & 6037 & -30.8\%\\
1500 GeV & C4 & 741 & 17.6\% & 934 & 48.3\% & 536 & -14.9\%\\
1500 GeV & C7 & 789 & 14.5\% & 1119 & 62.4\% & 601 & -12.8\%\\
\hline
\hline
$\pt^{lead}$ cut & Cone & $B$ (0\% JES) & $\Delta_{0}$ & +5\% JES & $\Delta_{5}$ & -5\% JES & $\Delta_{-5}$ \\
\hline
1000 GeV & C4 & 107661 & -5.4\% & 122291 & 7.5\% & 90232 & -20.7\%\\
1000 GeV & C7 & 192710 & -2.7\% & 224666 & 13.5\% & 154733 & -21.8\%\\
1500 GeV & C4 & 13615 & 23.9\% & 18144 & 65.2\% & 10108 & -8.0\%\\
1500 GeV & C7 & 18712 & 33.7\% & 25361 & 81.2\% & 13407 & -4.2\%\\
\hline
\end{tabular}%
\caption{\label{t:single_tag_xfx}
Acceptance of signal and background for the single tag method, relative to truth-level analysis, 
accounting for the leading effects of detector resolution and jet energy scale (JES).
The $\ttbar$ signal is represented in the top half; the QCD jet background is represented in the bottom half.
The statistics reflect $100 \, \invfb$ of integrated luminosity. 
$\Delta_{{\rm JES}}$ is the relative change in background for the indicated JES, 
relative to truth-level analysis in table~\ref{t:single_tag} (cf Eq.~(\ref{eqn:delta_acceptance})).
}
\end{center}
\end{table}

\subsubsection{Results for single tagging}
\begin{figure}[fhptb]
\begin{center}
\includegraphics[width=.74\hsize]{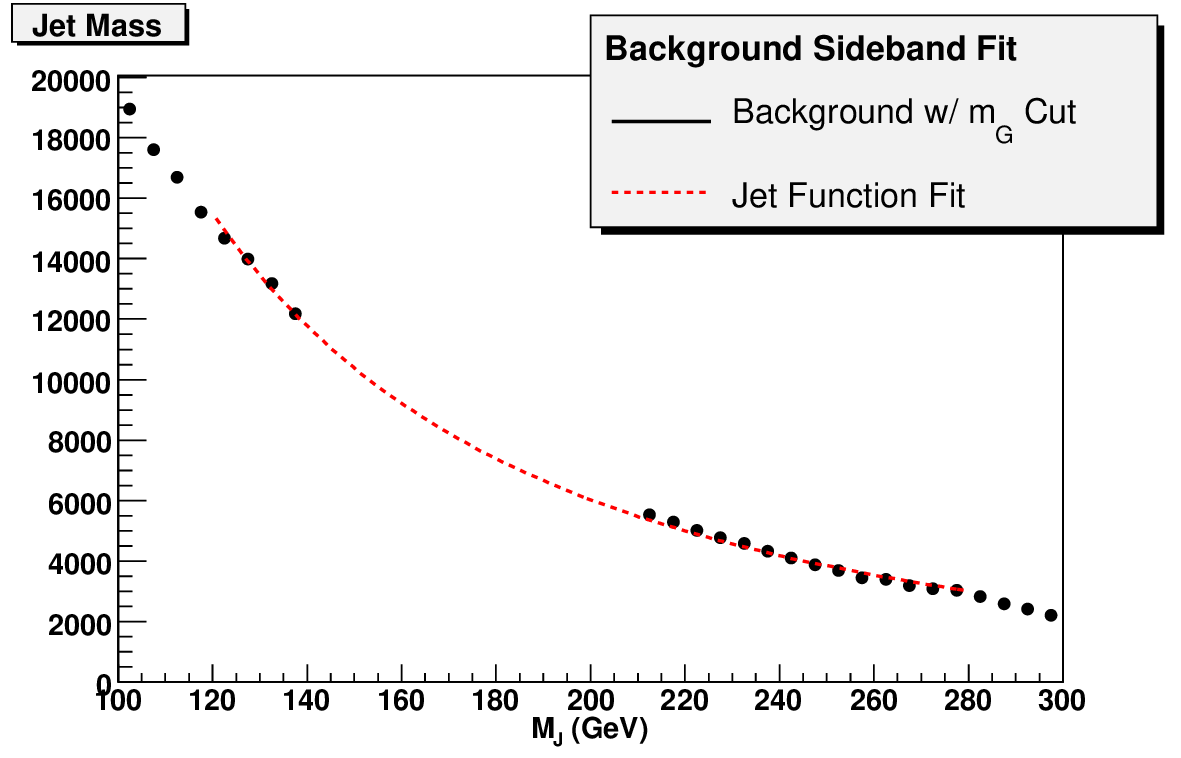} 
\caption{\label{p:background_fit}
A typical example of fitting jet functions to the jet mass distribution in the sideband regions 
$ \left( 120 \gev \le m_J \le 140 \gev \right) \cup \left( 210 \gev < m_J < 280 \gev \right)$.
This plot corresponds to a single-tag analysis with C7 jets with $\pt \ge 1000 \gev$.
}
\end{center}
\end{figure}

\begin{figure}[fhptb]
\begin{center}
\includegraphics[width=.44\hsize]{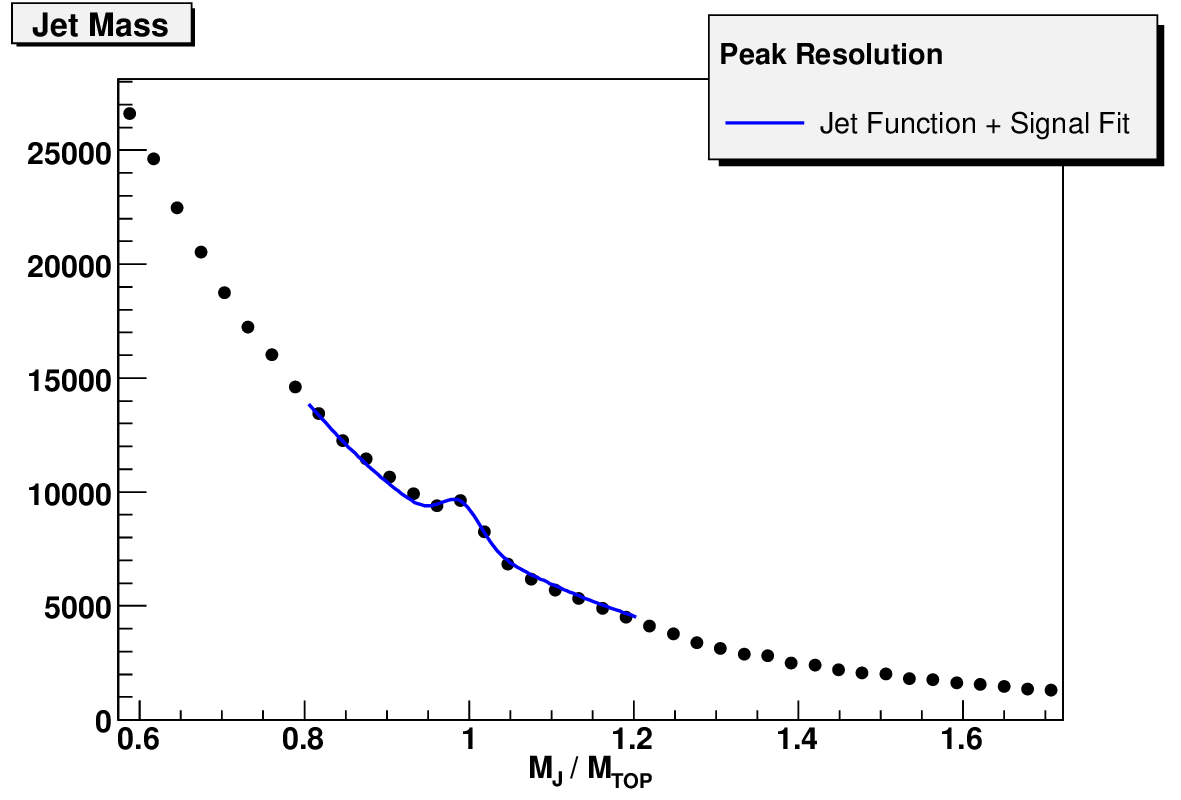} 
\includegraphics[width=.44\hsize]{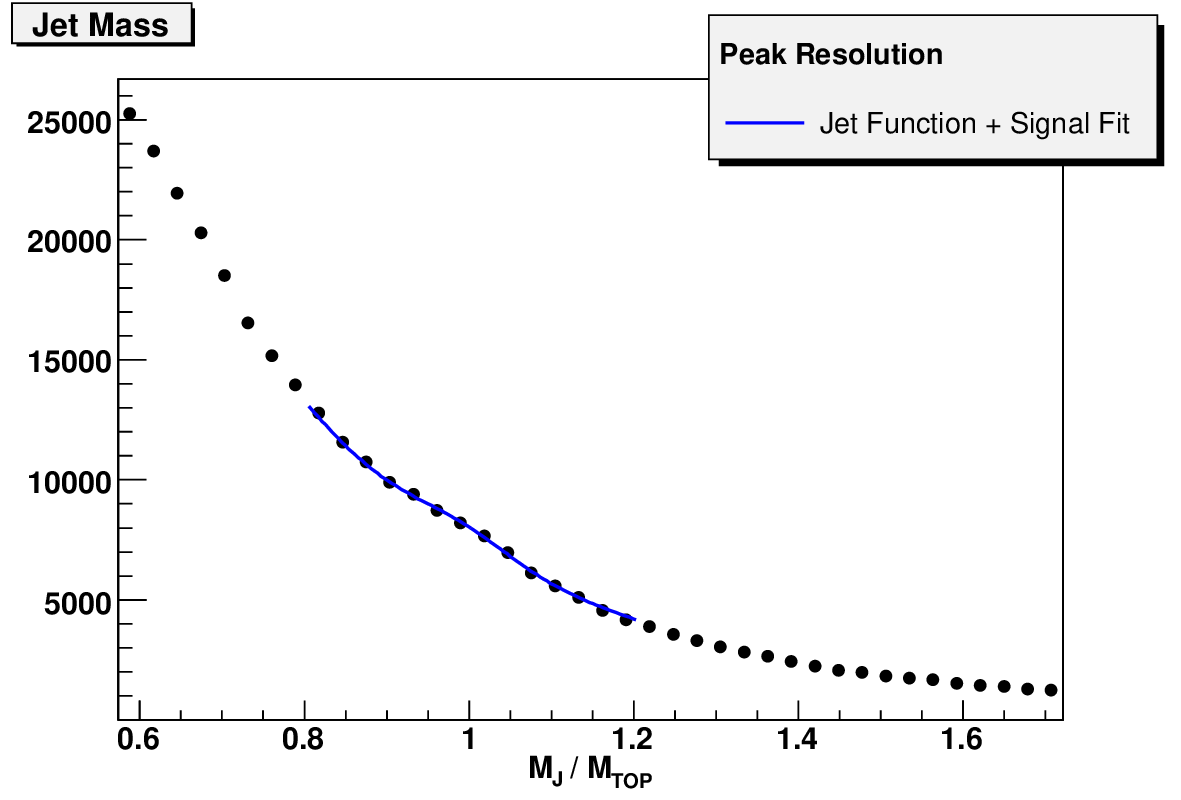} 
\caption{\label{p:signal_fit}
The results of fitting jet functions + signal shape to the jet mass distribution in the top mass window. 
The plot on the left corresponds to a truth-jet analysis.
The plot on the right depicts the effects of detector smearing.
The statistics reflect 100 $\invfb$ of integrated luminosity.
}
\end{center}
\end{figure}

We now apply the analysis described in the previous sections to resolve the peak related to the top quark in the signal region, the top mass window.
First we perform a sideband background analysis, to resolve the shape of the background. 
After applying the cuts described in Sec.~\ref{section:peak_resolution}, we fit the background to our Ansatz. 
Fig.~\ref{p:background_fit} shows an example of such background fit to our Ansatz.
%Finally one can perform a fit to background plus signal as described by Eq. (\ref{eqn:totaldistribution}).
The results of this fit described by Eq.~(\ref{eqn:totaldistribution}) and below are shown in Fig.~\ref{p:signal_fit}, 
which demonstrates how the detector affects the signal resolution. 

%For one of the cases, 
%We also compare results to those obtained by using the known shape of the (MC generated) background,
%also allowing the background and signal components to float
%This puts the results of the jet function method into perspective.
%

%%%%%%%%%%%%%%%%%
Our main results have been summarized in the tables below.
The results of the fitting procedures for the different $p_T$ cuts and cone sizes are shown in
tables~\ref{t:single_tag_significance} and \ref{t:single_tag_significance_low_luminosity}
for integrated luminosities of 100 $\invfb$ and 25 $\invfb$ respectively, and subsequently in tables~\ref{t:double_tag_significance} 
and \ref{t:double_tag_significance_low_luminosity} for the double top tagging case which is discussed in the following subsection.  
Our model for the background in these analyses was already introduced in subsection~\ref{subsec_significance}.
Apart from the cone size and $\ptmin$, for each JES, we show the result of the fit regarding the number of background ($B_{\rm FIT}$) and signal ($S_{\rm FIT}$) events in the mass window
and their ratio. $\Delta S$ is the error on $S_{\rm FIT}$  and p-value and $\chi^2 / ndf$ are given to describe the quality of the fit in each case ~\cite{pdg}. 
For our analysis, the total number of degree of freedom is 14 (18 bins $-$ 4 fit parameters: $\beta_0$, $\beta_1$, $N_B$ and $N_S$).

Most importantly, we give the statistical significance, $n_\sigma$ (defined in Eq.~(\ref{eqn:significance})), 
which is a measure of the probability that fluctuations of the {\it{proposed}} background yield in the observed data. 
The significance value is only as good as the p-value which indicates the goodness of fit.
We point out that for entries in which the p-value is lower than, say 5\%, 
the significance figure is probably not reliable. 
The fitting procedure on that data sample requires further examination,
for residuals and bias analyses, for example, but this falls outside the scope of this work.
We find two such instances of failed fits, both in table~\ref{t:single_tag_significance}. 
This also suggests how we are to interpret the results of the tables.
The relatively large background to signal ratio means that small errors in the background induce relatively large errors in the signal.
Furthermore, we have not quantified correlations between the background and signal shapes.
Similarities in the shapes can lead to small ambiguities, which are reflected in the fitting parameter errors.
The combination of these two difficulties gives rise to an effect, which, strictly speaking, is a defect in the analysis.
We remind the reader that we have a large uncertainty in the $\ttbar$ signal cross section (see table ~\ref{t:cross_section}), 
which we have not accounted for in the analysis.
We have singled out Sherpa MC data for use in our analyses, 
and the reader should bear this in mind when interpreting the results in tables ~\ref{t:single_tag_significance}, ~\ref{t:single_tag_significance_low_luminosity},
~\ref{t:double_tag_significance}, ~\ref{t:double_tag_significance_low_luminosity}.
Small errors in the background shape can yield good fits with high significance figures,
and still have relatively large errors in the signal.
We are led to interpret the results in the tables 
as the significance of the peak, relative to the indicated background shape hypothesis (the jet functions in our case).
We find that our single tagging method allows us to resolve the $\ttbar$ signal from the QCD background with $\ptmin\sim 1\, \tev$ and $25\,\invfb$ of data.
This jet mass analysis does not include any $b$-tagging or jet-shapes (to be discussed in the following section). 

\begin{table}[fhptb]
\begin{center}
\begin{tabular}{cc}
$\pt^{lead} \ge 1000 \gev$ & Cone $R$ = 0.4 \\
\end{tabular}
\begin{tabular}{|c|c|c|c|c|c|c|c|c|}
\hline
JES & $B_{\rm FIT}$ & $S_{\rm FIT}$ & $\Delta S$ &  $n_{\sigma}$ & p-value & $\chi^2 / ndf$ & $(S /  B)_ {\rm FIT}$ \\ 
\hline
\hline
%0\% & 107661 & 5778 & 658 & MC & 8.6 & 1.00 & 0.00 & 0.054 \\
0\% & 106571 & 6868 & 671 & 10.3 & 0.73 & 0.74 & 0.064 \\
\hline
%5\% & 122291 & 6562 & 681 & MC & 9.4 & 1.00 & 0.00 & 0.054 \\
5\% & 120717 & 8137 & 715 &  11.4 & 0.01 & 2.01 & 0.067 \\
\hline
%-5\% & 90232 & 4798 & 568 & MC & 8.3 & 1.00 & 0.00 & 0.053 \\
-5\% & 89136 & 5895 & 615 &  9.6 & 0.95 & 0.46 & 0.066 \\
\hline
%%%
\end{tabular}
\begin{tabular}{cc}
\\ $\pt^{lead} \ge 1000 \gev$ & Cone $R$ = 0.7 \\
\end{tabular}
\begin{tabular}{|c|c|c|c|c|c|c|c|c|}
\hline
JES & $B_{\rm FIT}$ & $S_{\rm FIT}$ & $\Delta S$ &  $n_{\sigma}$ & p-value & $\chi^2 / ndf$ & $(S /  B)_ {\rm FIT}$ \\ 
\hline
\hline
%0\% & 192710 & 7367 & 812* & MC & 9.1 & 1.00 & 0.00 & 0.038 \\
0\% & 189185 & 10892 & 800 &  13.7 & 0.09 & 1.52 & 0.058 \\
\hline
%5\% & 224666 & 8543 & 845 & MC & 9.9 & 1.00 & 0.00 & 0.038 \\
5\% & 219189 & 14020 & 859  & 16.4 & 0.02 & 1.87 & 0.064 \\
\hline
%-5\% & 154733 & 6037 & 705 & MC & 8.4 & 1.00 & 0.00 & 0.039 \\
-5\% & 151556 & 9214 & 720  & 12.9 & 0.63 & 0.83 & 0.061 \\
\hline
\end{tabular}
%%%
\begin{tabular}{cc}
\\ $\pt^{lead} \ge 1500 \gev$ & Cone $R$ = 0.4 \\
\end{tabular}
\begin{tabular}{|c|c|c|c|c|c|c|c|c|}
\hline
JES & $B_{\rm FIT}$ & $S_{\rm FIT}$ & $\Delta S$ &  $n_{\sigma}$ & p-value & $\chi^2 / ndf$ & $(S /  B)_ {\rm FIT}$ \\  
\hline\hline
%0\% & 13615 & 741 & 220 & MC & 3.3 & 1.00 & 0.00 & 0.054 \\
0\% & 13562 & 794 & 224 &  3.6 & 1.00 & 0.26 & 0.059 \\
\hline
%5\% & 18144 & 934 & 259 & MC & 3.5 & 1.00 & 0.00 & 0.051 \\
5\% & 17803 & 1275 & 256 &  5.0 & 0.89 & 0.58 & 0.072 \\
\hline
%-5\% & 10108 & 536 & 188 & MC & 2.8 & 1.00 & 0.00 & 0.053 \\
-5\% & 10155 & 489 & 193 & 2.5 & 0.94 & 0.49 & 0.048 \\
\hline
\end{tabular}
%%%
\begin{tabular}{cc}
\\ $\pt^{lead} \ge 1500 \gev$ & Cone $R$ = 0.7 \\
\end{tabular}
\begin{tabular}{|c|c|c|c|c|c|c|c|c|}
\hline
JES & $B_{\rm FIT}$ & $S_{\rm FIT}$ & $\Delta S$ &  $n_{\sigma}$ & p-value & $\chi^2 / ndf$ & $(S /  B)_ {\rm FIT}$ \\ 
\hline\hline
%0\% & 18712 & 789 & 253* & MC & 3.1 & 1.00 & 0.00 & 0.042 \\
0\% & 18456 & 1045 & 252 &  4.2 & 0.75 & 0.72 & 0.057 \\
\hline
%5\% & 25361 & 1119 & 287 & MC & 3.8 & 1.00 & 0.00 & 0.044 \\
5\% & 24921 & 1559 & 284 &  5.4 & 0.96 & 0.45 & 0.063 \\
\hline
%-5\% & 13407 & 601 & 213 & MC & 2.8 & 1.00 & 0.00 & 0.045 \\
-5\% & 13315 & 693 & 213 &  3.3 & 1.00 & 0.20 & 0.052 \\
\hline
\end{tabular}
\caption{\label{t:single_tag_significance}
Estimate of upper limit on significance of peak resolution via single tag method, accounting for detector smearing. 
$S_{\rm FIT} \, \mbox{and } \, B_{\rm FIT}$ are the results of an extended maximum likelihood fit.
$\Delta S$ is the error on $S_{\rm FIT}$.
Significance $n_{\sigma}$ is defined in Eq.~(\ref{eqn:significance}).
These results are derived with 100 $\invfb$ of integrated luminosity.
}
\end{center}
\end{table}

\begin{table}[fhptb]
\begin{center}
\begin{tabular}{cc}\\ 
$\pt^{lead} \ge 1000 \gev$ & Cone $R$ = 0.4 \\ 
\end{tabular}
\begin{tabular}{|c|c|c|c|c|c|c|c|c|c|}
\hline
JES & $B_{\rm FIT}$ & $S_{\rm FIT}$ & $\Delta S$ &  $n_{\sigma}$ & p-value & $\chi^2 / ndf$ & $(S /  B)_ {\rm FIT}$ \\ 
\hline
\hline
%0\% & 26915 & 1444 & 338 & MC & 4.3 & 1.00 & 0.00 & 0.054 \\
0\% & 26642 & 1712 & 335 &  5.1 & 1.00 & 0.19 & 0.064 \\
\hline
%5\% & 30573 & 1640 & 350 & MC & 4.7 & 1.00 & 0.00 & 0.054 \\
5\% & 30206 & 1995 & 346 &  5.8 & 0.96 & 0.45 & 0.066 \\
\hline
%-5\% & 22558 & 1200 & 291 & MC & 4.1 & 1.00 & 0.00 & 0.053 \\
-5\% & 22371 & 1379 & 288 &  4.8 & 1.00 & 0.11 & 0.062 \\
\hline
\end{tabular}
\begin{tabular}{cc}\\ 
$\pt^{lead} \ge 1000 \gev$ & Cone $R$ = 0.7 \\ 
\end{tabular}
\begin{tabular}{|c|c|c|c|c|c|c|c|c|c|}
\hline
JES & $B_{\rm FIT}$ & $S_{\rm FIT}$ & $\Delta S$ &  $n_{\sigma}$ & p-value & $\chi^2 / ndf$ & $(S /  B)_ {\rm FIT}$ \\ 
\hline
\hline
%0\% & 48178 & 1842 & 406 & MC & 4.5 & 1.00 & 0.00 & 0.038 \\
0\% & 47277 & 2730 & 399 &  6.8 & 0.98 & 0.38 & 0.058 \\
\hline
%5\% & 56166 & 2136 & 433 & MC & 4.9 & 1.00 & 0.00 & 0.038 \\
5\% & 54870 & 3419 & 424 &  8.1 & 0.87 & 0.60 & 0.062 \\
\hline
%-5\% & 38683 & 1509 & 361 & MC & 4.2 & 1.00 & 0.00 & 0.039 \\
-5\% & 37910 & 2274 & 354 &  6.4 & 1.00 & 0.21 & 0.060 \\
\hline
\end{tabular}
\begin{tabular}{cc}\\ 
$\pt^{lead} \ge 1500 \gev$ & Cone $R$ = 0.4 \\ 
\end{tabular}
\begin{tabular}{|c|c|c|c|c|c|c|c|c|c|c|}
\hline
JES & $B_{\rm FIT}$ & $S_{\rm FIT}$ & $\Delta S$ &  $n_{\sigma}$ & p-value & $\chi^2 / ndf$ & $(S /  B)_ {\rm FIT}$ \\ 
\hline
\hline
%0\% & 3404 & 185 & 112 & MC & 1.6 & 1.00 & 0.00 & 0.054 \\
0\% & 3381 & 201 & 112 &  1.8 & 1.00 & 0.06 & 0.059 \\
\hline
%5\% & 4536 & 234 & 133 & MC & 1.8 & 1.00 & 0.00 & 0.051 \\
5\% & 4418 & 346 & 130 & 2.7 & 1.00 & 0.07 & 0.078 \\
\hline
%-5\% & 2527 & 134 & 96 & MC & 1.4 & 1.00 & 0.00 & 0.053 \\
-5\% & 2519 & 136 & 96 & 1.4 & 1.00 & 0.09 & 0.054 \\
\hline
\end{tabular}
\begin{tabular}{cc}\\ 
$\pt^{lead} \ge 1500 \gev$ & Cone $R$ = 0.7 \\ 
\end{tabular}
\begin{tabular}{|c|c|c|c|c|c|c|c|c|c|}
\hline
JES & $B_{\rm FIT}$ & $S_{\rm FIT}$ & $\Delta S$ &  $n_{\sigma}$ & p-value & $\chi^2 / ndf$ & $(S /  B)_ {\rm FIT}$ \\ 
\hline
\hline
%0\% & 4678 & 197 & 126 & MC & 1.6 & 1.00 & 0.00 & 0.042 \\
0\% & 4609 & 259 & 125 & 2.1 & 1.00 & 0.18 & 0.056 \\
\hline
%5\% & 6340 & 280 & 147 & MC & 1.9 & 1.00 & 0.00 & 0.044 \\
5\% & 6231 & 382 & 144 & 2.6 & 1.00 & 0.12 & 0.061 \\
\hline
%-5\% & 3352 & 150 & 109 & MC & 1.4 & 1.00 & 0.00 & 0.045 \\
-5\% & 3320 & 174 & 99 &  1.6 & 1.00 & 0.06 & 0.052 \\
\hline
\end{tabular}
\caption{\label{t:single_tag_significance_low_luminosity}
Estimate of upper limit on significance of peak resolution via single tag method, accounting for detector smearing. 
$S_{\rm FIT} \, \mbox{and } \, B_{\rm FIT}$ are the results of an extended maximum likelihood fit.
$\Delta S$ is the error on $S_{\rm FIT}$.
Significance $n_{\sigma}$ is defined in Eq.~(\ref{eqn:significance}).
These results are derived with 25 $\invfb$ of integrated luminosity.
}
\end{center}
\end{table}
%Eq. ~\ref{eqn:significance}

%%%%%%%%%%%%%%%%%%%%%%%%%%%%%%%%%%%%%%%%%%%%%%%%%%%%%%%%%%%
%%%%%%%%%%%%%%%%%%%%%%%%%%%%%%%%%%%%%%%%%%%%%%%%%%%%%%%%%%%
%%%%%%%%%%%%%%%%%%
%%%%%%%%%%%%%%%%%%
\subsection{Double Top-Tagging}

\begin{figure}[hptb]
\begin{tabular}{cc}
\includegraphics[width=.48\hsize]{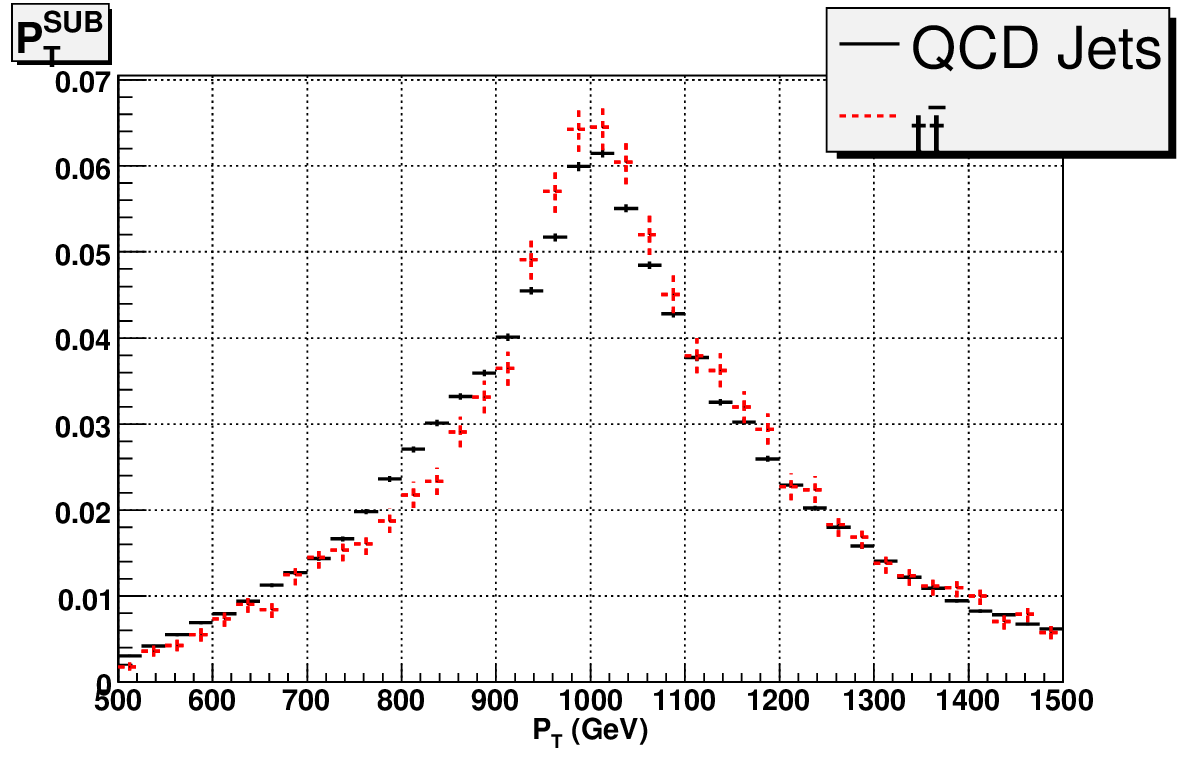}  &
\includegraphics[width=.48\hsize]{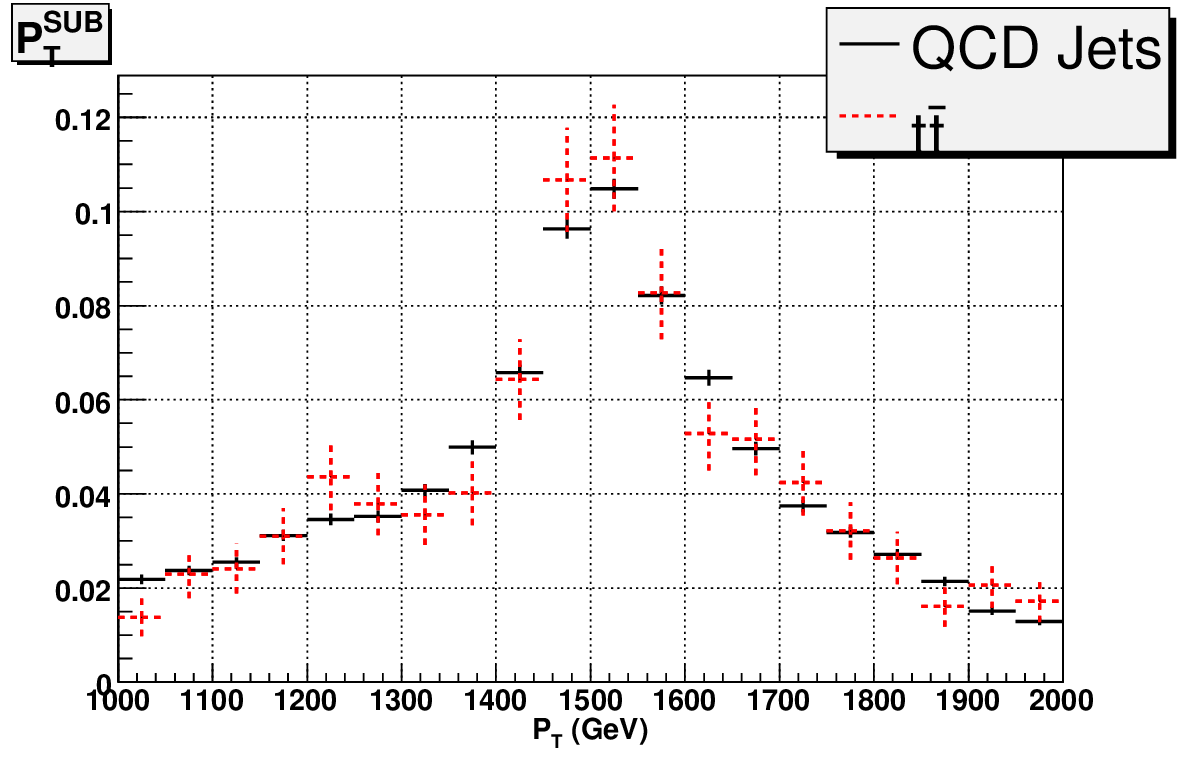} \\ 
\end{tabular}
\caption{
We compare the $p_T$ distribution of the subleading jet for the $\ttbar$ 
signal with (the red dotted curve) and without (the black solid curve) leading detector effects.
The plot on the left, right corresponds to C4 jets with $\left( \pt^{lead} \ge 1000,1500 \gev \right)$ respectively.
} \label{p:pt_2nd}
\end{figure}

The above analyses related to single top-tagging are useful not only for $\ttbar$ production, 
but rather for general cases in which we expect to have at least one very energetic top jet.
However, for the cases where there is more than one heavy high-$\pt$ particle, 
we certainly have more information which can be used to distinguish signal from the QCD background. 
Clearly, $\ttbar$ events contain more information than what is encoded in a single top jet mass.
We augment the single-tag analysis for the $\ttbar$ signal,
by simply requiring that the subleading jet mass be in the top mass window, without imposing a $\pt$ cut.
This cut preferentially removes more background events than the signal events, without biasing the distributions.
The sideband analysis, applied to the leading jet, remains the same as for the single top-tagging case.
As we shall see even this simple treatment yields a sizable improvement in the significance.
Roughly half of the events have smaller $p_T$ than the minimum $p_T$ for the leading jet as shown in Fig.~\ref{p:pt_2nd}. 
Although, by definition, a subleading jet has smaller $\pt$ than the leading one, its $\pt$ distribution 
is peaked at the $\pt^{min}$, and only small portion of events are in the smaller $\pt$ tail region.
The number of events for the signal and background, at the truth-level, are presented in table~\ref{t:double_tag}.
To get an idea on how the subleading  mass cut affects our signal and background samples, 
one can compare the numbers given in table~\ref{t:single_tag} with the ones in~\ref{t:double_tag}.
For example, we see that at truth level for $R=0.4$ and $\pt^{min}=1\,$TeV the size of the signal sample is decreased by 50\% while the background sample
by roughly 12\%.
This is consistent with the results shown in Figs.~\ref{p:collimation_rate} and \ref{p:fraction} in which the analysis is done for a fixed $\pt$.

In principle, one could apply a sideband analysis to the subleading jet. 
However, due to the fact that the $\pt$ is allowed to float, the required analysis would necessarily be more complicated.
The double-tagging method increases the signal-to-background ratio,
and the significance of the measurements increases.
The leading effects of detector resolution and jet energy scale on the signal and background acceptance can be seen in Tables ~\ref{t:double_tag_significance} and  ~\ref{t:double_tag_significance_low_luminosity}.
We find that our double tagging method yields a reach of up to $\pt^{min}\sim 1.5\,$TeV with 100~fb$^{-1}$, without relying
on $b$-tagging or jet-shapes (to be discussed in the following section).

\begin{table}[hptb]
\begin{center}
\begin{tabular}{|l|c|c|c|c|}
\hline
$\pt^{lead}$ cut & Cone Size & $\ttbar$ ($S$) & Background ($B$) & $S / B$ \\ 
\hline
1000 GeV & C4 & 3430 & 13505 & 0.254 \\
1000 GeV & C7 & 6302 & 36765 & 0.171 \\
1500 GeV & C4 & 403 & 1874 & 0.215 \\
1500 GeV & C7 & 458 & 2724 & 0.168 \\
\hline
\end{tabular}%
\caption{\label{t:double_tag}
Truth-level (no detector effects) results for double-tag jet mass method using, reflecting $100 \, \invfb$ of integrated luminosity.}
\end{center}
\end{table}

\begin{table}[fhptb]
\begin{center}
\begin{tabular}{|c|c||c|c||c|c||c|c|}
\hline
$\pt^{lead}$ cut & Cone & $S$ (0\% JES) & $\Delta_{0}$ & +5\% JES & $\Delta_{5}$ & -5\% JES & $\Delta_{-5}$ \\ 
\hline
\hline
1000 GeV & C4 & 2601 & -24.2\% & 2868 & -16.4\% & 2228 & -35.0\%\\
1000 GeV & C7 & 4563 & -27.6\% & 5351 & -15.1\% & 3765 & -40.3\%\\
1500 GeV & C4 & 403 & 0.0\% & 489 & 21.3\% & 292 & -27.5\%\\
1500 GeV & C7 & 487 & 6.3\% & 688 & 50.2\% & 352 & -23.1\%\\
\hline
$\pt^{lead}$ cut & Cone & $B$ (0\% JES) & $\Delta_{0}$ & +5\% JES & $\Delta_{5}$ & -5\% JES & $\Delta_{-5}$ \\
\hline
1000 GeV & C4 & 13680 & 1.3\% & 15187 & 12.5\% & 12054 & -10.7\%\\
1000 GeV & C7 & 39361 & 7.1\% & 45596 & 24.0\% & 32192 & -12.4\%\\
1500 GeV & C4 & 2373 & 26.6\% & 3109 & 65.9\% & 1746 & -6.8\%\\
1500 GeV & C7 & 4195 & 54.0\% & 5651 & 107.5\% & 3014 & 10.6\%\\
\hline
\end{tabular}%
\caption{\label{t:double_tag_xfx}
Acceptance of signal and background for the double tag method, relative to truth-level analysis, 
accounting for the leading effects of detector resolution and jet energy scale (JES).
The $\ttbar$ signal is represented in the top half; the QCD jet background is represented in the bottom half.
The statistics reflect $100 \, \invfb$ of integrated luminosity. 
$\Delta_{\rm JES}$ is the relative change in background for the indicated JES, 
relative to truth-level analysis in table~\ref{t:double_tag} (cf Eq.~(\ref{eqn:delta_acceptance})).
}
\end{center}
\end{table}

%%%%%%%%%%%%%%
%%%%%%%%%%%%%%

%%%%%%%%%%%%%%%%%%%%The new modified double tagger%%%%%%%%%%%%%%%%%%%%%%%%%%%%%%%%%%%%%%%%
\begin{table}[fhptb]
\begin{center}
%%%
\begin{tabular}{cc}\\ 
$\pt^{lead} \ge 1000 \gev$ & Cone $R$ = 0.4 \\ 
\end{tabular}
\begin{tabular}{|c|c|c|c|c|c|c|c|c|c|}
\hline
JES & $B_{\rm FIT}$ & $S_{\rm FIT}$ & $\Delta S$ &  $n_{\sigma}$ & p-value & $\chi^2 / ndf$ & $(S /  B)_ {\rm FIT}$ \\ 
\hline
\hline

0\% & 13488 & 2789 & 237 & 11.8 & 0.99 & 0.33 & 0.207 \\
\hline
5\% & 14653 & 3395 & 255 & 13.3 & 0.94 & 0.50 & 0.232 \\
\hline
-5\% & 11762 & 2516 & 212 & 11.9 & 0.99 & 0.31 & 0.214 \\
\hline
\end{tabular}
\begin{tabular}{cc}\\ 
$\pt^{lead} \ge 1000 \gev$ & Cone $R$ = 0.7 \\ 
\end{tabular}
\begin{tabular}{|c|c|c|c|c|c|c|c|c|c|}
\hline
JES & $B_{\rm FIT}$ & $S_{\rm FIT}$ & $\Delta S$ &  $n_{\sigma}$ & p-value & $\chi^2 / ndf$ & $(S /  B)_ {\rm FIT}$ \\  
\hline
\hline
0\% & 38101 & 5813 & 358 & 16.2 & 0.72 & 0.76 & 0.153 \\
\hline
5\% & 43993 & 6943 & 386 & 18.0 & 0.66 & 0.81 & 0.158 \\
\hline
-5\% & 31290 & 4655 & 320 & 14.6 & 0.57 & 0.89 & 0.149 \\
\hline
\end{tabular}
\begin{tabular}{cc}\\ 
$\pt^{lead} \ge 1500 \gev$ & Cone $R$ = 0.4 \\ 
\end{tabular}
\begin{tabular}{|c|c|c|c|c|c|c|c|c|c|}
\hline
JES & $B_{\rm FIT}$ & $S_{\rm FIT}$ & $\Delta S$ &  $n_{\sigma}$ & p-value & $\chi^2 / ndf$ & $(S /  B)_ {\rm FIT}$ \\  
\hline
\hline
0\% & 2341 & 430 & 94 & 4.6 & 0.99 & 0.35 & 0.184 \\
\hline
5\% & 2968 & 624 & 110 & 5.7 & 0.96 & 0.45 & 0.210 \\
\hline
-5\% & 1593 & 436 & 79 & 5.5 & 0.82 & 0.66 & 0.274 \\
\hline
\end{tabular}
\begin{tabular}{cc}\\ 
$\pt^{lead} \ge 1500 \gev$ & Cone $R$ = 0.7 \\ 
\end{tabular}
\begin{tabular}{|c|c|c|c|c|c|c|c|c|c|}
\hline
JES & $B_{\rm FIT}$ & $S_{\rm FIT}$ & $\Delta S$ &  $n_{\sigma}$ & p-value & $\chi^2 / ndf$ & $(S /  B)_ {\rm FIT}$ \\  
\hline
\hline
0\% & 4053 & 625 & 129 & 5.2 & 1.00 & 0.28 & 0.154 \\
\hline
5\% & 5532 & 801 & 128 & 6.3 & 0.93 & 0.50 & 0.145 \\
\hline
-5\% & 2965 & 399 & 100 & 4.0 & 1.00 & 0.14 & 0.135 \\
\hline
\end{tabular}
\caption{\label{t:double_tag_significance}
Estimate of upper limit on significance of peak resolution via double tag method, 
accounting for detector smearing, and jet energy scale (JES).
$S_{\rm FIT} \, \mbox{and } \, B_{\rm FIT}$ are the results of an extended maximum likelihood fit.
$\Delta S$ is the error on $S_{\rm FIT}$.
Significance $n_{\sigma}$ is defined in Eq.~(\ref{eqn:significance}).
These results are derived with 100 $\invfb$ of integrated luminosity.
}
\end{center}
\end{table}
%%%%%%%%%%%%%%
%%%%%%%%%%%%%%

%%%%%%%%%%%%%%%%%%%%The new modified double tagger%%%%%%%%%%%%%%%%%%%%%%%%%%%%%%%%%%%%%%%%
\begin{table}[fhptb]
\begin{center}
%%%
\begin{tabular}{cc}\\ 
$\pt^{lead} \ge 1000 \gev$ & Cone $R = 0.4$ \\ 
\end{tabular}
\begin{tabular}{|c|c|c|c|c|c|c|c|c|c|}
\hline
JES & $B_{\rm FIT}$ & $S_{\rm FIT}$ & $\Delta S$ &  $n_{\sigma}$ & p-value & $\chi^2 / ndf$ & $(S /  B)_ {\rm FIT}$ \\ 
\hline
0\% & 3367 & 696 & 119 & 5.9 & 1.00 & 0.08 & 0.207 \\
\hline
5\% & 3658 & 848 & 128 & 6.7 & 1.00 & 0.12 & 0.232 \\
\hline
-5\% & 2931 & 631 & 106 & 6.0 & 1.00 & 0.07 & 0.215 \\
\hline
\end{tabular}
\begin{tabular}{cc}\\ 
$\pt^{lead} \ge 1000 \gev$ & Cone $R = 0.7$ \\ 
\end{tabular}
\begin{tabular}{|c|c|c|c|c|c|c|c|c|c|}
\hline
JES & $B_{\rm FIT}$ & $S_{\rm FIT}$ & $\Delta S$ &  $n_{\sigma}$ & p-value & $\chi^2 / ndf$ & $(S /  B)_ {\rm FIT}$ \\  
\hline
\hline
0\% & 9521 & 1452 & 181 & 8.1 & 1.00 & 0.19 & 0.152 \\
\hline
5\% & 10997 & 1732 & 193 & 9.0 & 1.00 & 0.20 & 0.158 \\
\hline
-5\% & 7817 & 1162 & 160 & 7.3 & 1.00 & 0.22 & 0.149 \\
\hline
\end{tabular}
\begin{tabular}{cc}\\ 
$\pt^{lead} \ge 1500 \gev$ & Cone $R = 0.4$ \\ 
\end{tabular}
\begin{tabular}{|c|c|c|c|c|c|c|c|c|c|}
\hline
JES & $B_{\rm FIT}$ & $S_{\rm FIT}$ & $\Delta S$ &  $n_{\sigma}$ & p-value & $\chi^2 / ndf$ & $(S /  B)_ {\rm FIT}$ \\  
\hline
\hline
0\% & 577 & 111 & 47 & 2.4 & 1.00 & 0.08 & 0.192 \\
\hline
5\% & 737 & 155 & 55 & 2.8 & 1.00 & 0.11 & 0.210 \\
\hline
-5\% & 393 & 109 & 40 & 2.8 & 1.00 & 0.16 & 0.277 \\
\hline
\end{tabular}
\begin{tabular}{cc}\\ 
$\pt^{lead} \ge 1500 \gev$ & Cone $R = 0.7$ \\ 
\end{tabular}
\begin{tabular}{|c|c|c|c|c|c|c|c|c|c|}
\hline
JES & $B_{\rm FIT}$ & $S_{\rm FIT}$ & $\Delta S$ &  $n_{\sigma}$ & p-value & $\chi^2 / ndf$ & $(S /  B)_ {\rm FIT}$ \\  
\hline
\hline
0\% & 1005 & 159 & 70 & 2.7 & 1.00 & 0.06 & 0.158 \\
\hline
5\% & 1376 & 200 & 64 & 3.1 & 1.00 & 0.12 & 0.145 \\
\hline
-5\% & 739 & 96 & 50 & 1.9 & 1.00 & 0.04 & 0.130 \\
\hline
\end{tabular}
\caption{\label{t:double_tag_significance_low_luminosity}
Estimate of upper limit on significance of peak resolution via double tag method, 
accounting for detector smearing, and jet energy scale (JES).
$S_{\rm FIT} \, \mbox{and } \, B_{\rm FIT}$ are the results of an extended maximum likelihood fit.
$\Delta S$ is the error on $S_{\rm FIT}$.
Significance $n_{\sigma}$ is defined in Eq.~(\ref{eqn:significance}).
These results are derived with 25 $\invfb$ of integrated luminosity.
}
\end{center}
\end{table}
%%%%%%%%%%%%%%

%%%%%%%%%%%%%%%%%%%%%%%%%%%%%%%%%%%%%%%%%%%%%%%%%%%%%%%%%%%%%%%%%%%%%%%%%%%%%
%%%%%%%%%%%%%%%%%%%%%%%%%%%%%%%%%%%%%%%%%%%%%%%%%%%%%%%%%%%%%%%%%%%%%%%%%%%%%

%%%%%%%%%%%%%%%%%%%%%%%%%%%%%%%%%%%%%%%%%%%%%%%%%%%%%%%%%%%%%%%%%%%%%%%%%%%%%
%%%%%%%%%%%%%%%%%%%%%%%%%%%%%%%%%%%%%%%%%%%%%%%%%%%%%%%%%%%%%%%%%%%%%%%%%%%%%

%%%%%%%%%%%%%%%%%%%%%%%%%%%%%%%%%%%%
\section{Jet Substructure}
\label{section_substructure}

\begin{figure}[fhptb]
\begin{center}
\includegraphics[width=.74\hsize]{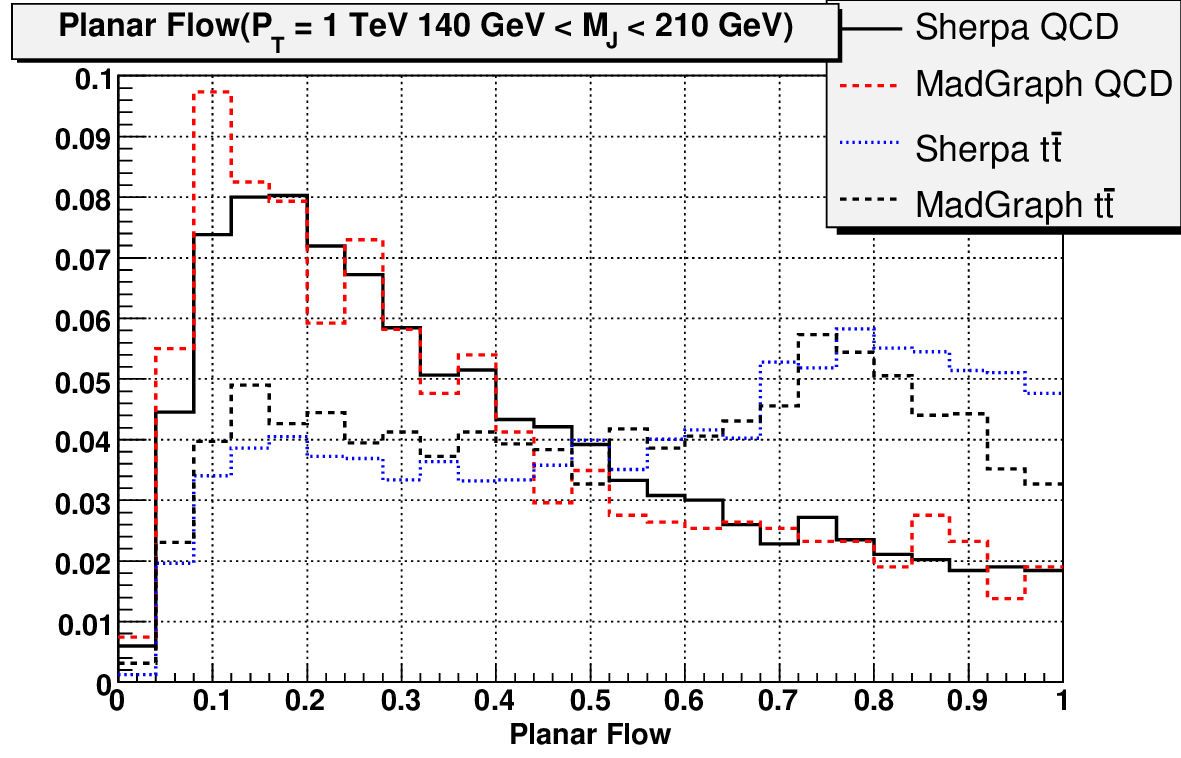}
\caption{The planar flow distribution is plotted for $\ttbar$ and QCD jets with mass in the top mass window, $\massmin \le m_J \le \massmax$. 
Sherpa and MG/ME distributions are represented.
} \label{p:planarity}
\end{center}
\end{figure}

We discussed simple single- and double-mass tagging methods, 
%and have shown that they are, by themselves, inadequate for resolving SM $\ttbar$ signals.
and we found that we may need additional handles in order to resolve SM $\ttbar$ signals for smaller integrated luminosities or higher $\pt$.
We discuss briefly the possibility of using substructure to further analyze energetic jets in the top mass window.
We defer the details to our recent work in~\cite{us} (see also~\cite{Thaler:2008ju}), where we derive simple analytic expressions to approximate the jet shape variable distributions.
For developing additional tools to reslove $\ttbar$ signals, there are approaches which exploit information outside of hadronic calorimeter~\cite{Stras_ph07} such as tracker or electromagnetic calorimeter. But we limit ourselves to the information encoded only within the hadronic calorimeter to develop significance for resolving
$\ttbar$ signals. We do not also discuss the possibility of $b$-tagging for high $p_T$ top-jet~\cite{BTagging}, which is still under speculation for the range of $p_T$
relevant for our analysis.

Jet shapes are the extensions of well-known event shapes, used at lepton colliders, applied to the analysis of energy flow inside single jets.
The fact that we consider only jets with high mass is crucial since it allows us to control the shape of various
distributions related to energy flow in a perturbative manner.
\begin{figure}[fhptb]
\begin{center}
\includegraphics[width=.74\hsize]{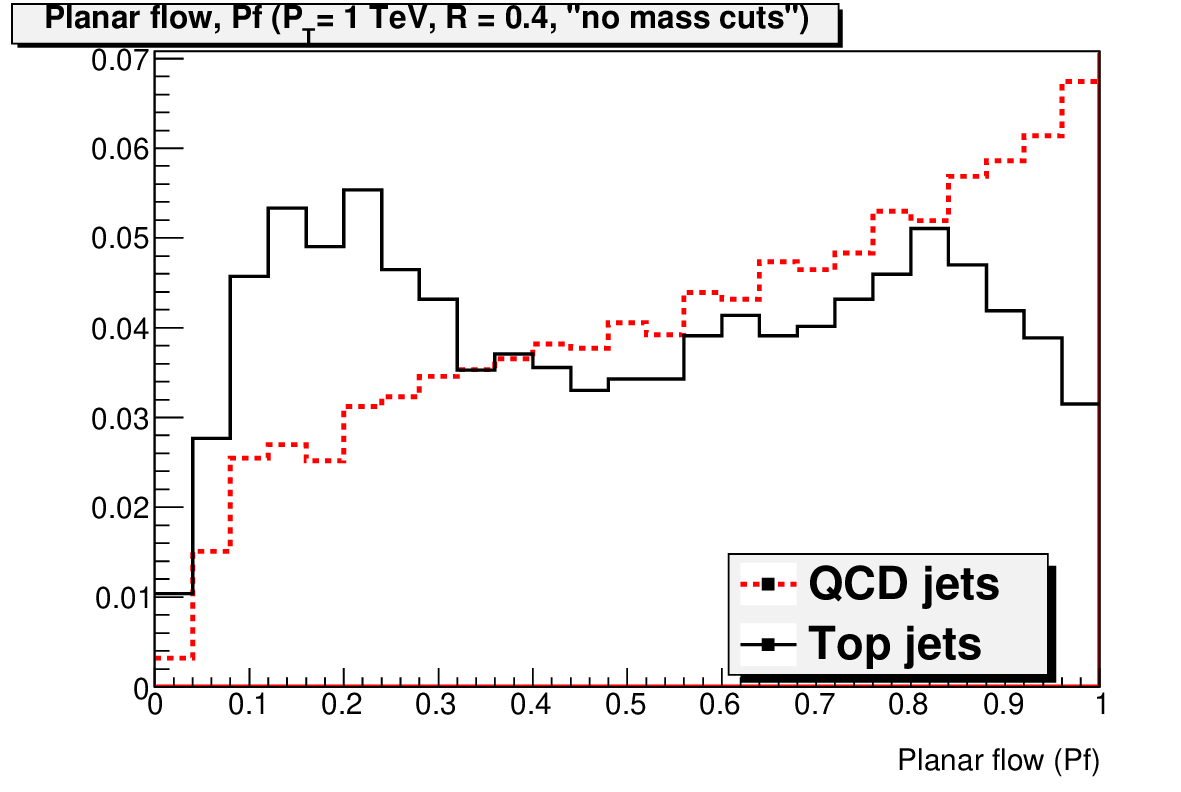} 
\end{center}
\caption{The planar flow distribution is plotted for $\ttbar$ and QCD jets without fixing jet mass. 
MG/ME distributions are represented.
}\label{p:planarity_no_fixing}
\end{figure}
As an example, we examine the {\it{planar flow}} variable, which measures the extent to which the energy flow inside the jet is linear or planar.
Planar flow ($Pf$) is defined as follows.
We first define an (unnormalized) event shape tensor $I_w$ as\footnote{The overall normalization is not important to this discussion.}
\be
I^{kl}_{w}= \sum_i w_i \frac{p_{i,k}}{w_i}\,\frac{p_{i,l}}{w_i}\, ,
\ee
where $w_i$ is the energy of particle $i$ in the jet,
and $p_{i,k}$ is the $k^{th}$ component of its transverse momentum relative to the thrust axis, which typically coincides with the jet axis.
Given $I_{w}$, we define $Pf$ as
\be
Pf ={4\,{\rm det}(I_w)\over{\rm tr}(I_w)^2}={4 \lambda_1 \lambda_2\over(\lambda_1 + \lambda_2)^2}\, ,
\ee
where $\lambda_{1,2}$ are the eigenvalues of $I_w$.
$Pf$ approaches zero for linear shapes and approaches unity for isotropic depositions of energy.
In Fig.~\ref{p:planarity}, we plot the planar flow distributions for QCD jets and $\ttbar$.
As can be seen by comparing Figs.~\ref{p:planarity} and~\ref{p:planarity_no_fixing}, it is crucial to consider only events in the top mass window. 
Without a jet mass cut, the jet shape analysis loses its rejection power.

\section{Top Quark Polarization Measurement}\label{top_polarization}

\begin{figure}[fhptb]
\begin{tabular}{cc}
\includegraphics[width=.48\hsize]{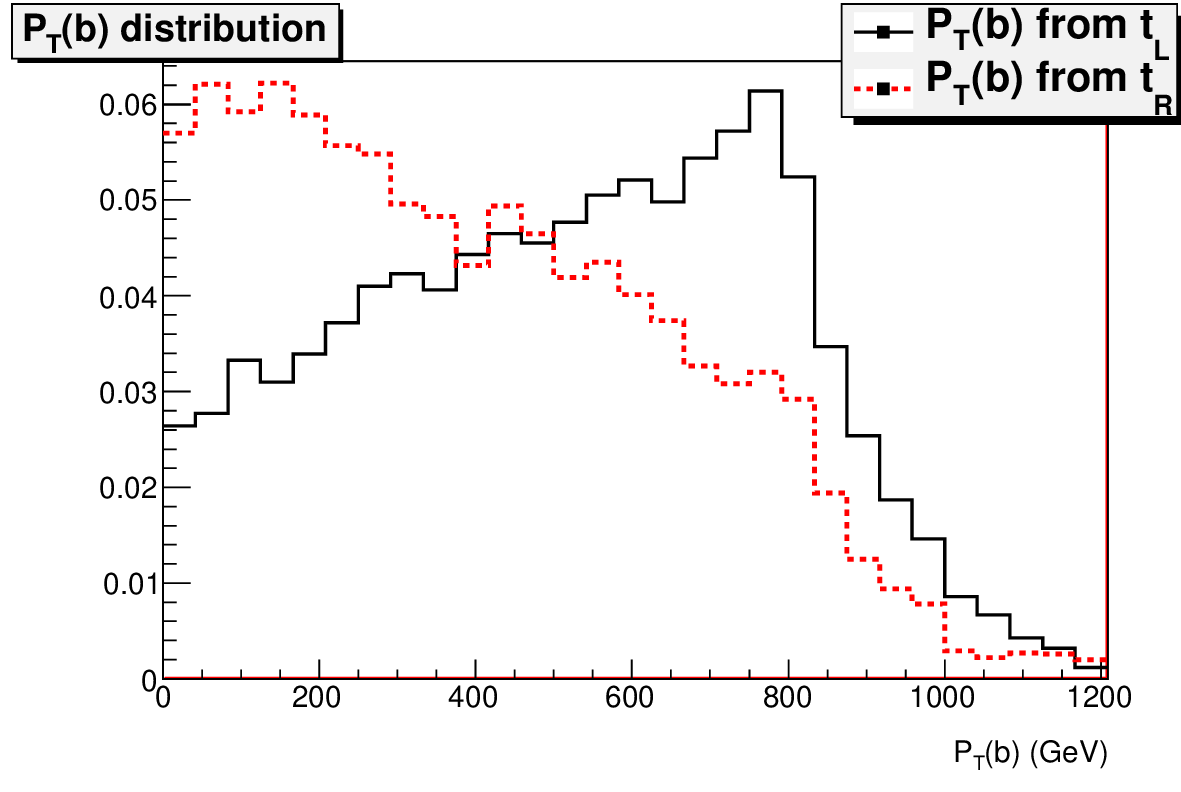} &
\includegraphics[width=.48\hsize]{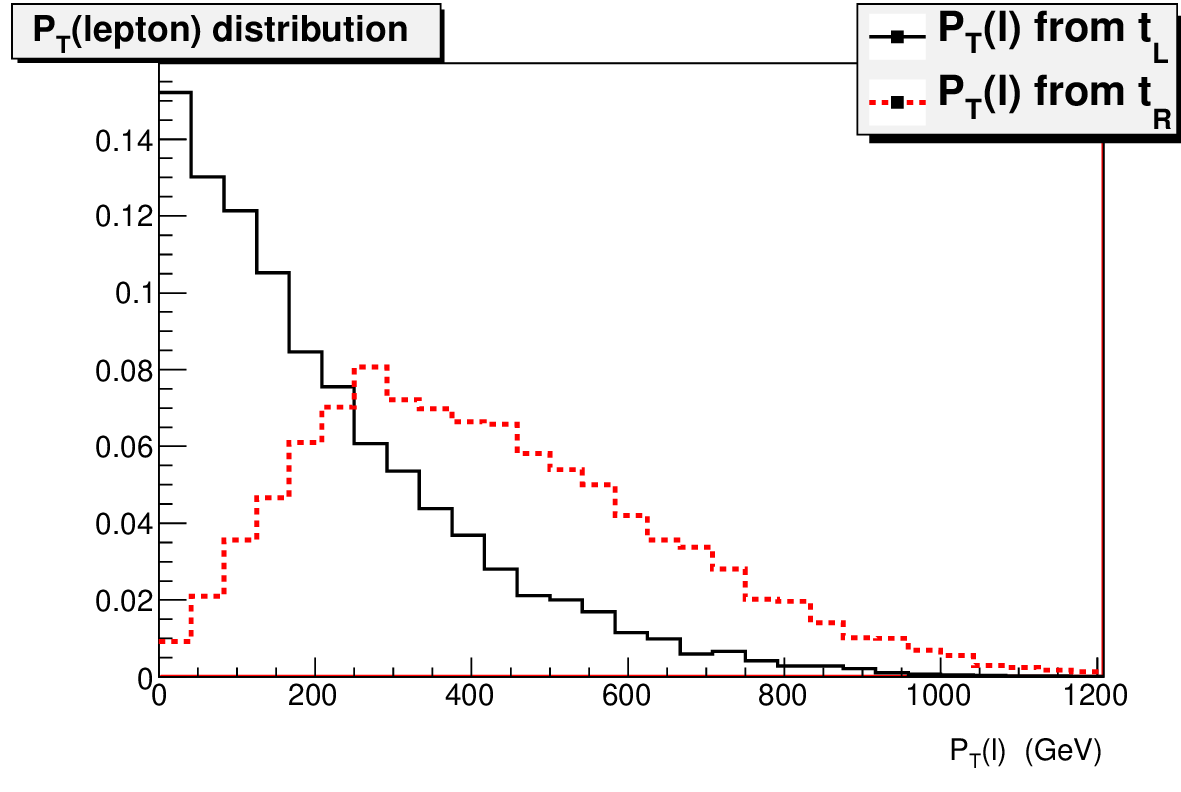} \\ 
\end{tabular}
\caption{In the plot on the left, we show a comparison of the $p_T$ distribution of $b$ quark from $t_L$ (the solid curve) vs. $t_R$ (the dashed curve). 
In the plot on the right we show the $p_T$ distribution of the charged lepton from $t_L$ (the solid curve) vs. $t_R$ (the dashed curve). 
We have imposed a lower cut, $p_T^{min}=1000$~GeV.
} \label{p:b_pt_comparison}
\end{figure} 

In this section, we consider how to exploit $b$-quarks to measure the polarization of highly boosted hadronic tops. 
Various new physics models have particle spectra which couple preferentially to one chirality, giving rise to parity violation.
Since chirality equals helicity for ultra-relativistic fermions, highly boosted top quarks can help us probe parity violation in the bottom/top quark sector.
As is well-known, the top quark decays before the hadronization process occurs,
and measurement of the top quark polarization from its leptonic daughters has been studied ~\cite{Agashe:2006hk,Fitzpatrick:2007qr}.
We propose using the transverse momentum of the $b$-quark, inferred from the $b$-tagged jet, to perform similar measurements.
The $\pt$ distribution for the $b$-quark depends on the chirality of the top-quark.
The $b$-tagged jets should, therefore, also act as good spin analyzers.
In Fig.~\ref{p:b_pt_comparison}, we compare the $\pt$ distributions for leptons and $b$-quarks, for both left- and right-handed top quarks.

As mentioned earlier, the issue of $b$-tagging at high $\pt$ is quite challenging at this time (for recent studies see~\cite{BTagging}),
and a fully quantitative study is not yet available.
The main idea is to examine the $\pt$ distribution of $b$-tagged jets, in events where we believe these jets originate from $t \to b\,W$.
In order to measure the $p_T$ of the $b$ quark, we need to require at least one of the top-jets should be resolved into more than two jets, since we cannot measure the $p_T$ of the $b$ quark inside a single top-jet. As shown in Fig.~\ref{p:collimation_rate}, even for high $p_T$ ($p_T\ge 1000 \gev $) top jet, with cone size $R=0.4$, $\sim30\%$ of top-jets can be resolved into more than two jets.
By fixing the cone size for jet reconstruction, it is important to understand any biases towards right-handed or left-handed top quarks.
Bottom quarks from $t_L$ have a harder $\pt$ distribution than those from $t_R$,
while the opposite is true for leptons from leptonic top quark decays.
If one uses a small reconstruction cone, the efficiencies for jet mass reconstruction between the $t_L$ and $t_R$ may differ.
We found a negligible bias using cone jets with $R = 0.4$.

\begin{figure}[fhptb]
\begin{center}
\includegraphics[width=.74\hsize]{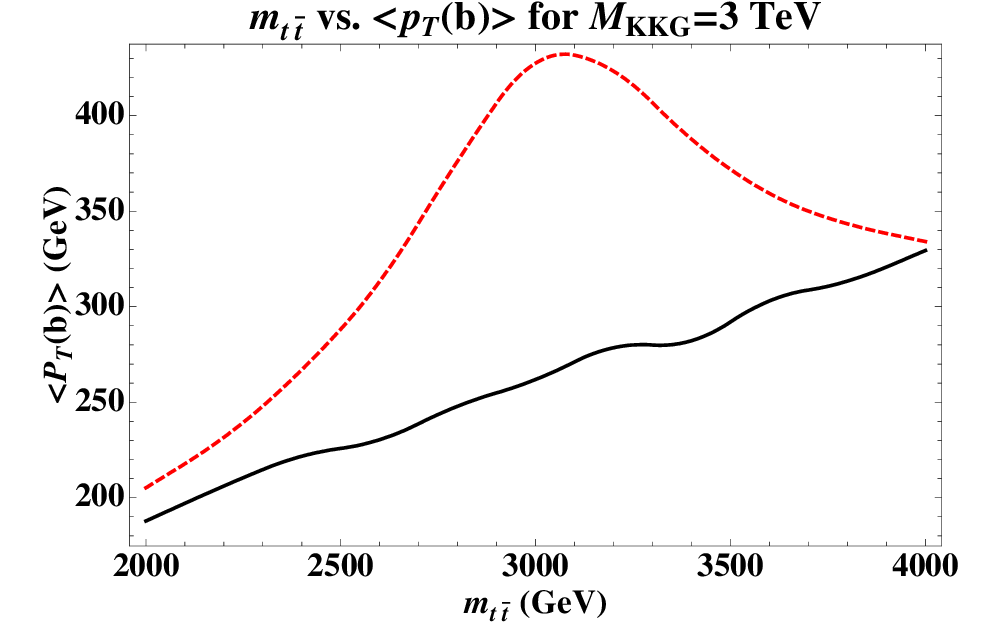}
\end{center}
\caption{We compare the $ \langle P_T \rangle $ distribution of the $b$ quark,
as predicted by the Standard Model (black solid curve) and by RS1 model with SM fields propagating in the bulk (red dashed curve). 
% and RScombination of SM top and RS top from KK gluon for $m_{KKG}=3 \tev$. 
} 
\label{p:bpt_in_rs}
\end{figure} 

\begin{figure}[fhptb]
\begin{center}
\includegraphics[width=.74\hsize]{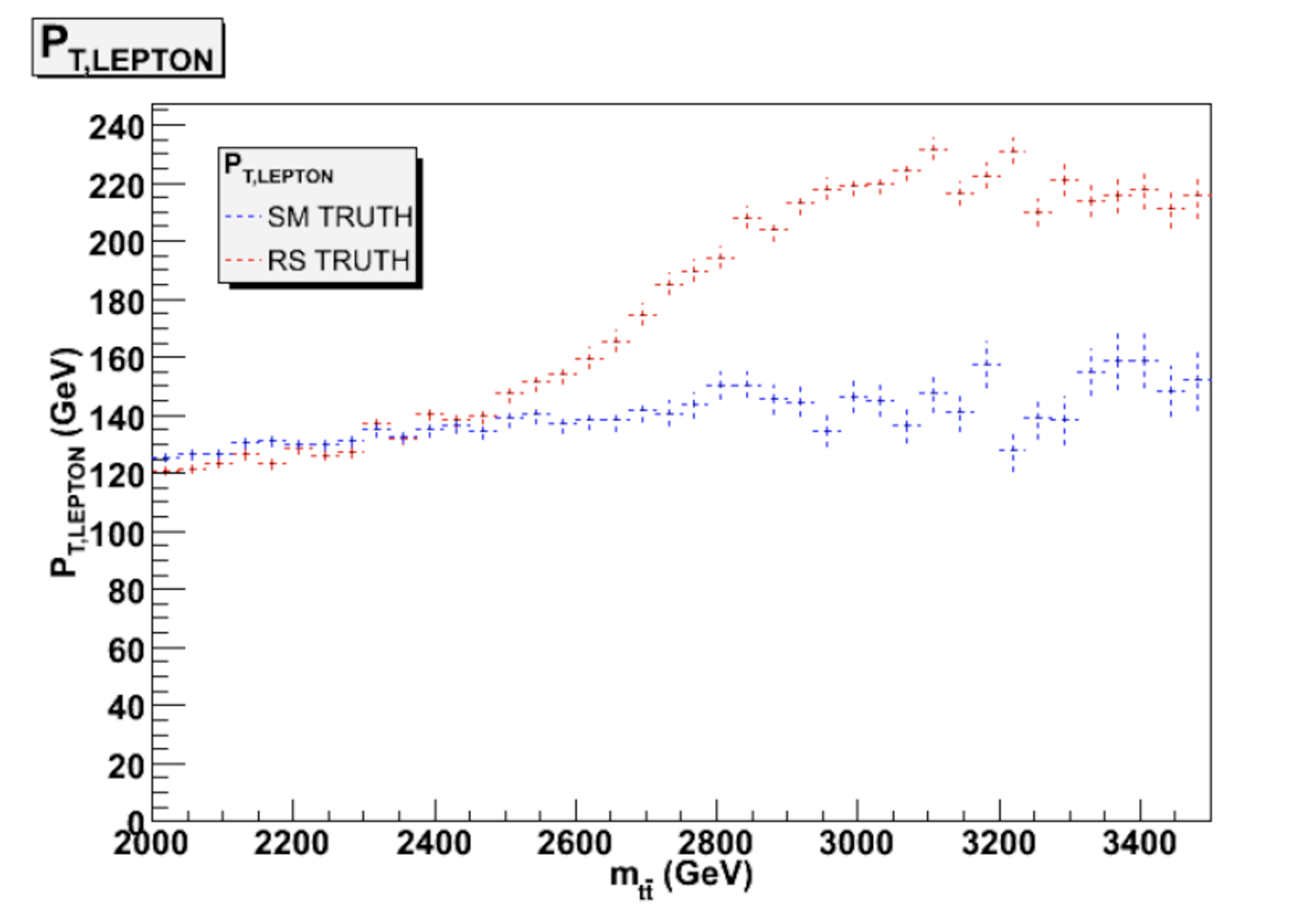}
\end{center}
\caption{
We compare the $ \langle p_T \rangle $ distributions of the lepton,
as predicted by the Standard Model (blue curve) and by RS1 model with SM fields propagating in the bulk (red curve). 
} 
\label{p:lpt_in_rs}
\end{figure}

We can develop this discussion further with an example, namely the Randall Sundrum (RS)~\cite{rs1} model with the SM fields propagating in the bulk.
We consider the case where the first Kaluza-Klein (KK) excitation of the gluon has a mass $M_{KKG} = 3 \, \tev$.
We perform this analysis at partonic level.
In the model we are considering for using $b$-quark $ \langle \pt \rangle $,
KK excitations of the gluon couple to left-handed top quarks $\sim5\times$ stronger than to right-handed top quarks.
Typical cross sections for KK gluon production are relatively small.
The (background) SM top quarks are produced dominantly via parity-invariant QCD processes, and tend to wash out the signal.
In order to resolve the signature, we are compelled to correlate deviations in the $b$-quark $ \langle \pt \rangle$ spectrum to an excess in KK gluon production.
In Fig.~\ref{p:bpt_in_rs}, we compare the mean value of the $b$-quark $\pt$ spectrum, for the Standard Model and RS1 scenarios with SM fields propagating in the bulk.
When correlated to the invariant mass of the KK gluon, we see a substantial deviation in the distribution of the $b$-quark $ \langle \pt \rangle $.
In Fig.~\ref{p:lpt_in_rs}, we compare the mean value of the lepton $\pt$ spectrum, for the Standard Model and RS1 scenarios with SM fields propagating in the bulk, where
KK excitations of the gluon couple to right-handed top quarks $\sim5\times$ stronger than to left-handed top quarks.

%%%%%%%%%%%%%%%%%%%%%%%%%%%%%%%%%%%%%%%%%%%%%%%%%%
%%%%%%%%%%%%%%%%%%%%%%%%%%%%%%%%%%%%%%%%%%%%%%%%%%
\section{Conclusions}\label{section_conclusions}

In this study we have mainly focused on high $\pt$, hadronically
decaying, tops in cases where they are fully collimated.
Above $\pt$ of 1~TeV the majority of the top daughters particles will
be found inside a single cone jet even when the cone size is as small
as $R=0.4$.
Therefore, they are simply denoted as top-jets.
The leading background for top-jets comes from high $\pt$ QCD jets.
We provided analytic expressions for the QCD jet functions which
approximate the background and show consistency with MC data.
As an example we consider the case of SM $t \bar t $ production,
and demonstrate how these jet functions, via side band analysis, allow
us to efficiently resolve $1 \, \mbox{TeV}$ top-jet from the QCD
background with $25 \, \mbox{fb}^{-1}$,
and $\sim 1.5 \, \mbox{TeV}$ top-jets with $100 \, \mbox{fb}^{-1}$.

A wide class of new physics models posits $\ttbar$ production
mechanisms which would significantly contribute to the mass
distributions,
possibly allowing resolution of excess production with less data.
To further improve the significance we consider jet shapes
(recently analyzed in~\cite{us} and also in~\cite{Thaler:2008ju}),
which resolve substructure
of energy flow inside cone jets.
Augmentation of the analysis, such as the use of jet substructure in
combination with a jet mass cut and $b$-tagging, may improve the
signal resolution,
allowing us to discover NP signal through top quark channel even with
lower luminosity or higher $p_T$ cut.
We provided such an example using the {\it{planar flow}} jet shape
variable, and a detailed analysis is presented in our recent
work~\cite{us}.
We also proposed using the transverse momentum of the bottom quarks to
measure top quark polarization as a probe of parity violation.

In this paper we mostly focused on fairly  extreme  (but not uncommon at the LHC) 
kinematical configurations where the tops are fully
collimated.
This has several advantages such as having direct contact with theoretical based calculation of the jet functions 
and also the ability to consider arbitrarily high top momenta (at least in principle).
However, it is clear that some fraction of the hadronic tops will be reconstructed in 2-jet (intermediate) or $\ge 3$-jet (conventional) topologies.
The fraction of events related to the different topologies is a function of the cone size and $\pt$.
Solid reconstruction algorithms and analyses must be flexible enough to interpolate between these different regions.
% which we refer to as jet-like (single top-jet), intermediate (2-jet) and conventional ($\ge 3$ jets ).
We note that our approach is complimentary to others that have been proposed
recently~\cite{LHCnotes,Butterworth:2002tt,Kaplan:2008ie,Thaler:2008ju}.
In most cases, the difference is related to the fact that the tops
considered are not fully collimated and a two-jet topology is
exploited to increase the signal to background ratio.
It would be very interesting and important to derive theoretically based techniques
to control the corresponding distribution of the background relevant to the intermediate region.
It is likely that there are overlaps between the different regions. 
Such issues are important to examine in detail. 
Mastering these complimentary methods may help to make potential new physics observations more robust, 
if verified via multiple and independent techniques.

Finally, we want to emphasize that the analysis proposed herein is also
applicable to other processes involving, highly boosted, heavy particles such
as electroweak gauge bosons, the Higgs and other new physics particles, to which QCD is a
leading background  as well.

%%%%%%%%%%%%%%%%%%%%%%%%%%%%%%%%%%%%%%%%%%%%%%%%%%%%%%%%%%%%%%%%%%%%%%%%%%%%%%%

\section*{Acknowledgments}

We especially thank George Sterman for suggestions, encouragement and comments on the manuscript.
Also, we thank Kaustubh Agashe, Johan Alwall, Gustaaf Brooijmans, Amanda Deisher, H.C. Fang, Shrihari Gopalakrishna, Michel Herquet, Ian Hinchliffe, Chung Kao, 
Fabio Maltoni, Konstantin Matchev, Patrick Meade, Johannes Muelmenstaedt, Frank Paige, Matthew Reece, Gavin Salam, 
Marjorie Shapiro, Jack Smith, Gregory Soyez, Iain Stewart, Jesse Thaler, Laurent Vacavant and Jan Winter for useful discussions. 
We appreciate the efforts of Jan Winter to provide customized Sherpa source code.
We also thank Johan Alwall to clarify the uncertainty involved in the matching procedure.
The work of L.A., S.L., G.P.\, and I.S.\ was supported by the National Science Foundation,
grants PHY-0354776, PHY-0354822, PHY-0653342 and PHY-06353354.
The work of J.V.\ was supported by the Director,
Office of Science, Office of High Energy Physics,
of the U.S. Department of Energy under Contract No.
DE-AC02-05CH11231.
S.L.\ and G.P.\ thank the hospitality of the theoretical physics group of Boston and Harvard universities where part of this work was done.
G.P.\ thanks the Aspen center for physics for hospitality while working on the project. S.L.\ and G.P.\ thank KITP (Santa Barbara) for hospitality while working on the project.

%%%%%%%%%%%%%%

\section*{Appendix}
\setcounter{section}{0}
\renewcommand{\thesection}{\Alph{section}}

%%%%%%%%%%%%%%%%%%%%%%%%%%%%
\section{Jets at Fixed Invariant Mass\label{sec:Jetfunction} } 
\renewcommand{\theequation}{A.\arabic{equation}}
\setcounter{equation}{0}
%%%%%%%%%%%%%%%%%%%%%%%%%%%%

Here we give details of the definitions and calculations for the jet functions that we employ in section 3. Single inclusive Jet cross sections have been studied intensively~\cite{Ellis:2007ib,Furman:1981kf,Aversa:1988vb, Kidonakis:2000gi}. Here, we are interested in computing the QCD background to jets of measured mass. The main background to the production of $t \bar{t}$ pairs is from dijet production from hadronic collisions,
 \beq
 H_a(p_a) +H_b(p_b) \to J_1(m_{J_1}^2,p_{1,T},\h_1,R) +J_2(m_{J_2}^2,p_{2,T},\h_2,R) +X,
 \eeq
\noindent
where the final states are jets in the directions of the outgoing partons, each with a fixed jet mass $m_J^2$ , a ``cone size" $ R^2 = \D \h^2 + \D \f^2$, and transverse momenta, $p_{i,T}$. For simplicity we choose the cone sizes equal for the two jets, although they can be different. For $R < 1$, we can isolate the leading $(R^0)$  dependence of such cross-sections in factorized ``jet'' functions, 
\beeq
\frac{d \s_{ H_A H_B \to J_1 J_2}}{dp_T d m_{J_1}^2 d m_{J_2}^2 d \h_1 d \h_2 } &=&  \sum_{abcd} \int  d x_a \, d x_b \, \f_a(x_a) \, \f_b(x_b) H_{ab\to cd} \left(x_a,x_b,p_T,\h_1, \h_2,\as (p_T) \right)  \nn \\
&&
 \times J_{1}^{c}(m_{J_1}^2,p_T \cosh \h_1,R,\as(p_T))  \; J_{2}^{d} (m_{J_2}^2,p_T \cosh \h_2, R, \as(p_T)),   \nn \\  
\label{Hfac} 
\eeeq
\noindent
with corrections that vanish as powers of $R$.
Here the $\f $'s are parton distribution functions for the initial hadrons,  $H_{ab\to cd} $ is a perturbative  $2 \to 2$ QCD hard-scattering function, equal to the dijet Born cross section at lowest order, and the $J_{i}$ are jet functions, which are defined below. Jet function $J_i$ summarizes the formation of a set of final state particles with fixed invariant mass  and momenta collinear to the $i^{\mbox{th}}$ outgoing parton.  Corrections to the cross section of order $R^0$ can only occur through collinear enhancements which factorize into these functions~\cite{Kidonakis:1998ur}.

%Since the partonic cross-section is only $2 \to 2$, the dependence in $p_T= \frac{ \sqrt{s} }{ 2 \cosh \h} $ is trivial.
%Initial State Radiation may contribute to the jet, but these contribution are sub-leading in $R^2$.

Following Ref.~\cite{Berger:2003iw}  we define jet function for quarks at fixed jet mass by
\beeq
J^{q}_i ( m_J^2, p_{0,J_i}, R)  &=& \frac{(2 \pi)^3}{2\sqrt{2} \, (p_{0,J_i})^2}  \frac{ \x_{\mu} }{N_c} \sum_{N_{J_i}} \mbox{Tr} \left\{ \g^\m \la 0 | q(0)  \F^{(\bar{q}) \dagger}_\x ( \infty,0 )  |N_{J_i}\ra \la N_{J_i}| \F^{(\bar{q})}_\x( \infty,0 ) \bar{q}(0) | 0\ra \right\} \nn \\
&& \times \d \left( m_J^2 - \tilde{m}_J^2( N_{J_i},R) \right) \d^{(2)} ( \hat{n} - \tilde{n}(N_{J_i})) \d( p_{0,J_i} - \w(N_{J_c}) ), \label{quarkjet1}
\eeeq
where $ \tilde{m}_J^2( N_{J_i},R)$ is the invariant mass of all particles within the cone centered on direction $\hat{n}$ in state $N_{J_i}$.
Correspondingly, gluon jet functions are defined by
\beeq
 J^{g}_i(  m_J^2, p_{0,J_i}, R)  &=& \frac{(2 \pi)^3}{2 (p_{0,J_i})^3}   \sum_{N_{J_i}}  \la 0 |  \x_\s F^{\s \nu}(0)  \F_{\x}^{(g) \dagger} \left(0,\infty\right)  | 
  N_{J_i} \ra \la  N_{J_i} |\F^{(g)}_\x \left(0,\infty\right)   F_{\nu}^{\,\rho}(0) \x_\rho | 0 \ra
\nn \\
&& \times\d \left( m_J^2 - \tilde{m}_J^2( N_{J_i},R) \right) \d^{(2)} ( \hat{n} - \tilde{n}(N_{J_i})) \d(  p_{0,J_i} - \w(N_{J_c}) ). \label{gluonjet}
\eeeq
These functions absorb collinear enhancements to the outgoing particles that emerge from the underlying hard perturbative process and fragment into the observed jets. The $\F$'s are path ordered exponentials (Wilson lines) defined by 
\beq
\F^{(f)}_\x (\infty,0;0)= \cp \left\{ e^{ -i g \int^{\infty}_0 d \h \, \x \cdot A^{(f)} ( \h\, \x^\m) }  \right\}, \label{wilson}
\eeq
where $\cp$ indicates ordering along the integral and where $\x$ is a direction with at least one component in the direction opposite to the jet.
The full hadronic cross-section is independent of the choice for $\x$. As indicated, the gauge field $A^{(f)}$ is a matrix in the representation of the generators for parton $f$. In general the jet function depends on $\vec{\x} \cdot \hat{n}$, but for simplicity we suppress this dependence below. Finally the jet functions in Eqs.~(\ref{quarkjet1}) and (\ref{gluonjet}) are normalized such that at lowest order we have 
\beq
J^{(0)}_i ( m_{J_i}^2, p_{0,J_i}, R) =\d ( m_{J_i}^2 ). 
\eeq
%In particular, our intent is to find the leading contributions for large $m_J^2$ and highly collimated jets, $R^2<1$.  Such leading contributions are given by the following factorization:

%%The leading logarithmic behavior can be found by taking a no recoil approximation.
%%The main result is that the leading logarithmic behaviour for the jet mass is given only by the Jet functions. At LL the hadronic cross-section factorizes as,
%\beeq
%\frac{d \s_{ H_A H_B \to J_1 J_2}}{d m_{J_1}^2 d \h } &=&\sum_{abcd} \int  \, \f_a\, \f_b \frac{d \hat{\s}_{ab\to cd} } { d p_T d \h} \left(p_T,\h,\as (p_T), \mu_F \right) \times \nn \\
%&&J_{1} (m_{J_1}^2,p_T,\h,R^2,\as(p_T))\, J_{2}(p_T,\h,R^2,\as(p_T))  + \co(R^2), \nn
%\eeeq
%where we suppressed the convolution and dependence in $x_a$ and $x_b$. Any contribution due to initial state radiation will be due to wide angle soft gluons, leading to subleading contributions in $R^2$. 
%At \NLO \,this factorization simplifies further to:
%\beeq
%\frac{d \s_{ H_A H_B \to J_1 J_2}}{d m_{J_1}^2 d \h } &=&\sum_{abcd} \int  \, \f_a\, \f_b \frac{d \hat{\s}_{ab\to cd} } { d p_T d \h} \left(p_T,\h,\as (p_T), \mu_F \right) 
%J_{1} (m_{J_1}^2,p_T,\h,R^2,\as(p_T)) + \co(R^2). \nn \\
%\eeeq
%The independence of the cross-section in the soft function, leads to an almost DY process, where the leading logs in the final state are given completely by a Sudakov like resummation.

\subsection{Jet Functions at Next-to-Leading Order}  

\begin{figure}[htbp]
\begin{center}
\includegraphics[width=.8\hsize]{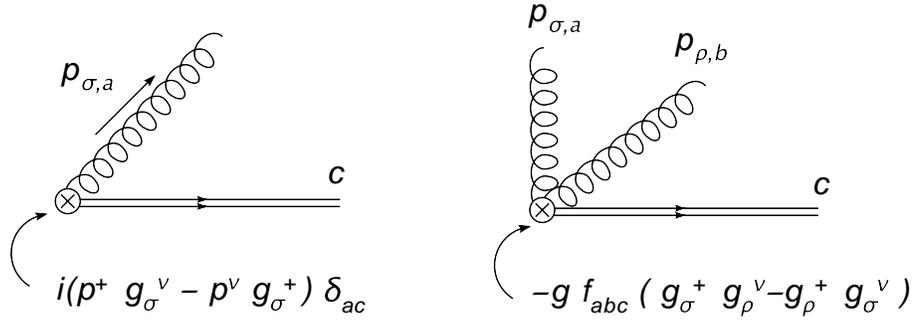} 	
\caption{Feynman rules associated with the $F^{+ \nu}$ operator at the end of a Wilson line. }
\label{gluerulz}
\end{center}
\end{figure}

\begin{figure}[htbp]
\begin{center}
\includegraphics[width=.5\hsize]{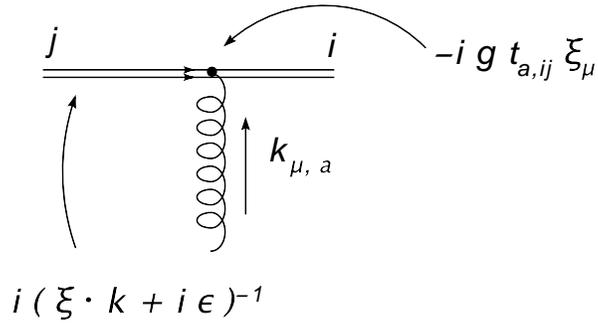} 	
\caption{Feynman rules associated with eikonal lines, from the expansion of the Wilson lines. }
\label{eikonal}
\end{center}
\end{figure}

\begin{figure}[htbp]
\begin{center}
\includegraphics[width=.8\hsize]{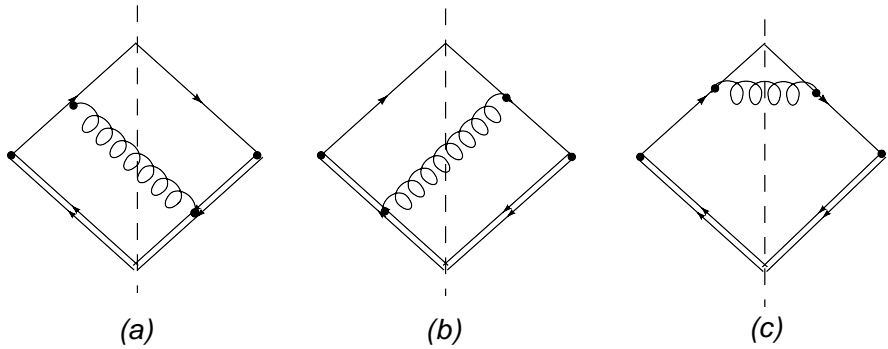} 	
\caption{Real contributions to the quark jet function at order $\as$.}
\label{quarkjet}
\end{center}
\end{figure}

\begin{figure}[tbp]
\begin{center}
\includegraphics[width=.8\hsize]{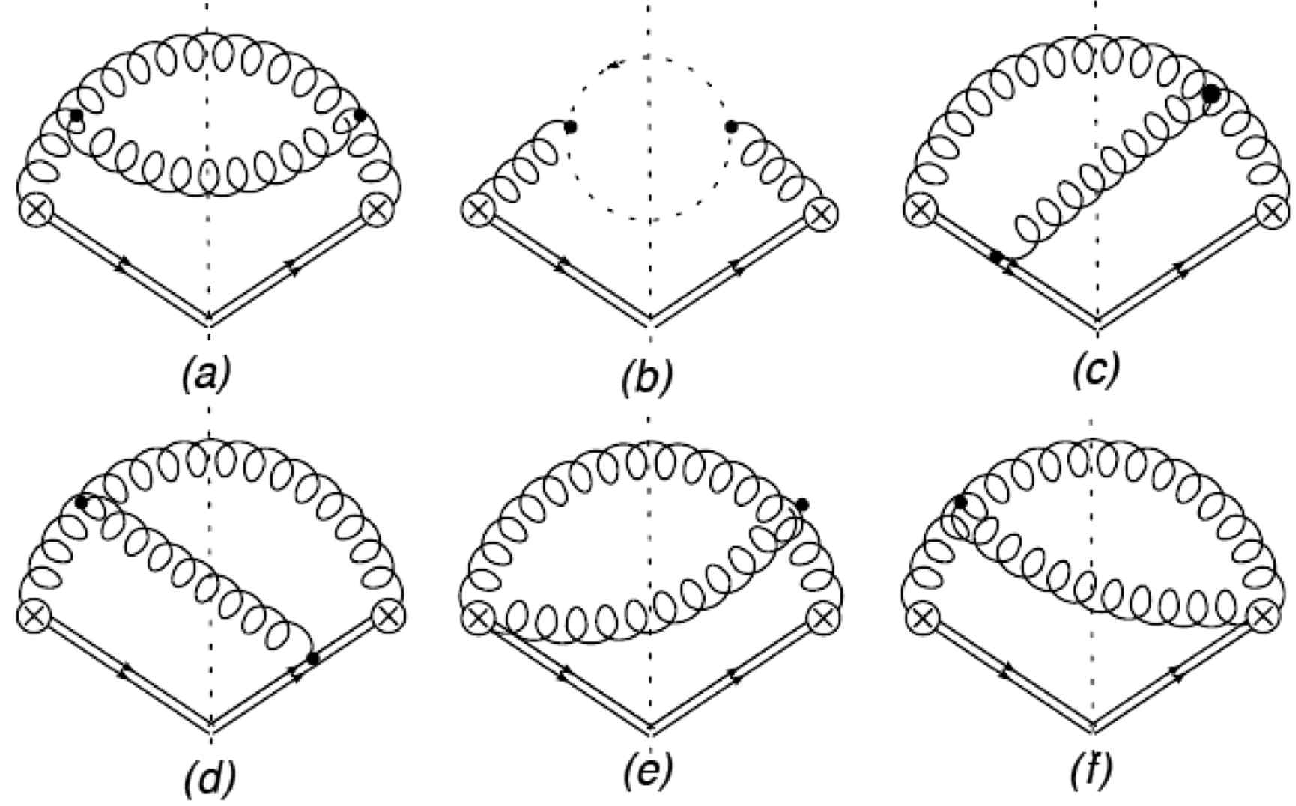} 	
\caption{Real non-vanishing contributions to the gluon jet function in Feynman gauge at NLO.}
\label{gluejet}
\end{center}
\end{figure}

At \NLO , contributions to the jet mass distributions for light quark or gluon jets have only two particles in their final states. 
For the quark jet we have the following matrix element which has to be calculated to $\co(g^2)$,
\beeq
J^{q}_i ( m_J^2,  p_{0,J_i}, R) &=& \frac{(2 \pi)^3}{2 (p_{0,J_i})^2}  \frac{ \x_\m }{N_c \sqrt{2} } \sum_{\s,\l} \int \frac{d^3 p}{ (2 \pi)^3 2 \w_p } \frac{d^3 k}{ (2 \pi)^3 2 \w_k }\Tr\left\{ \g^\m \la 0 |q(0)  \F^{(\bar{q}) \dagger}_\x ( \infty,0 ) |p,\s ; k,\l\ra  \right. \nn \\
&& \left.  \hspace{-24mm}  \times \la p,\s ; k,\l| \F^{(\bar{q})}_\x( \infty,0 ) \bar{q}(0) | 0\ra \right\} \,  
 \d \left( m_J^2 - (p+k)^2 \right)\, \d^{(2)} ( \hat{n} -\hat{n}_{\vec{p}+\vec{k}}) \, \d( p_{0,J_i}- p^0 - k^0), \nn \\
\eeeq
where $\s$ and $\l$ denote the polarizations , and $p$ and $k$ the momenta of the final-state quark and gluon respectively with $\hat{n}_{\vec{p}+\vec{k}} \equiv \frac{\vec{p}+\vec{k}}{|\vec{p}+\vec{k}|}$. Similarly, for the gluon jet we have
\beeq
J^{g}_i ( m_J^2, p_{0,J_i}, R)&=&  \frac{(2 \pi)^3}{4 (p_{0,J_i})^3}   \sum_{N_{J_i}} \int \frac{d^3 p}{ (2 \pi)^3 2 \w_p } \frac{d^3 k}{ (2 \pi)^3 2 \w_k } 
 \la 0 |   \x_\s F^{\s \nu}(0)  \F_{\x}^{(g) \dagger} \left(0,\infty\right)  |p,\s ; k,\l\ra \nn   \\
&&   \hspace{-22mm}  \times \la p,\s ; k,\l| \F^{(g)}_\x \left(0,\infty\right)   F_{\nu}{}^{\r}(0) \x_\r | 0 \ra
 \d \left( m_J^2 - (p+k)^2 \right)\, \d^{(2)} ( \hat{n} -\hat{n}_{\vec{p}+\vec{k}}) \, \d(p_{0,J_i}- p^0 - k^0), \nn \\
\eeeq
where $p$ and $k$ are the final state momenta within the cone size, $R$. To evaluate these matrix elements, we need the rules for vertices shown in Fig.~\ref{gluerulz} for the field strengths. The double lines represent  the perturbative expansion of the Wilson lines (\ref{wilson}) in the $\x$-direction
(see Eq.~(\ref{wilsonrule})), whose vertices and propagators are shown in Fig.~\ref{eikonal}. The resulting diagrammatic contributions to the quark and gluon jet functions at \NLO \, are shown in Fig.~\ref{quarkjet} and Fig.~\ref{gluejet} respectively.

%The result is dominated by the leading logarithmic correction, thus we simply use the approximation of no recoil, thus the final state quark retains the original direction of the outgoing quark from the hard process. The only diagrams that survive this approximation are $(a)$ and $(b)$, and the quark line becomes eikonal.

We choose a frame where the jet is in the $\eta_J=\f_J=0$ direction and the vector $\x$ is  light-like  and in a direction opposite to the jet,
 \beq
 p_{J_i} = p_{0,J_i} ( 1,\beta_i,0,0)  \;\;\;\;  \x = \frac{1}{\sqrt{2}} (1 ,-1,0,0)\, ,
\eeq
where $\beta_i=\sqrt{ 1 - m_{J_i}^2/p^2_{0,J_i} }$ is the velocity of the jet. In this frame we parametrize the momenta $p$ and $k$ above by
\beq
 p = p^0 ( 1,\cos \theta_p , \sin \theta_p,0)  \;\;\;\;  k= k^0 (1, \cos \theta_k , - \sin \theta_k,0)\, ,
 \eeq
where $\theta_{p,k}$ represents the angle of each particle to the jet axis $\hat{n}$. The path ordered exponentials are expanded order-by-order in $g_s$, related to the rules in Fig.~\ref{eikonal}  by the expansion,
\beeq
\F_\x (\infty,0;0)&=& \cp \left\{ e^{ -i g \int^{\infty}_0 d \h \, \x \cdot A ( \h\, \x^\m) }  \right\} \nn \\
&=& 1 -i g \int \frac{d^4k}{(2 \pi)^4} \frac{i}{ \x \cdot k + i\e} \x \cdot A (k ) + \ldots \;. \label{wilsonrule}
\eeeq

We begin with the calculation of the quark jet function, which readily reduces to an integral over the energy and angle of one of the particles, 
\beeq
J^{q(1)}_i ( m_J^2,p_{0,J_i}, R) &=&\frac{\beta_i}{  8 \sqrt{2} }\int  \frac{  d \cos \theta_k d k_0 k_0}{  \pi (p_{0,J}-k_0)} | \cm (p,k) |^2 \nn \\
&& \hspace{2mm} \times \delta (m_j^2 -2 k^0 p_{0,J} \left( 1 -  \beta_i \cos \theta_k \right) )  \Theta(R-\theta_k) \, ,  
\eeeq
where we choose $k$ to represent the gluon and $p$ the quark.
For $k$ the softer momentum, we easily see that $\theta_k \geq \theta_p$. Therefore, $p_0=k_0$ fixes the minimum angle for the softest particle, and we find $\cos (\theta_{S,min})=\beta_i$.
The region $\w_p<\w_k$ is found by simply interchanging $p$ and $k$ in $| \cm (p,k) |^2$ so that 

\beeq
J^{q(1)}_i ( m_J^2,p_{0,J_i}, R) &=&\frac{\beta_i}{ 16 \sqrt{2}  }\int_{\cos(R)}^{\beta_i}  \frac{  d \cos \theta_S}{ (2 \pi)^2} \frac{m^2_{J_i}/p^2_{0,J} }{\left(2 (1 - \beta_i \cos \theta_S) - \frac{m^2_{J_i}}{p^2_{0,J}}\right)} \frac{1}{ p_{0,J} ( 1 -\beta_i \cos \theta_S)}   \nn \\
&& \hspace{5mm} \times \left( | \cm_{q_i} (p,k) |^2+ |\cm_{q_i}(k,p)|^2\right)\, . 
 \eeeq
%%%%%%%%%%%%%%%%
%quark matrix elements
%
%\beeq
%| \cm_{q_i} (p,k) |^2+| \cm_{q_i} (k,p) |^2&=&  \as 4 \sqrt{2}   C_F \left( 
%\frac{z^2}{p_{0,J}} \frac{(1+\cos  \theta_S)^2}{(1-\b_i \cos  \theta_S)} \frac{1}{ \left( 2 (1+\b_i) (1-\b_i \cos  \theta_S) -z^2 (1+\cos  \theta_S) \right)} +\right. \nn \\
%&&\hspace{-30mm} \left. \frac{ 3 (1+\beta_i)}{p_{0,J} z^2} +
%\frac{1}{p_{0,J} z^4 } \frac{1}{(1+\cos  \theta_S) (1 -\b_i \cos  \theta_S)} \left( 2 (1+\b_i) (1-\b_i \cos  \theta_S) -z^2 (1+\cos  \theta_S) \right)^2 
% \right) \nn
%\eeeq
The evaluation of $ | \cm_{q_i} (p,k) |^2$  is straightforward from the diagrams of Fig.~\ref{quarkjet}, and we find 
\beeq
J^{q(1)}_i ( m_J^2,p_{0,J_i}, R) &=&\frac{ C_F \beta_i}{ 4 m^2_{J_i} }\int_{\cos(R)}^{\beta_i}  \frac{  d \cos \theta_S}{ \pi} \frac{ \as(k_0)\, z^4}{\left(2 (1 - \beta_i \cos \theta_S) - z^2\right) ( 1 -\beta_i \cos \theta_S)} \times \nn \\
&& \hspace{-20mm} \left\{
z^2\frac{(1+\cos  \theta_S)^2}{(1-\b_i \cos  \theta_S)} \frac{1}{ \left( 2 (1+\b_i) (1-\b_i \cos  \theta_S) -z^2 (1+\cos  \theta_S) \right)} + \right.\nn \\
&& \hspace{-20mm} \left. \frac{ 3 (1+\beta_i)}{z^2} +
\frac{1}{ z^4 } \frac{\left( 2 (1+\b_i) (1-\b_i \cos  \theta_S) -z^2 (1+\cos  \theta_S) \right)^2 }{(1+\cos  \theta_S) (1 -\b_i \cos  \theta_S)} 
 \right\}, \nn \\
  \label{eqn:jfq}
\eeeq
where $z= \frac{m_{J_i}}{p_{0,J_i}}$, $p_{0,J_i}=\sqrt{ m_{J_i}^2+p_{T}^2}$, and $k_0 =\frac{p_{0,J_i}}{2} \frac{z^2}{1 - \beta_i \cos \theta_S}$. 

The calculation of the gluon jet function proceeds along the same lines, with the exception that both particles in the final states are now identical, and the presence of the field strengths, which appear at the end of each Wilson line. The rules for these vertices, as mentioned before, are shown in Fig.~\ref{gluerulz}. Once again, we can write the gluon jet function as an integral over the angle of the softer particle,
\beeq
J^{g(1)}_i ( m_J^2,p_{0,J_i}, R) &=&\frac{\beta_i}{ 16 m_{J_i}^2 }\int_{\cos(R)}^{\beta_i}  \frac{  d \cos \theta_S}{(2 \pi)^2 p^2_{0,J_i}} \frac{z^2}{\left(2 (1 - \beta_i \cos \theta_S) - z^2 \right) ( 1 -\beta_i \cos \theta_S)}    |\cm_{g_i} (p,k)|^2 \, , \nn \\
\eeeq
where $|\cm_{g_i} (p,k)|^2 $ is symmetric under the interchange of $p$ and $k$. We find  from the diagrams shown in Fig.~\ref{gluejet}, the result
\beeq
J^{g(1)}_i ( m_J^2, p_{0,J_i}, R) &=& \frac{C_A \beta_i}{16 m_{J_i}^2}\int_{\cos(R)}^{\beta_i}  \frac{  d \cos \theta_S}{\p}\frac{\as(k_0) }{(1-\beta \cos \theta_S)^2 (1- \cos^2 \theta_S )( 2(1+\beta)-z^2)} \nn \\
&& \hspace{-25mm} \times \big( z^4 (1+\cos \theta_S)^2+z^2 (1-\cos^2 \theta_S) (2 (1+\beta_i)-z^2) + (1-\cos \theta_S)^2 (2(1+\beta_i)-z^2)^2 \big)^2. \nn \\  
\label{eqn:jfg}
\eeeq
%\beeq
%|\cm_{g_i} (p,k)|^2&=&  \as ( 4 \pi) C_A  \frac{p^2_{0,J_i}}{z^2} \frac{1}{(1-\beta \cos \theta_S)^2 (1- \cos^2 \theta_S )( 2(1+\beta)-z^2)} \nn \\
%&& \hspace{-15mm} \times \big( z^4 (1+\cos \theta_S)^2+z^2 (1-\cos^2 \theta_S) (2 (1+\beta_i)-z^2) + (1-\cos \theta_S)^2 (2(1+\beta_i)-z^2)^2 \big)^2. \nn \\ 
%\eeeq
These one-loop expressions have been used to generate the comparisons to event generator output given in Section 3.

\section{R-dependence}

It is of interest to isolate the leading logarithmic contributions in both gluon and quark jets, which can be found from eikonal graphs in the adjoint and fundamental representations respectively,
\beeq
J^{(eik),c} ( m_{J_1}^2, p_T, R) &=&  \frac{2 \,C_c}{ \sqrt{2} p_T} g^2 \int \frac{d^3 k}{ (2 \pi)^3 2 \w_k } \frac{\x \cdot p_J}{ \x \cdot k } \,   \,   \frac{\x \cdot p_J \,}{ 2 p_J \cdot k}  \nn \\
&& \hspace{-10mm} \times \d \left( m_{J_1}^2 - (p_1+k)^2 \right)  \Theta \left ( p_T -k_T\right) \, .
\eeeq
Parametrizing $k$ as
\beq
 k =k_T \left(\cosh \h_k, \cos \f_k, \sin \f_k, \sinh \h_k\right) \, , \label{ketaphi}
 \eeq
this leads to
\beeq
J^{(eik),c} ( m_{J_1}^2, p_T, R) &=&   g^2 \frac{\,C_c}{  (2 \pi)^3} \int d k_T \, k_T  \; d \f_k d \h_k \,      \frac{1}{  k^2_T \,(  \cosh^2 \eta_k - \cos^2 \phi_k )}\nn \\ 
&& \times \d\left(2 p_T \,k_T \,(  \cosh \eta_k - \cos \phi_k )-m_{J_1}^2\right) \Theta \left ( p_T -k_T\right). \nn \\
&=&g^2 \frac{C_c}{  (2 \pi)^3} \int  d \f_k d \h_k \,    \frac{1}{m_{J_1}^2}  \frac{1}{  k^2_T \,(  \cosh^2 \eta_k - \cos^2 \phi_k )}\nn \\ 
&& \times \d\left(2 p_T \,k_T \,r-m_{J_1}^2\right)  \Theta \left ( \cosh \h_k - \cos \f_k - \frac{m_{J_1}^2}{p_T^2} \right) \, .
\eeeq
In this expression we can change the variables to 
\beq
\h_k = r \cos \theta, \;\;\;\; \f_k = r \sin \theta \, .
\eeq
Since we are dealing with highly collimated jets we can expand the integrand in $r$ and integrate over $\theta$, finding
\beeq
J^{(eik),c} ( m_{J_1}^2, p_T, R) &\simeq&   g^2 \frac{2 C_c}{  (2 \pi)^2} \int_{m_{J}/p_T}^R d r \,     \frac{1}{m_{J_1}^2}  \left\{  \frac{1}{ r} + \co(r^3) \right\} \nn  \\
&\simeq& \as(p_T)    \frac{ \, C_c \, }{\, m_{J_1}^2 \pi} \left\{ \log \left( \frac{R^2 \, p^2_T}{m^2_J} \right)  + \co(R^4) \right\} ,
\eeeq
which shows explicitly the logarithmic behavior in $R$. Leading logarithmic contributions can be exponentiated, giving us a qualitative description of lower jet masses,
\beq
J^{(eik),c} ( m_{J_1}^2, p_T, R) \simeq \frac{\as}{  \pi} C_c \frac{1}{m_J^2} \log \left( \frac{ R^2 p_T^2}{m_J^2}\right) \exp \left\{ -\frac{\as}{2 \pi} C_c \log^2 \left( \frac{R^2 p_T^2 }{m_J^2} \right) \right\}.
\eeq
Without the above approximations, the eikonal jet function is given by
\beeq
J^{(eik),c} (m_J,p_T,R) = \as(p_T)  \frac{  4\,  C_c}{\pi m_J} \log\left( \frac{p_T}{m_J} \tan\left(\frac{R}{2}\right)   \sqrt{4-\left({m_J \over p_T}\right)^2} \right) \, .
\eeeq
As we have observed above, all $R^0$ behavior in the cross section can be found from the jet functions. We can also estimate the contribution of soft initial-state radiation on the cone-jet masses. Here we verify that such radiation is sub-leading in powers of  $R^2$. Contributions due to wide angle gluons come from a ``soft function''~\cite{Kidonakis:1998ur}, which is defined in terms of an eikonal cross section,
\beq
S (m^2_{J_i}) \sim \sum_{N_{s}} \s^{(eik)}(N_s) \d( m_{J_1}^2 -  \tilde{m}_J^2(N_s,R) ).
\eeq
\begin{figure}[t]
\begin{center}
\includegraphics[width=.7\hsize]{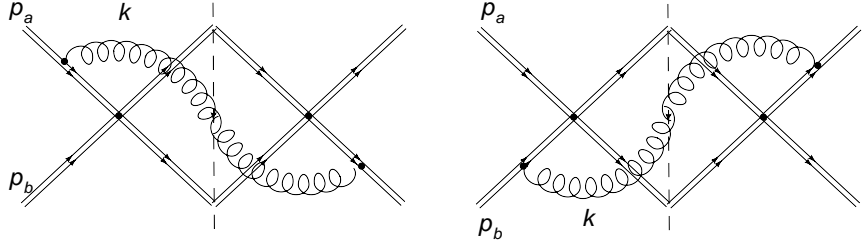}
\caption{Contributions to the jet mass from the soft function. }
\label{SoftFunction}
\end{center}
\end{figure}
Diagrams that can contribute to the jet mass are illustrated in Fig.~\ref{SoftFunction}. The initial state radiation shown behaves as 
\beeq
S \sim  \int d^4 k \d(k^2)  \frac{ p_a \cdot p_b}{ ( p_a \cdot k ) ( p_b \cdot k) } \d( m_{J_1}^2 -  2 p_1 \cdot k )  \Theta (R-\tilde{R}(\h_k,\f_k)) \, ,
\eeeq
with $p_a$ and $p_b$ the momenta of incoming partons, neither of which is in the direction of the observed jets. Choosing a frame where the initial momenta are given by
 \beeq
 p_a &=& \frac{ \sqrt{s} }{ 2} ( 1,0,0,1)\, , \;\;\;
 p_b = \frac{ \sqrt{s} }{ 2} ( 1,0,0,-1)\, , 
 \eeeq
and parametrizing the radiated gluon's momentum $k$ as in Eq.~(\ref{ketaphi}) above, we find
\beeq
S &\sim&  \int  d k_T d\phi_k d\h_k  \frac{ 1}{ k_T}\frac{1}{ 2 p_T (\cosh \eta_k - \cos \phi_k )}   \d\left( k_T - \frac{m_{J_1}^2}{  2 p_T (\cosh \eta_k - \cos \phi_k )}\right)  \nn \\
&\sim& \frac{2 \pi}{m_{J_1}^2} \int^R_0 d r \, r = \frac{ \pi R^2}{m_{J_1}^2} \, ,
\eeeq
which is, as expected, power-suppresed in $R$ compared to the logarithmic dependence we get from the jet function.

\end{document}